%% file: main.tex
\documentclass[12pt]{iopart}
\usepackage[utf8]{inputenc}
\usepackage{amssymb}
\usepackage{xcolor}%for changing text color
\usepackage{graphicx}% Include figure files
\usepackage{hyperref}
\usepackage{aasmacros}
\usepackage{floatrow}
\usepackage[sort&compress,numbers]{natbib}
\def\iso#1{$^{#1}$}
\def\msun{M$_\odot$}

\makeatletter
\long\def\@makefntext#1{\parindent 1em\noindent 
 \makebox[1em][l]{\footnotesize\rm$\m@th{\arabic{footnote}}$}%
 \footnotesize\rm #1}
\def\@makefnmark{\hbox{$\textsuperscript{{\arabic{footnote}}}\m@th$}}
\def\@thefnmark{\arabic{footnote}}
\makeatother

\begin{document}

\pagenumbering{roman}

\title[Direct Measurements for Stellar Burning]{The Status and Future of Direct Nuclear Reaction Measurements for Stellar Burning}
\author{M.~Aliotta\textsuperscript{1,2,3},
R.~Buompane\textsuperscript{4,5},
M.~Couder\textsuperscript{6},
A.~Couture\textsuperscript{7},
R.J.~deBoer\textsuperscript{6},
A.~Formicola\textsuperscript{8},
L.~Gialanella\textsuperscript{4,5},
J.~Glorius\textsuperscript{3},
G.~Imbriani\textsuperscript{9,10},
M.~Junker\textsuperscript{11},
C.~Langer\textsuperscript{3,12},
A.~Lennarz\textsuperscript{13},
Yu.A.~Litvinov\textsuperscript{3},
W.-P.~Liu\textsuperscript{14},
M.~Lugaro\textsuperscript{15,16,17},
C.~Matei\textsuperscript{18},
Z.~Meisel\textsuperscript{19,*},
L.~Piersanti\textsuperscript{20,21},
R.~Reifarth\textsuperscript{2,7},
D.~Robertson\textsuperscript{6},
A.~Simon\textsuperscript{6},
O.~Straniero\textsuperscript{11,21},
A.~Tumino\textsuperscript{22,23},
M.~Wiescher\textsuperscript{6},
Y.~Xu\textsuperscript{18}
}

\address{\textsuperscript{1}
SUPA, School of Physics and Astronomy, University of Edinburgh, EH9 3FD, UK} %1
\address{\textsuperscript{2} Goethe-Universit{\"a}t Frankfurt, Frankfurt am Main, Germany} %2
\address{\textsuperscript{3} GSI Helmholtzzentrum f{\"u}r Schwerionenforschung, Darmstadt, Germany} %3
\address{\textsuperscript{4} Dipartimento di Matematica e Fisica, Universit\`{a} degli Studi della Campania ``Luigi Vanvitelli", Viale Lincoln 5, 81100 Caserta, Italy}
\address{\textsuperscript{5} Istituto Nazionale di Fisica Nucleare, Sezione di Napoli, Complesso Universitario di Monte S. Angelo, Via Cintia ed. G, 80126 Naples, Italy}
\address{\textsuperscript{6} Department of Physics, The Joint Institute for Nuclear Astrophysics,  University of Notre Dame, Notre Dame, Indiana 46556 USA} %3
\address{\textsuperscript{7} Los Alamos National Laboratory, Los Alamos, NM 87545, USA.}
\address{\textsuperscript{8}Istituto Nazionale di Fisica Nucleare, Sezione di Roma, Piazzale A. Moro 2, 00185 Roma, Italy}
\address{\textsuperscript{9} Universit\`{a} degli Studi di Napoli ``Federico II", Dipartimento di Fisica E. Pancini, Via Cintia 21, 80126 Napoli}
\address{\textsuperscript{10} Istituto  Nazionale  di  Fisica  Nucleare,  Sezione  di  Napoli, Strada  Comunale  Cinthia,  80126  Napoli  NA,  Italy}
\address{\textsuperscript{11} Istituto Nazionale di Fisica Nucleare, Laboratori Nazionali del Gran Sasso (LNGS), Via G. Acitelli 22, 67100 L'Aquila -- Assergi, Italy} 
\address{\textsuperscript{12} University of Applied Sciences Aachen, 52066 Aachen, Germany} %3
\address{\textsuperscript{13} TRIUMF, 4004 Wesbrook Mall, Vancouver, British Columbia, Canada V6T 2A3}
\address{\textsuperscript{14} China Institute of Atomic Energy, P.O. Box 275(1), Beijing 102413, China}
\address{\textsuperscript{15} Konkoly Observatory, Research Centre for Astronomy and Earth Sciences, E\"otv\"os Lor\'and Research Network (ELKH), Konkoly Thege Mikl\'{o}s \'{u}t 15-17, H-1121 Budapest, Hungary} 
\address{\textsuperscript{16} ELTE E\"{o}tv\"{o}s Lor\'and University, Institute of Physics, Budapest 1117, P\'azm\'any P\'eter s\'et\'any 1/A, Hungary} 
\address{\textsuperscript{17} School of Physics and Astronomy, Monash University, VIC 3800, Australia} 
\address{\textsuperscript{18} Extreme Light Infrastructure - Nuclear Physics, Horia Hulubei National R$\&$D Institute for Physics and Nuclear Engineering, 077125 Magurele, Romania}
\address{\textsuperscript{19} Institute of Nuclear \& Particle Physics, Department of Physics \& Astronomy, Ohio University, Athens, Ohio 45701, USA} %2
\address{\textsuperscript{20} Istituto  Nazionale  di  Fisica  Nucleare, Sezione di Perugia, Via A. Pascoli snc, 06123 Perugia, Italy}
\address{\textsuperscript{21} Istituto Nazionale di Astrofisica, Osservatorio Astronomico d’Abruzzo, Via Mentore Maggini snc, 64100 Teramo, Italy}
\address{\textsuperscript{22} Istituto  Nazionale  di  Fisica  Nucleare, Laboratori Nazionali del Sud, Via S. Sofia 62, 95123 Catania, Italy}
\address{\textsuperscript{23} Facolt\`{a} di Ingegneria e Architettura, Universit\`{a} degli Studi di Enna ``Kore", Cittadella Universitaria, 94100, Enna, Italy}

\ead{\textsuperscript{*}meisel@ohio.edu}

\date{\today}

\begin{abstract}
    The study of stellar burning began just over 100 years ago. Nonetheless, we do not yet have a detailed picture of the nucleosynthesis within stars and how nucleosynthesis impacts stellar structure and the remnants of stellar evolution. Achieving this understanding will require precise direct measurements of the nuclear reactions involved. This report summarizes the status of direct measurements for stellar burning, focusing on developments of the last couple of decades, and offering a prospectus of near-future developments.
\end{abstract}

\newpage
\tableofcontents
\newpage
\title[Direct Measurements for Stellar Burning]{}
\vspace{-3.5cm}

\pagenumbering{arabic}
\setcounter{page}{1}

\section{Introduction}
\input{Introduction}

\section{Key Stages of Stellar Evolution and Relevant Reactions}
\label{sec:motivation}
\input{Intro-MassiveStars}

\input{HeBurning}
\input{SProcess}
\input{MassiveStars}

\section{Low-Background Measurements with Accelerators Deep Underground} \label{sec:DUA}
\input{LUNA}
\input{CASPAR}
\input{JUNA}

\section{Recoil Separators - Selectivity and Access to Reactions Involving Radioactive Nuclei}\label{sec:separators}

\input{intro-rings}

\input{ERNA}

\input{DRAGON}
\input{StGeorge}
\input{SECAR}

\section{Overcoming Beam Intensity Limitations with Storage Rings}
\input{intro-rings2}
\input{ESR}

\section{Neutron-Capture Reaction Studies with Neutron Beams}

\input{NeutronActivation}
\input{NeutronTOF}

\section{Photodisintegration Reactions Using \texorpdfstring{$\gamma$}{gamma}-ray Sources}
\input{Photon-Induced}

\section{Outlook}
\input{Conclusion}

\newpage
\addcontentsline{toc}{section}{Acknowledgements}
\section*{Acknowledgements}
\input{Acknowledgements}

\newpage
%the \newcommand line is needed to make latex happy with a natbib bibliography
\newcommand{\newblock}{}
\addcontentsline{toc}{section}{References}
\bibliographystyle{ieeetr}
\bibliography{References}

\end{document}

%% file: Introduction.tex
The seminal paper by Burbidge, Burbidge, Fowler, Hoyle (B$^2$FH)~\cite{RevModPhys.29.547} on the synthesis of the elements in stars, has led to the emergence of nuclear astrophysics as one of the most important disciplines in the field of nuclear physics. 
It is concerned with the nuclear engine driving stellar evolution and explosion and with a quest for determining the origin of the elements in our universe. 
New ideas and developments have emerged through this effort: the field of neutrino physics was born out of the desire to confirm the burning conditions of our sun; radioactive beam physics, the desire to understand nuclear structure and reactions near the limits of stability was driven by the need for mapping nucleosynthesis events in supernovae and other cataclysmic stellar environments. 
Today the field is driven by a wealth of new observational data, from telescopes measuring the entire electromagnetic spectrum~(e.g. Keck~\cite{2018SPIE10702E..07K}, SDSS~\cite{2000AJ....120.1579Y}, FAST~\cite{2011IJMPD..20..989N}, NICER~\cite{Gend12}),
%have IR, Optical,Radio,X)
 to the analysis of minuscule inclusions in meteoritic and lunar materials (with e.g. LION~\cite{Savi17}, CHILI~\cite{STEPHAN20161}), reflecting the abundances of condensed material formed in the stellar winds and eruptions. 
The present multi-messenger era including neutrino observations (with e.g. BOREXINO~\cite{2018Natur.562..505B}, Super-Kamiokande~\cite{ABE201639}) and the ringing of spacetime by gravitational waves (with e.g. LIGO and VIRGO~\cite{PhysRevX.9.031040}) provide a wealth of complementary observational signatures. 
These efforts are accompanied by new computational developments and capabilities modeling the dynamics of stellar evolution in 3-D over long periods of evolutionary epochs~\cite{10.1093/mnras/stx1962,10.1093/mnras/stz2979}, providing a better understanding of complex convection and mixing processes for nuclear fuel in stellar environments. 
Last but not least there have been a number of new developments and methods in attacking the experimental challenges associated with the understanding of the underlying nuclear reaction processes, several of which are highlighted in Sections~3-7.

Today the field manifests itself in three major directions. The goal of understanding nuclear processes far from stability is to provide  reliable interpretations of the light curves and production patterns of heavy elements observed in multi-generational stars or those associated with stellar explosions, namely through the rapid neutron-capture ($r$-)process \cite{Horowitz_2019, cowan2020origin}, the $p$-process \cite{Rapp_2006, ARNOULD20031} and the rapid proton-capture ($rp$)-process \cite{Woosley_2004, PARIKH2013225}. 
A new generation of radioactive beam facilities (e.g. FRIB~\cite{2018EPJWC.17801001S}, FAIR~\cite{2019RScI...90l3309B}, RIBF~\cite{2015aris.confc0135M}, ARIEL~\cite{2014AcPPB..45..503H}) has been developed to measure nuclear reaction and nuclear structure parameters that are important for improved simulations of the reaction paths far from stability mapping the observational results. 
The measurement of nuclear reactions on unstable nuclei is a challenge because of difficulties producing sufficiently intense radioactive ion beams at energies relevant for stars and stellar explosions.
Nonetheless, when intensities are limited, radioactive ion beams are useful for probing relevant nuclear quantities, such as masses and decay characteristics, towards the limits of stability.

The second direction is the study of neutron-induced reactions and reaction chains through which the production of heavy elements in our universe has been facilitated. It is not only the slow neutron-capture ($s$)-process~\cite{RevModPhys.83.157} and the $r$-process as it was initially argued but a multitude of different neutron-driven reaction chains associated with specific astrophysical environments and neutron sources of different intensities that are considered responsible for the build-up of the heavy elements. Only recently a clearer picture based on new observations of the particular signatures of these processes has emerged. The $i$-process is reflected in the abundance distribution associated with carbon-enhanced metal-poor (CEMP) stars~\cite{Aoki_2007}, while the weak $s$-process is part of the nucleosynthesis in the core of red giant stars and the main $s$-process is associated with intershell burning in asymptotic giant branch stars as discussed in section 2.1. While the main $r$-process now seems to be clearly identified to take place in neutron star mergers, a weak $r$-process~\cite{1992ApJ...395..202W} and the $n$-process~\cite{1976ApJ...209..846B} are associated with core collapse supernovae at the onset of the shock and the shock traversing the helium burning shell of the presupernova star. Each of these processes had its own particular neutron source and reaction path along or far away from the line of stability. The s-process path is being mapped at the n\_Tof/CERN and LANSCE/LANL neutron sources as well as a number of smaller facilities \cite{rei14}. Measuring neutron-capture reactions far off stability provides its own unique challenges since beam and target are both short-lived. New ideas and initiatives are emerging to address these challenges~\cite{rei17}. 

The third direction is the exploration of low energy nuclear processes associated with various evolutionary stages of stellar life. These are driven by charged particle interactions at very low energies due to the relatively low temperatures in the stellar environment. The cross sections are extremely low, in the sub-femto-barn range and rely strongly on the low energy extrapolation of the laboratory data. The detailed structure is characterized by the single particle and cluster configurations of compound and final nuclear stages as well as possible Coulomb and quantum effects near the threshold. To explore these configurations in the required detail and accuracy, direct reaction measurements are frequently complemented by indirect transfer or other reaction techniques. These depend on normalization to existing direct data and on a proper theoretical treatment of the Coulomb barrier and very low energy threshold effects for extrapolations. Reliable model predictions for the solar neutrino flux from the $pp$-chains and the CNO cycles~\cite{Serenelli2018} as well as for the nucleosynthesis patterns in late stellar evolution \cite{El_Eid_2004} and the seed distribution for subsequent explosive events require an uncertainty range of 10\% in the reaction rate predictions, a goal that has only been achieved in a very limited number of cases. There are numerous open questions associated with stellar nucleosynthesis, which require improved data and improved modeling to reliably address the underlying physics of stellar evolution. These data are characterized by the quantum dynamic physics at very low energies in the vicinity of particle thresholds, in particular in the case of two or more asymptotically closed channels at comparable excitation energies affecting each other. Little attention has been given to these kinds of phenomena, but they may affect the presently most urgent problems in the field.

The main questions for stellar hydrogen burning are reactions associated with the solar neutrino productions in our sun. 
This mainly includes processes that produce radioactive nuclei whose decay adds to the neutrino flux, such as $^7$Be$(p,\gamma)^8$Be for the $pp$-chains as well as $^{12}$C$(p,\gamma)^{13}$N and  $^{14}$N$(p,\gamma)^{15}$O for the CNO cycles. In particular the latter one is important for the first detection of solar $^{15}$O neutrinos by Borexino~\cite{agostini2020direct}.

Understanding of nucleosynthesis in first stars raises critical challenges. 
During their short lifetimes, the first generation stars successfully convert the primordial abundances produced in the Big Bang into light nuclei from the CNO range up to calcium~\cite{10.1093/mnras/staa3328}. 
The question remains what is the exact interplay between reaction and convective and mixing dynamics in such an early environment. 
Its interplay is critical for bridging the abundance gaps at mass numbers $A=5$ and $A=8$ and may also provide a solution for the still unsolved lithium abundance problem~\cite{doi:10.1146/annurev-nucl-102010-130445}. 
The connection between the CNO range to higher Z nuclei in the sd-shell may be facilitated through low energy quantum effects in the $^{19}$F$(p,\gamma)^{20}$Ne reaction that have not been taken into account in traditional $R$-matrix techniques~\cite{2021PhRvC.103e5815D}.

In terms of stellar helium burning, the $^{12}$C$(\alpha,\gamma)^{16}$O reaction still holds its fascination for the scientific community. 
It not only determines the ratio of carbon to oxygen in our universe, the third and fourth most abundant elements in the solar system and the two most critical elements for the evolution of life, but it also dictates the final stages of stellar life. These in turn set the the conditions for the different kinds of supernova explosions from type Ia, which serve as standard candles for our universe, to pair instability supernovae triggered in massive stars and the associated gap in the black hole mass distribution. 
While the $R$-matrix analysis of all experimental data on the $^{16}$O compound nucleus yields a reaction rate with only 25\% uncertainty in the stellar burning range~\cite{2017RvMP...89c5007D}, reverse engineering techniques from observational data suggest a slightly higher rate than predicted from the existing data from white dwarf seismology~\cite{Metcalfe_2002}. 
On the other hand, such deviations could also be caused by unaccounted mixing mechanisms in the late evolution of the star which could influence the white dwarf chemical composition~\cite{Straniero_2003}.

We still do not know reliably the neutron production rate for the various neutron induced nucleosynthesis processes such as the $s$-processes in red giant branch and asymptotic gian branch stars~\cite{RevModPhys.83.157}, the $i$-process in early stars~\cite{1977ApJ...212..149C}, and the $n$-process in the supernova shock front~\cite{1976ApJ...209..846B, 1978ApJ...222L..63T}. New efforts are underway to address the study of the complex reaction mechanisms leading to the release of the neutron flux necessary for the production of heavy elements parallel to explosive mechanisms such as the $r$-process in neutron star mergers or type II supernova environments. The uncertainties are in the interplay between $(\alpha,n)$ and $(\alpha,\gamma)$ reactions and the impact of very low energy cluster resonance or subthreshold structures and the associated interference effects~\cite{1994ApJ...437..396K}.

Finally, the question of sub-Coulomb fusion reactions still remains as an enormous challenge. There is the fundamental and yet unanswered question as to what drives the fusion cross section towards lower energies: is it just the Coulomb barrier or are there other phenomena such as the hindrance observed for fusion of heavier nuclei, where the incompressibility of nuclear matter causes a further reduction in the fusion probability~\cite{PhysRevC.73.014613, PhysRevC.76.035802}?

A further question is the nature of resonance phenomena as observed for $^{12}$C+$^{12}$C, $^{12}$C+$^{16}$O, and $^{16}$O+$^{16}$O, the three most important fusion reactions in massive stars; can these be explained by the traditional compound model as emergence of pronounced carbon or oxygen cluster configurations at high excitation energies near the fusion thresholds, or are these reflections of new quantum mechanical transition mechanisms which emerge with the matching of the wave functions in the initial and final quantum configurations~\cite{PhysRevC.97.055802}? 
We do not know and therefore any extrapolation towards low energies remains highly unreliable. 
This kind of uncertainty does not only affect our interpretation of late stellar evolution or the ignition of type Ia supernovae through carbon fusion, but it also expands to include our interpretation of pycnonuclear fusion of very neutron rich carbon to magnesium ions in the deep layers of the neutron star crust~\cite{BEARD2010541}. 

We do not know exactly the dynamical behavior of fusion in quantum systems near the threshold; that is most obvious in terms of low energy heavy-ion fusion, but may also play a role in low-energy alpha-capture reactions, in particular when several reaction channels are open such as in the $^{22}$Ne$(\alpha,n)$ versus the $^{22}$Ne$(\alpha,\gamma)$ reactions, which dictates the efficiency of neutron sources.
Quantum effects may even affect low-energy proton-capture processes. 
The community had focused on glaring examples such as electron screening modifying the Coulomb barrier by the negative charges of the surrounding electrons, but quantum effects that may emerge at the threshold have not really been considered or implemented in the nuclear reaction models.  
Other yet unaccounted variations in the Coulomb barrier or quantum effects in the fusion process near the threshold associated with the theoretical treatment of the nuclear potential may also introduce unexpected changes of the non-resonant reaction rate contributions \cite{TOKIEDA2020168005, 10.3389/fphy.2020.00008}.

In terms of resonant behavior in fusion reactions, besides yet unobserved compound states which may interfere with other reaction contributions at very low energies, the open entrance channel may strongly couple with some asymptotically closed channels near the reaction zone, influencing the reaction rates at near threshold energies. 
In the case of the $^{24}$Mg compound nucleus, the $^{12}$C+$^{12}$C collision represents the entrance channel. 
Possible asymptotically closed channels are: $^{23}$Na+$p$, $^{20}$Ne+$\alpha$, and even $^{23}$Mg+$n$. 
In the case of the $^{22}$Ne+$\alpha$ reactions, the asymptotically closed channels may be $^{21}$Ne+$n$ and $^{22}$Ne+$\gamma$. 
To evaluate the possible importance of such threshold effects new data are needed, not only at very low energies, but also over a wide energy range to map the various reaction components and contributions and to evaluate their behavior in the vicinity of the Coulomb barrier towards the threshold energy. 
This effort needs to be accompanied by a strong initiative in nuclear reaction theory to study the dynamics and evaluate conditions at extremely low energies~\cite{doi:10.1146/annurev-nucl-020620-063734}.

While several experimental and theoretical methods and techniques have been developed over the years to measure and model the reaction mechanisms towards stellar energies, there are significant shortfalls. 
Direct measurements are primarily pursued by deep underground accelerator measurements to reduce the cosmic-ray induced background in the detector materials. 
However, that still necessitates a substantial reduction in radiogenic backgrounds in parallel with measurements over extensive periods of time to acquire sufficient statistics. 
These techniques are complemented by inverse kinematics methods, which requires high resolution mass separators to separate the few reaction products from the intense primary beam particles with a sensitivity considerably better than required for radioactive beam experiments. 
The third approach, labeled as indirect techniques, seeks to populate the compound nucleus by transfer reactions or Coulomb-, electron-, or photon-disintegration processes to explore the nuclear structure near the threshold and deduce the reaction mechanism from these results. 
Here the question is primarily with the quality of the theoretical models and functions applied for reliably transforming the observed nuclear structure results into a low energy cross section. 
The predictions are only as good as the theoretical models utilized.
This means that the experimental effort needs to be complemented by improved reaction models, considering not only the standard reaction modes but also yet unexplored effects that alter the Hamiltonians near the threshold and may change the reaction cross section substantially. 
We have indications for such effects, but both experimental data as well as theoretical reaction models remain insufficient for reliable predictions.

In the following we provide a review of the field of nuclear astrophysics associated with the present questions and challenges in stellar burning, its challenges and the various experimental and theoretical tools the community is developing. 
Nuclear astrophysics with stable beams may be seen as a mature field, but it is far from understanding the intricacies of the reactions at stellar burning conditions. 
A clearer understanding of the Coulomb barrier at low energies, and the ambiguities of the quantum dynamical effects of quantum transitions at threshold energies, 
have been largely neglected until now. 
The goal here is to fully investigate these two aspects to yield a more reliable approach for predicting stellar reaction rates.

%% file: Intro-MassiveStars.tex
The chemical evolution of the Universe is governed by an intricate pattern of nuclear processes that take place in stars, both during quiescent evolution and explosive scenarios. 
The initial chemical composition and mass of a star govern which reactions in turn dominate the burning processes, thus affecting and regulating the star's evolutionary fate. In this Section we discuss particularly consequential stages of nuclear burning in stellar evolution and highlight a few of the important nuclear reactions that have seen substantial recent progress in the area of direct measurements.

All stars start their evolution by fusing primordial hydrogen into helium in a stage known as hydrogen burning. 
For stars with masses lower than about 1.5 solar masses, hydrogen burning proceeds through the proton-proton chain; for more massive stars, it proceeds through the CNO-cycle, provided that pre-existing CNO material is available in the star \cite{Ili2007}. Following a period of hydrogen burning in a spherical shell surrounding the core, helium burning ensues, producing a stellar core comprised of carbon and oxygen. This is accomplished via the reactions $3\alpha\rightarrow^{12}{\rm C}$ and $^{12}{\rm C}(\alpha,\gamma)^{16}{\rm O}$ . The former is a three-body process, meaning that it cannot be measured directly in the laboratory and is therefore not discussed further here (See Ref.~\cite{Nguy12} for a recent discussion). The latter is the focus of Section~\ref{sec:HeBurning}. The next nuclear burning stage involves a helium burning shell nested within a hydrogen burning shell and is beset by thermal instabilities that lead to significant neutron fluxes. These asymptotic giant branch stars and the $^{22}{\rm Ne}(\alpha,n)^{25}{\rm Mg}$ reaction, the source of neutrons during helium shell flashes, are the focus of Section~\ref{sec:sprocess}. For stars with an initial mass above 11~M$_{\odot}$, core burning moves to higher-$Z$ fuels in episodes known as C-, Ne-, O-, and Si- burning. Section~\ref{sec:massive-stars} focuses on massive star evolution, highlighting the importance of the $^{12}{\rm C}+^{12}$C reaction.

%% file: HeBurning.tex
\subsection{Helium burning and  the \texorpdfstring{$^{12}{\rm C}(\alpha,\gamma)^{16}{\rm O}$}{12C(a,g)16O} reaction
}
\label{sec:HeBurning}

When a low-mass star like the Sun runs out of hydrogen at its center, it becomes a red giant converting H to He via the CNO cycle in a shell surrounding the inert He core.
When the core temperature reaches 100 MK, the He nuclei in the core have sufficient kinetic energy to fuse together (helium burning) and form $^{12}$C, via the so-called triple-$\alpha$ process. Here,  two $\alpha$ particles ($^4$He nuclei) first combine to form unstable $^8$Be ($t_{1/2} \simeq 10^{-16}$~s). Equilibrium between the production of $^8$Be and its decay allows a second alpha capture reaction to occur, $^8$Be($\alpha$,$\gamma$)$^{12}$C, which is primarily facilitated through a pronounced alpha cluster state in $^{12}$C, the so-called Hoyle state.
Subsequent fusion of $^{12}$C with another He nucleus produces $^{16}$O  through the $^{12}$C($\alpha$,$\gamma$)$^{16}$O reaction.

Together with the 3$\alpha$-process, the $^{12}$C($\alpha,\gamma$)$^{16}$O reaction determines the $^{12}$C/$^{16}$O ratio at the end of helium burning. 
This in turn affects the onset of the next stages of stellar burning, that of carbon and oxygen fusion. 
Because of the cluster structure of $\alpha$-conjugate nuclei, it is at first somewhat surprising that helium burning does not continue via the $^{16}$O$(\alpha,\gamma)^{20}$Ne reaction. 
Yet, while there is a high probability of a cluster state close to threshold, in this rare case, no such state exits. 
Thus the star must turn to a more exotic reaction, although still one of an $\alpha$-cluster nature, the fusion of $^{12}$C with itself (see section \ref{sec:massive-stars}), as the next source of energy generation.

The $^{12}$C$(\alpha,\gamma)^{16}$O reaction almost presents the same problem as the $^{16}$O$(\alpha,\gamma)^{20}$Ne reaction, as there are no resonances close to the $\alpha$-particle threshold in $^{16}$O ($S_\alpha$~=~7.16~MeV). 
The closest resonance is the result of the broad 1$^-$ state in the $^{16}$O system at $E_x$~=~9.59~MeV. 
Instead, the low energy cross section is enhanced by two subthreshold states, a 2$^+$ at $E_x$~=~6.92~MeV and a 1$^-$ at $E_x$~=~7.12~MeV, together with their interference with higher energy resonances and direct capture. 
While the $\gamma$-ray decay of this reaction can proceed to any of the five bound states of $^{16}$O, the low energy portion of the cross section is dominated by decay to the ground state. 
One reason for this ground state transition dominance is that all five bound states in $^{16}$O decay with nearly 100\% probability directly to the ground state, therefore there is very little subthreshold enhancement for any of the so called ``cascade" transitions.

Greatly complicating the precise determination of the reaction rate is that the $^{12}$C$(\alpha,\gamma)^{16}$O cross section at the Gamow energy ($E_{\rm c.m.}\approx$~0.3~MeV) that corresponds to core helium burning temperatures ($T\approx$~0.2~GK) is $\approx$2$\times$10$^{-17}$ barn, well below the measurement sensitivity of any existing facility. Thus measurements are made at higher energies, and the cross section is extrapolated down. As the energy range of interest occurs very near to the $S$-factor interference minimum, the determination of the cross section depends on a precise modeling of the resonance and direct component interferences as well as possible weak background contributions from higher lying resonances. Because of its importance and rather insurmountable obstacles in its determination, Nobel laureate Willy Fowler was said to have dubbed this reaction the ``Holy Grail of nuclear astrophysics".

Since the modeling of the reaction cross section must include the reproduction of broad resonance interference, phenomenological $R$-matrix~\cite{1958RvMP...30..257L} has been the tool of choice for evaluating and extrapolating the experimental data. 
The use of such a model has proven crucial, as it provides a reaction framework where otherwise disjointed pieces of experimental information can be combined. 
This includes other compound nucleus reaction data such as $\alpha$-scattering cross sections on $^{12}$C and the $\beta$-delayed $\alpha$-emission spectrum of $^{16}$N as well as level parameters from transfer studies. 
The latter have provided stringent constraints on the properties of the subthreshold resonances in recent years

A lengthy review of this reaction has been recently provided by \cite{2017RvMP...89c5007D}, so the present work will be limited to recent developments since that publication. 
Despite the short amount of intervening time, several new investigations have been made and several more are planned or are already underway, emphasizing the consistent interest in this reaction.

\begin{figure} 
  \centering
\includegraphics[width=1.0\columnwidth]{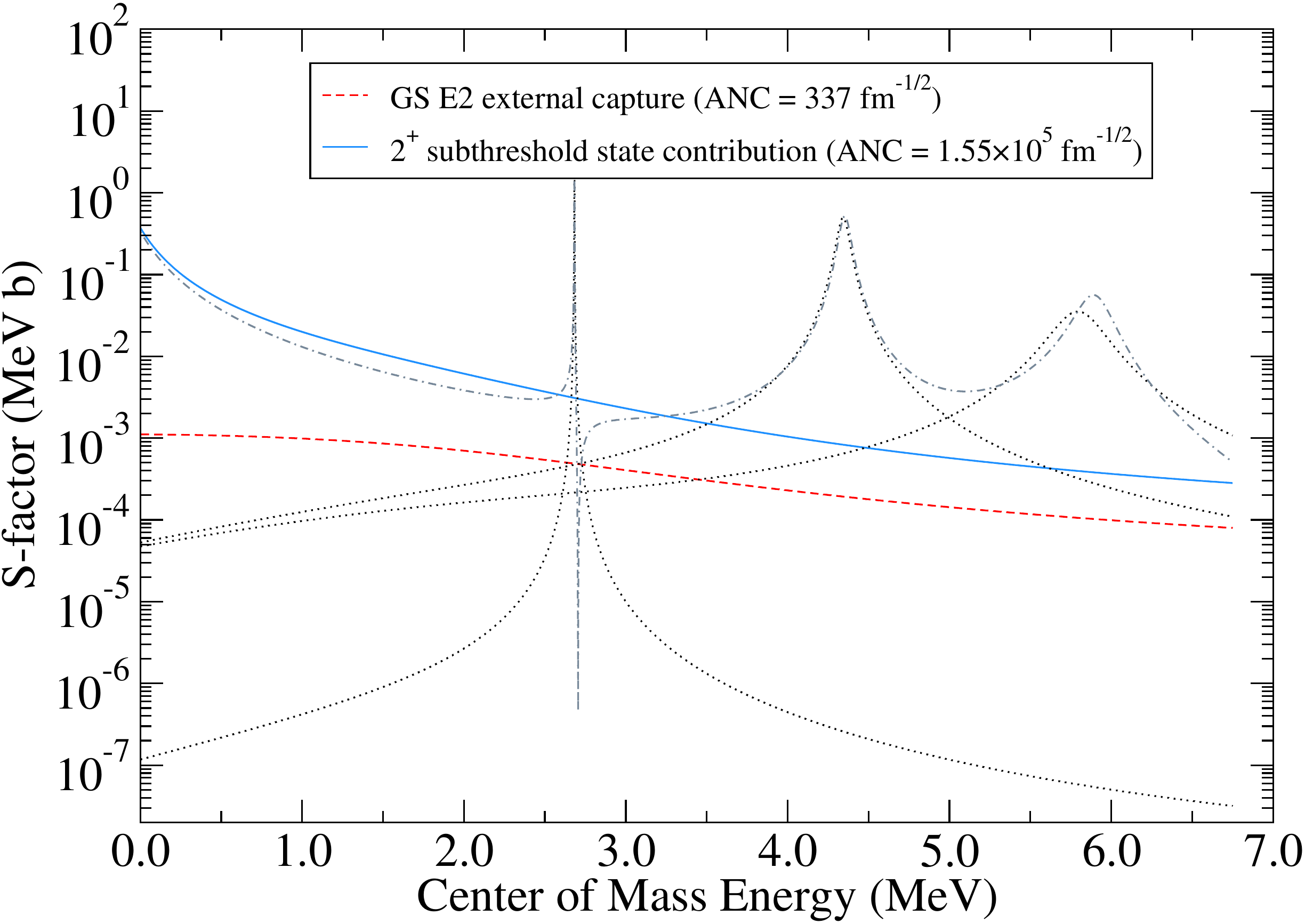}
\caption{Comparison of the $S$-factor energy dependence of the hard sphere external capture (red dashed line) and subthreshold contribution (solid blue line) for the $E2$ component of the $^{12}$C$(\alpha,\gamma_0)^{16}$O reaction. Resonance contributions are indicated by the grey dotted line, the total $E2$ $S$-factor by the grey dashed-dotted line. Calculations were made using the JINA $R$-matrix code \texttt{AZURE2}~\cite{AZUM2010, EBER2015}, based on the work of \cite{2020PhRvL.124p2701S}. \label{fig:12Cag_E2_components}} 
\end{figure}

In Ref.~\cite{2017RvMP...89c5007D}, the sensitivity of the $E2$ ground state cross section to the direct capture contribution was investigated. In most works, this component has been neglected, as it is a weaker contribution. In addition, there are few measurements, and they are wildly discrepant~\cite{ADHIKARI2009216, MORAIS20111, Adhikari_2016}. 
However, as the uncertainties have decreased on the experimental data, this secondary contribution may now be significant. 
This was highlighted in the $E2$ ground state capture fit of Ref.~\cite{PhysRevLett.109.142501}. 
To further investigate, Ref.~\cite{2020PhRvL.124p2701S} has used the $^{12}$C$(^{11}$B,$^{7}$Li$)^{16}$O transfer reaction to make a new measurement of the ground state $\alpha$-particle asymptotic normalization coefficient (ANC), a measure of its $\alpha$-particle cluster configuration. 
That work further investigates the role of the direct capture contribution to the ground state, and shows that through interference with the 2$^+$ subthreshold resonance, the ANC of the direct capture and the ANC of the 2$^+$ subthreshold state are highly correlated (see Fig.~\ref{fig:12Cag_E2_components}). While this new measurement of the ground state ANC is promising, further investigations are needed in order to establish a consistent value.

Finally, it has recently been found \cite{2020arXiv200606678F} that the uncertainty in the $^{12}$C$(\alpha,\gamma)^{16}$O reaction rate is one of the main uncertainties associated with the calculation of the upper mass gap for black holes, the mass-range in which massive stars are expected to self-destruct as pair-instability supernovae rather than leave behind a black hole remnant. This provides an exciting new area of direct overlap between nuclear astrophysics and gravitational wave measurements from LIGO~\cite{LIGO2015}.

Because of the overarching importance of the reaction rate multiple groups are still working on direct or indirect ways to inform the low energy cross section or reduce the present uncertainties in the extrapolation. New plans for a direct study exist for the recently installed INFN 3.5 MV accelerator in the Gran Sasso underground environment, taking advantage of the reduced cosmic ray background conditions. Complementary to that, multiple studies of additional reaction channels to the $^{16}$O compound nucleus are being investigated to improve the R-matrix approach in reducing the uncertainty in the extrapolation~\cite{2021PhRvC.103f5801D}. 
But besides these efforts there has been also rapidly growing interest in alternative techniques such as the measurement of the ground state transition of the $^{12}$C$(\alpha,\gamma)^{16}$O reaction using the inverse, the photo-dissociation reaction $^{16}$O$(\gamma,\alpha)^{12}$C. Efforts have been underway at the HI$\gamma$S facility~\cite{Weller2009257}, using both a bubble chamber~\cite{Ugalde201374} and a TPC~\cite{1742-6596-337-1-012054} (see Sec.~\ref{sec:HIgS}), and more recently at Jefferson Laboratory (see Sec.~\ref{sec:HIgS}). 
Measurements are also planned at the new ELI-NP facility as will be discussed in Sec.~\ref{sec:ELI-NP}. 
There is also renewed interest in using virtual photo-dissociation use the $^{16}$O$(e,e'\alpha)^{16}$O reaction, which is being pursued by the MIT group~\cite{2019PhRvC.100b5804F} and by the A1 and MAGIX collaborations at the MESA facility in Mainz. For both of these reactions, Holt \textit{et al.}~\cite{2019PhRvC..99e5802H, 2019PhRvC.100f5802H} have investigated the possible improvement in the low energy cross section determination. If achievable, these measurements will both push significantly lower in energy and provide a different set of systematic uncertainties that can be compared with more traditional previous measurements. A collaboration between groups at the University of Frankfurt and GSI Darmstadt mounted a new complementary effort in using Coulomb-dissociation to investigate the reaction towards lower energies~\cite{2020JPhCS1668a2016G}.

%% file: SProcess.tex
\subsection{AGB stars, the \texorpdfstring{$s$}{s}-process, and the \texorpdfstring{$^{22}{\rm Ne}(\alpha,n)^{25}{\rm Mg}$}{22Ne(a,n)25Mg} reaction}

\label{sec:sprocess}

Asymptotic Giant Branch (AGB) stars represent the final ``death-by-wind'' of stars of mass lower than roughly 10 \msun. 
After core H burning, stars like the Sun evolve onto the giant branch, where they burn H in a shell, and afterwards through core He burning which is followed by the AGB phase, where both H and He burn alternately in thin shells that surround the inert, degenerate C-O core. 
During the AGB phase, strong stellar winds powered by pulsation and dust formation drive most of the material off the star, until the core is left as a white dwarf. 
It is this wind, combined with mixing from the deep layers to the surface of the star, that carries new elements synthesised in the stellar interiors out in the interstellar medium, contributing to the galactic enrichment of many elements from the light C, N, and F to those heavier than iron, such as strontium, barium, and lead (see Ref.~\cite{Kara14} for a review).

\begin{figure}
    \centering
     \includegraphics[width=0.8\columnwidth]{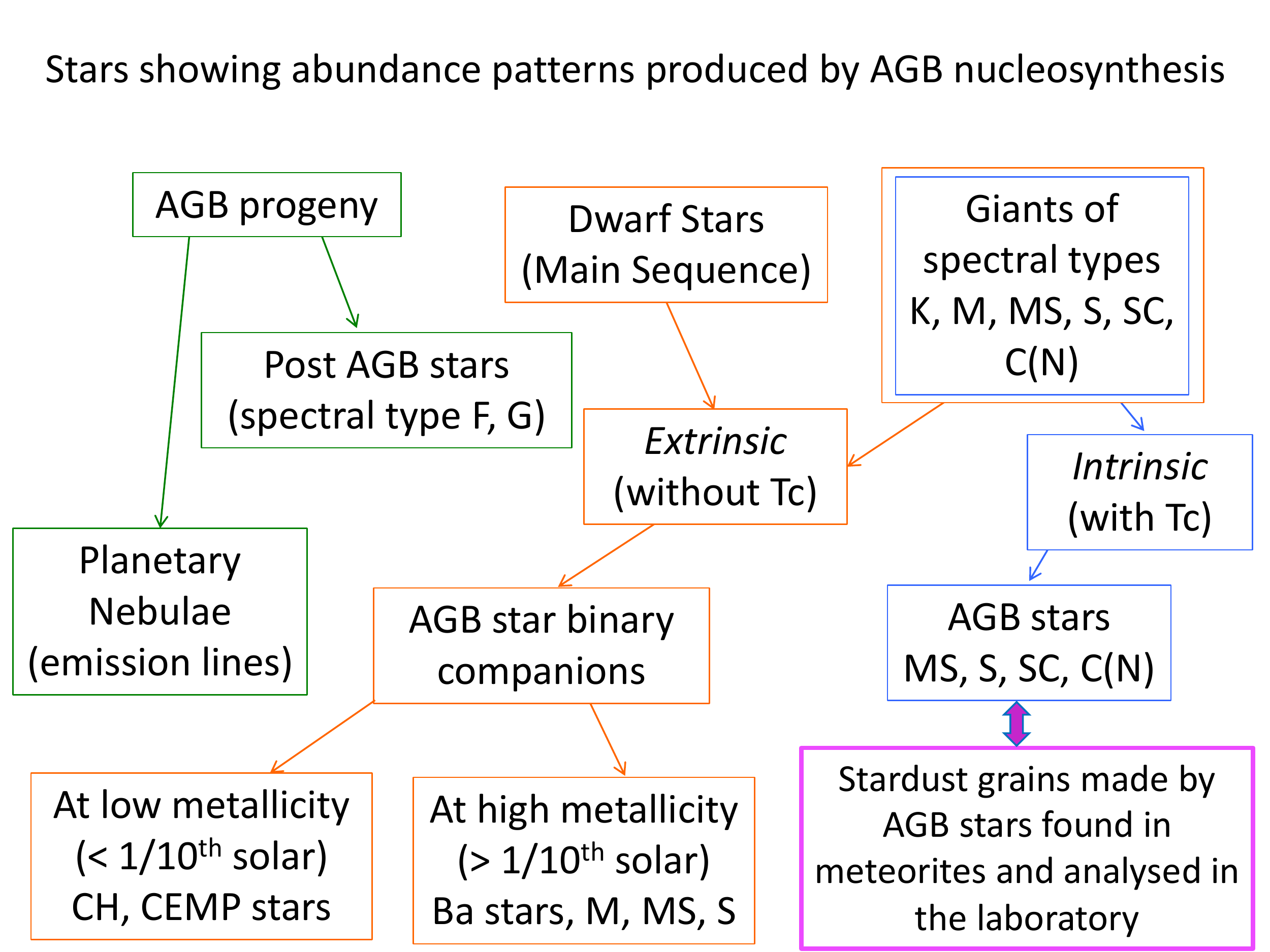}
    \caption{Schematic illustration of the different types of stars that show $s$-process enhancements: direct observation of AGB stars (blue arrows), or their progeny (green arrows), and binary enrichment (orange arrows). See Ref.~\cite{walker_2017} for a map between the spectral type and evolutionary phase of an AGB star. Stardust grains represent another direct link, as they formed around AGB stars. \label{fig:startypes}
    }
\end{figure}

Heavy nuclei produced during the ABG phase are the result of the $slow$ neutron-capture process ($s$ process) and can be observed via spectroscopic analysis both directly and indirectly in the progeny of these stars or in their binary companions, and via the laboratory analysis of the stardust grains that formed in the external layers of the AGB stars and were trapped inside meteorites (Figure.~\ref{fig:startypes}).

Nuclear reactions occur in the deep layers near the core, where it is hot (up to a few MK) and dense (up to 10$^4$ g/cm$^{3}$). 
The time evolution of the internal AGB structure shows recurrent shell He-burning thermal instabilities (thermal pulses, TPs), during which shell H-burning stops, and the material within the whole He-rich intershell, located between the H and the He burning shells, is mixed by convection. Many TPs can happen during the AGB evolution, depending on how long it takes the mass loss to remove the whole envelope. 
For example, in an AGB star of initial mass around 3 \msun, the AGB phase lasts for about 1 Myr, and roughly 20 TPs may occur at intervals of about 50,000 yr, while in an AGB star of higher mass, the TPs occur much more often (down to a few thousand years apart) and many more of them can happen, even if the AGB lifetime is shorter.
These numbers are, however, model dependent, especially given that the mass loss rate is one of the main physical uncertainties of the models.

Two neutron sources are active for the $s$ process in AGB stars \cite{Gal98,Gor00,Lug03s,Cri11,Bis15,Bat19}. The main neutron source is the \iso{13}C($\alpha$,n)\iso{16}O reaction, which is activated on \iso{13}C nuclei assumed to be produced by a partial mixing zone (PMZ) where protons from the envelope are mixed into the He- and C-rich intershell. 
This process is assumed to recurrently lead to the formation of a thin (10$^{-3}-10^{-4}$ \msun) ``\iso{13}C pocket'' via proton captures on \iso{12}C. 
The second neutron source is the \iso{22}Ne($\alpha$,n)\iso{25}Mg reaction, which is activated on the abundant \iso{22}Ne nuclei produced by double $\alpha$-capture on the \iso{14}N ingested into in the convective TPs. 
The two neutron sources operate in opposite ways. 
The \iso{13}C($\alpha$,n)\iso{16}O reaction occurs in the \iso{13}C pocket, which is typically within the radiative layer of the stars, not affected by mixing \cite{Str95}, at low temperatures (from 90 MK) during the relatively long time scales between TPs (roughly 10$^4$ y), and produces low neutron densities, of the order of 10$^7$ cm$^{-3}$, but high neutron exposures, $> 0.2$~mbarn$^{-1}$. 
The \iso{22}Ne($\alpha$,n)\iso{25}Mg reaction instead operates inside the convective TPs, at high temperature (from 270 MK) over short long time scales (roughly 10 yr), and produces high neutron density, up to 10$^{13}$ cm$^{-3}$, but low neutron exposure $< 0.1$~mbarn$^{-1}$ \cite{Van12}.
Therefore, while the \iso{13}C($\alpha$,n)\iso{16}O reaction is responsible for the production of the bulk of the $s$-process elements from AGB stars in the Galaxy, the  high neutron densities produced by the \iso{22}Ne($\alpha$,n)\iso{25}Mg reaction can affect the distribution via the activation of branching points on the $s$-process path, located at unstable nuclei where the neutron-capture cross section and the $\beta$-decay rate are comparable.
Therefore, AGB $s$-process predictions for any isotope affected by branching points crucially depend on the \iso{22}Ne($\alpha$,n)\iso{25}Mg reaction rate. 
As detailed in the Appendix of Ref. \cite{Lug11}, there are roughly 65 branching points along the $s$-process path, with more than 100 isotopes potentially affected. 
Here, we discuss four famous examples, their nuclear physics, and the observational constraints related to them.

\subsubsection{Branching ratios}
\label{sec:branching}

\begin{figure}
    \centering
     \includegraphics[width=0.6\columnwidth]{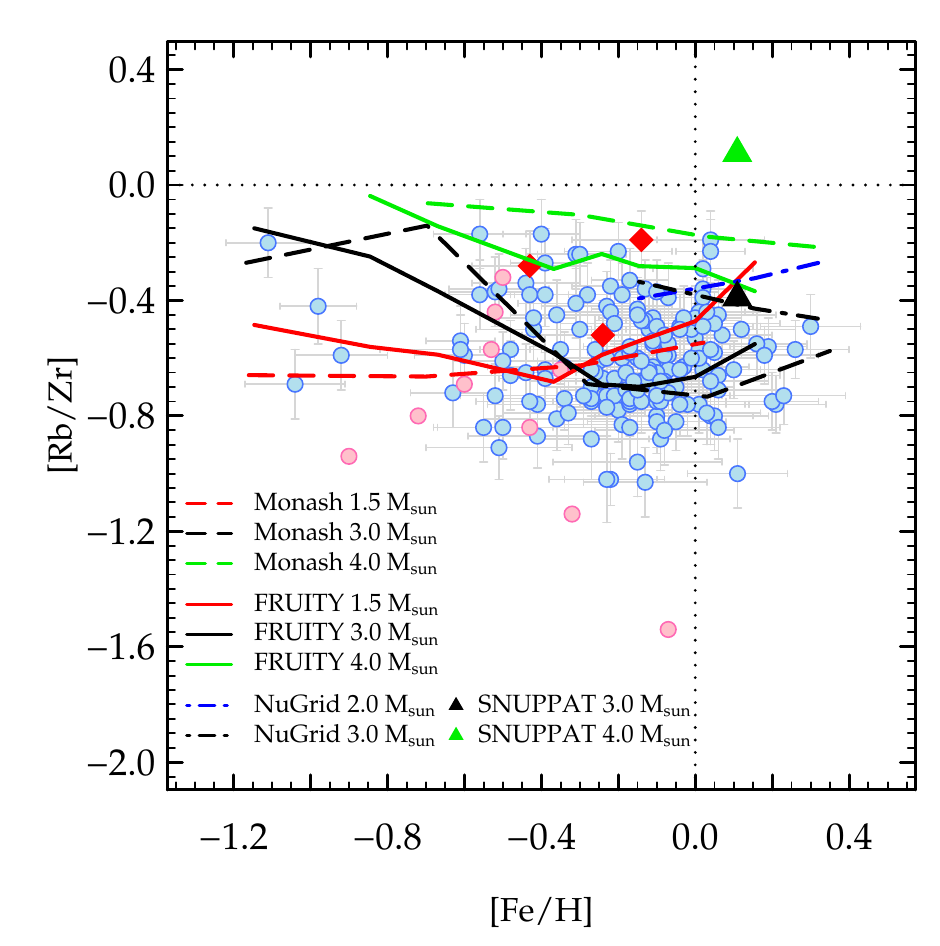}
    \caption{Comparison between [Rb/Zr] ratio observed in Ba stars and theoretical predictions. 
    Different abundance ratios for the Ba stars Blue dots represent the 180 Ba stars reported by Ref. \cite{Ror21}, with the [Zr/Fe] and [Fe/H] ratios were taken from Ref. \cite{Dec16}.  The three most Rb-rich stars ([Rb/Fe]$\simeq$1) from this sample are indicates as red diamonds. Magenta dots represent the sample stars analyzed by Ref. \cite{Kari18}. 
   The Monash models are from Ref. \cite{Fis14,Kar16,Kar18}, the FRUITY models from the FRUITY database \cite{Cri11}, the NuGrid models from Ref. \cite{Bat19}, and the SNUPPAT models from Yag\"ue L\'opez et al. (2021, submitted). Figure reproduced from Ref.~\cite{Ror21}: Monthly Notices of the Royal Astronomical Society, Volume 501, Issue 4, pp.5834-5844, 2021, Roriz, M. P., Lugaro, M., Pereira, C. B., Sneden, C., Junqueira, S., Karakas, A. I., Drake, N. A., ``Rubidium in Barium stars'', Oxford University Press and the Royal Astronomical Society.
   \label{fig:Rb} 
   }
\end{figure}

%\begin{enumerate}
%    \item 
{\bf The case of \iso{86}Rb.} The first famous branching point is related to the element Rb, and results in the opportunity to use observations of this element in AGB stars and their family  as an indicator of the stellar mass, and as a strong constraint to the AGB models.  This is because the $\sim$19~day half-life of \iso{86}{Rb} provides sufficient time for $^{86}{\rm Rb}(n,\gamma)^{87}{\rm Rb}$ to occur, depending on the local neutron flux and the $^{86}{\rm Rb}(n,\gamma)^{87}{\rm Rb}$ reaction rate.
Since \iso{87}Rb has a magic number of neutrons, and there are only two relatively stable isotopes of Rb, an increased flux toward \iso{87}Rb has a strong impact on the overall abundance of the element Rb, which is observable in stars. 
In particular, it is useful to compare the Rb abundance to that of neighbouring elements also belonging to the first $s$-process peak at N=50, such as Sr or Zr, to highlight the impact of the activation of the branching point rather than the impact of the relative distribution of the first to second $s$-process peaks. 
In the past, the branching point at \iso{85}Kr, where there is a similar competition between $\beta$-decay and neutron-capture, has also been considered for the production of \iso{87}Rb. However, this branching point produces \iso{86}Kr (before the flux reaches \iso{87}Rb), which also has magic number of neutrons. 
Therefore the \iso{85}Kr branching point leads to a decrease in \iso{85}Rb, which makes the overall abundance of Rb decrease (see discussion in, e.g., Ref \cite{Van12}). 

Early observations of negative [Rb/Sr] or [Rb/Zr] ratios in AGB stars demonstrated that the main neutron source in these stars must be the \iso{13}C($\alpha$,n)\iso{16}O reaction, because the high neutron densities of the  \iso{22}Ne($\alpha$,n)\iso{25}Mg would result in positive values instead \cite{Lam95}. Abia et al. \cite{Abi01} also used Rb to demonstrate that the C-rich AGB stars should be of relatively low mass, since temperatures increase in AGB stars as function of mass, and from around $4-5$ \msun the models predict a significant production of Rb.
This was in fact observed in these high-mass AGB stars, where [Rb/Zr] ratios are typically positive \cite{Gar06,Zam14,Per17}, which has been considered as proof of the activation of the \iso{22}Ne($\alpha$,n)\iso{25}Mg in AGB stars of relatively high initial mass \cite{Van12}.
Finally, strong independent constraints have been inferred from a large sample of Ba stars, the binary companions of AGB stars \cite{Ror21} (see Fig.~\ref{fig:Rb}). 
All these stars present negative [Rb/Zr] ratios, in agreement with models (such as the Monash and the FRUITY models shown in Fig.~\ref{fig:Rb}) where the \iso{22}Ne($\alpha$,n)\iso{25}Mg reaction is only marginally activated. 
Interestingly, no Ba stars were found that could represent the companions of the more massive AGB stars observed by, e.g., Ref. \cite{Per17}, with positive [Rb/Zr] ratios. 
For these missing Ba stars no obvious explanation exists yet.

\begin{figure}
    \centering
     \includegraphics[width=0.8\columnwidth]{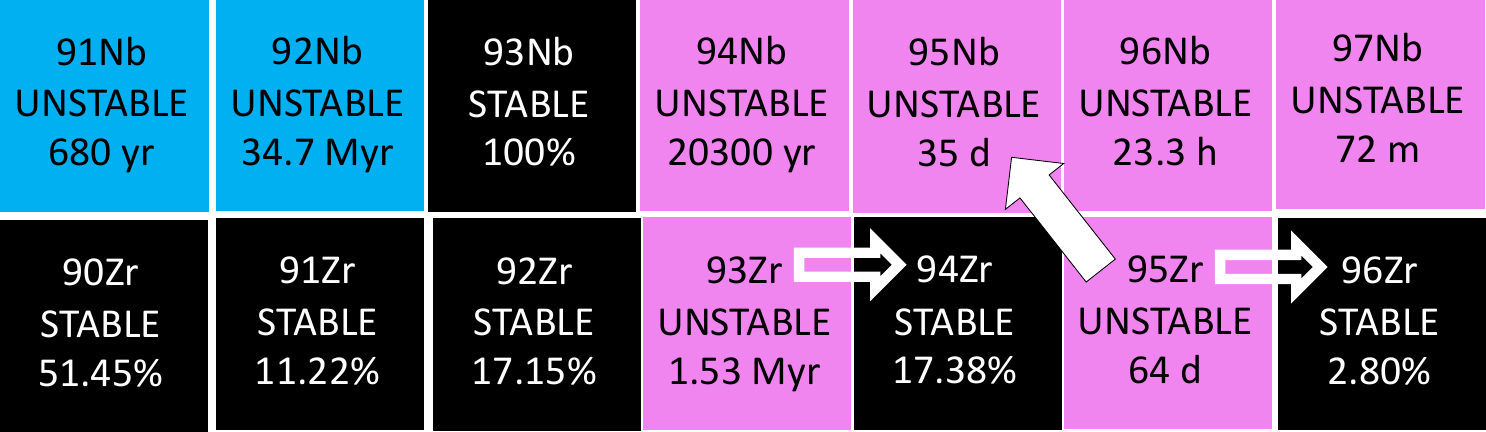}
    \caption{Section of the nuclide chart showing the $s$ process path at the unstable \iso{93}Zr and \iso{95}Zr isotopes (with thick, empty arrows representing neutron captures and the solid arrow $\beta^-$-decay): \iso{93}Zr lives too long to represent a branching point, while \iso{95}Zr can either decay to \iso{95}Nb, and then \iso{95}Mo, or capture a neutron to produce \iso{96}Zr under AGB $s$-process conditions during the activation of the \iso{22}Ne($\alpha$,n)\iso{25}Mg reaction. \label{fig:Zr_branch}}
\end{figure}

\begin{figure}
    \centering
     \includegraphics[width=\columnwidth]{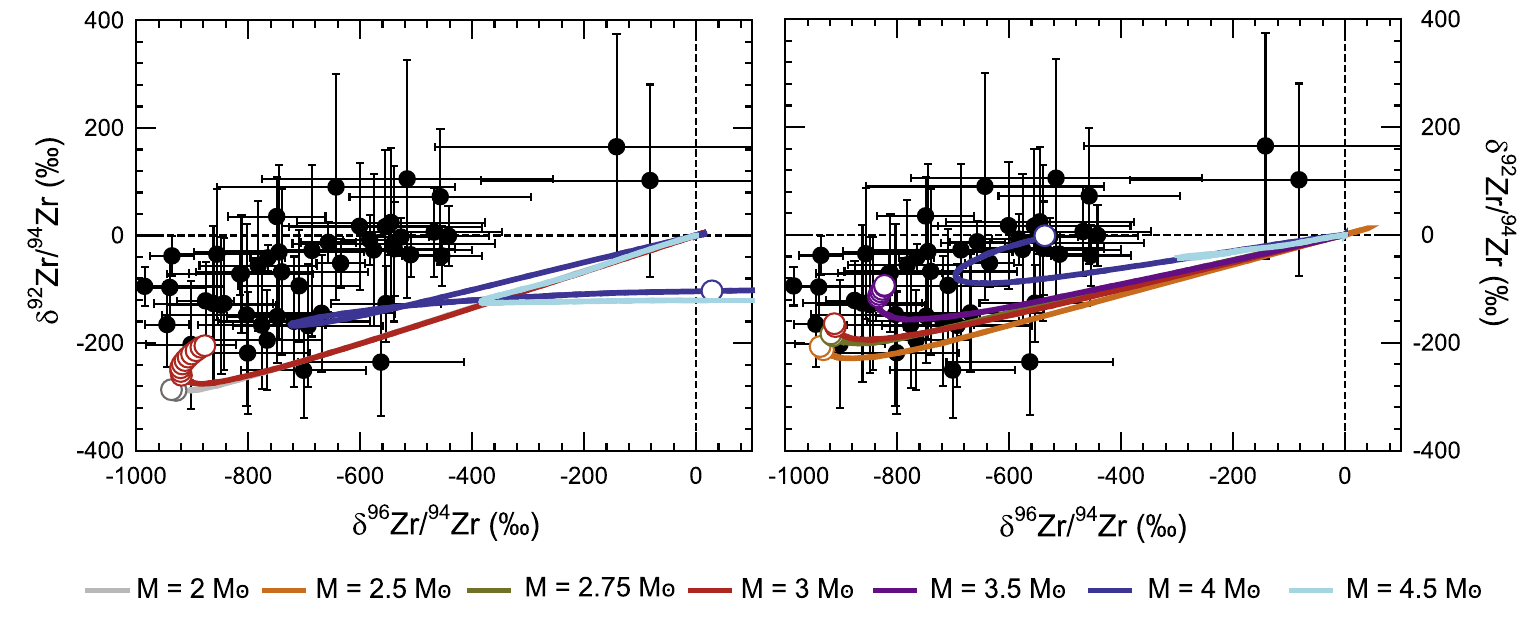}
    \caption{The RIMS SiC grains data for Zr (black circles with 2$\sigma$ error bars, for references see Ref. \cite{Liu14}) are compared to the surface evolution of stellar models of solar metallicity (left panel) and of twice-solar metallicity (right panel) of different masses from 2 to 4.5 \msun (as indicated by the different line colors). The $\delta$ notation indicates variation from solar, permil. For instance, $\delta$=0 represents the solar value of the ration (dashed lines), and $\delta$=+200 or $-200$ represents ratio 20\% higher or lower than solar. Each solid line represents the evolution of the corresponding initial mass and open circles on the lines represent the phase when C/O ~$> 1$ in the envelope, the condition necessary to produce SiC grains.
    Reprinted from Ref \cite{Lug18sic}: Geochimica et Cosmochimica Acta, Vol 221, Lugaro, M., Karakas, A. I., Pet\H{o}, M., Plachy, E., ``Do meteoritic silicon carbide grains originate from asymptotic giant branch stars of super-solar metallicity?'', Pages No. 6--20, Copyright (2018), with permission from Elsevier. \label{fig:Zr_grains}}
\end{figure}

{\bf The case of \iso{95}Zr.} Another important branching point is related to the \iso{95}Zr isotope (with a half life of roughy 64 days), where the relevant nuclear reaction sequence is shown in Fig.~\ref{fig:Zr_branch}. This branching determines the \iso{96}Zr/\iso{94}Zr ratio that has been measured in Chicago with high precision in large ($\sim$ 1 $\mu$m) silicon carbide (SiC) stardust grains using Resonant Ionisation Mass Spectrometry (RIMS, \cite{Liu14}). 
The grains show deficits relative to solar in this ratio, which can be explained by considering that the C-rich AGB parent stars of the grains should have mass below roughly 4-4.5 \msun\ (Fig.~\ref{fig:Zr_grains}) -- in agreement with the results of Ref. \cite{Abi01} from Rb mentioned above. 
Interestingly, it has been difficult to also match the high \iso{92}Zr/\iso{94}Zr ratio measured in many grains. 
A solution has been found by considering AGB stars of metallicity higher than solar 
\cite{Lug18sic,Lug20} 
As the metallicity increases, AGB stars become cooler, which results in a less efficient activation of the \iso{22}Ne($\alpha$,n)\iso{25}Mg reaction, and therefore a less efficient production of \iso{96}Zr for initial stellar masses up to 4 \msun. 
For these masses, the size of the \iso{13}C pocket is also found to be smaller than for the lower masses (see discussion, e.g., in Ref. \cite{Kar16}), therefore, here the \iso{22}Ne($\alpha$,n)\iso{25}Mg reaction has a stronger relative impact on the final abundance distribution. 
Because this neutron source operates at higher temperatures, and the neutron-capture rate of \iso{92}Zr decreases with temperature more significantly than that of \iso{94}Zr, the final result is an increase in the \iso{92}Zr/\iso{94}Zr ratio in these stars and a natural match to the data points.

\begin{figure}
    \centering
     \includegraphics[width=0.4\columnwidth]{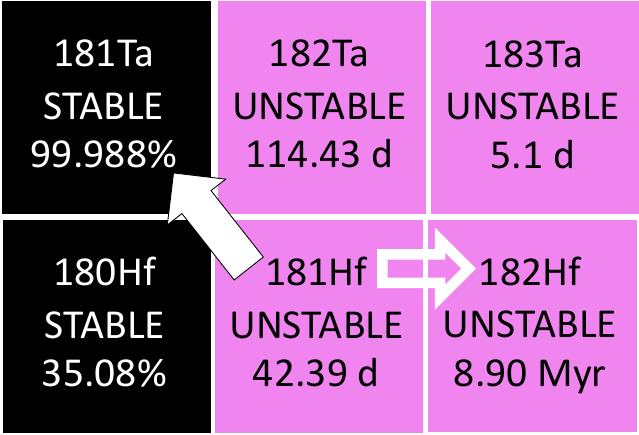}
     \hspace{1cm}
          \includegraphics[width=0.4\columnwidth]{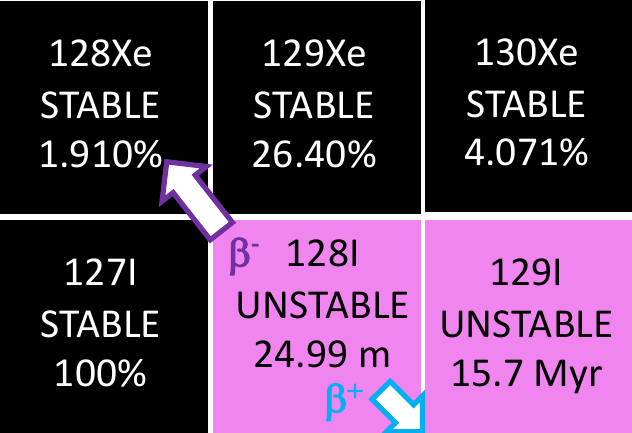}
    \caption{Sections of the nuclide chart showing the $s$ process path at the unstable \iso{181}Hf (left) and \iso{128}I (right). The thick, empty arrow in the left panel represent neutron captures leading to the creation of the long-lived isotopes \iso{182}Hf, and the solid arrow the potentially temperature dependent $\beta^-$-decay producing \iso{181}Ta instead. \label{fig:Hf} In the right panel, the solid arrows represent the decay of \iso{128}I: the purple arrow is the dominant ($\simeq$93\%) $\beta^-$ channel to \iso{128}Xe and the blue arrow represents the marginal ($\simeq$7\%) $\beta^+$-decay channel to \iso{128}Te.}
\end{figure}

{\bf The case of \iso{181}Hf.} The existence of a significant $s$-process branching point at \iso{181}Hf (Fig.~\ref{fig:Hf}) is a relatively new discovery.
Here, unlike in the previous two cases, the temperature dependence of the $\beta^-$-decay rate of \iso{181}Hf is crucial to control the production of the long-lived isotopes \iso{183}Hf (half life 8.9 Myr), which is known to have been present in the early Solar System. 
The most recent evaluation of the $\beta^-$-decay rate of \iso{181}Hf based on the latest information on the level structure of \iso{181}Hf \cite{Bon02}
does not show a strong temperature dependence; therefore, the terrestrial value of the rate (corresponding to a half life of roughly 43 days) is appropriate to be used in stellar model calculations and because of this it is possible to produce \iso{182}Hf via the $s$ process in AGB stars \cite{Lug14}.
Using the abundance of \iso{182}Hf at the time of the formation of the Sun, we can use the decay of \iso{182}Hf as a cosmic clock to provide a time range of 10 to 35 Myr for the time that elapsed from the birth of the stellar nursery where the Sun was born and the formation of the first solid bodies in the Solar System \cite{Cot19}.

{\bf The case of \iso{128}I.} This is not a standard branching point because there is no neutron capture involved, as the half life of \iso{128}I of roughly 25 minutes is always too short to produce the long-lived isotope \iso{129}I (half life 15.6 Myr), which is also observed to be present in the early Solar System. 
Because of this, \iso{129}I can only be produced by the $r$ process. 
Therefore, we can use it as a pure indicator of the last $r$-process event that polluted the material from which the Solar System formed and derive constraints on the nature of such an event \cite{Cot21}. 
Still a branching point is present at \iso{128}I, due to its double decay. 
The $\beta^+$-decay channel (to \iso{128}Te) competes with the $\beta^-$-decay channel, which produces \iso{128}Xe, an $s$-only isotope with a minor $p$-process component. 
Investigation of this peculiar branching point therefore can allow us to determine accurately the $s$-process component of \iso{128}Xe, and therefore its $p$-process component, providing constraints for $p$-process models \cite{Rei04}.
Note that in this case, the dependence on the \iso{22}Ne($\alpha$,n)\iso{25}Mg reaction rate is indirect since no neutrons are captured by \iso{128}I. 
However, the activation of the $\beta^+$-decay channel depends on the temperature and density at which the \iso{22}Ne($\alpha$,n)\iso{25}Mg reaction operates.

%\end{enumerate}

\subsubsection{Neutron sources} 
%\label{sec:n-sources}
\label{sssec:Ne22}
As demonstrated by the examples above, many of the effects from $s$-process branching points are highly consequential, the results of such effects crucially depend on the 
$^{22}{\rm Ne}(\alpha,n)^{25}{\rm Mg}$ reaction rate. Most of the model predictions shown above used the rate from Ref. \cite{Ili10} and they appear to be generally good in matching the observations\footnote{The highest observed $[$Rb/Zr$]$ around 1 in massive AGB stars \cite{Per17} are still unmatched. However, they cannot be matched by neutron captures. This is because higher neutron densities result in the flux reaching equilibrium, so that there is a maximum amount of $^{87}{\rm Rb}$ that can be produced before neutron captures on it become efficient. Therefore, the problem is not due to nuclear physics and may instead be observational.}. 
From general considerations derived from the comparison between models and observations, we can reach some conclusions on the \iso{22}Ne($\alpha$,n)\iso{25}Mg reaction rate. 
Specifically, if its rate was much higher than currently used in the models, the \iso{96}Zr in stardust grains may be over-produced compared to observations. 
Indeed, some models already struggle to keep it as low as observed (see e.g. Ref. \cite{Bat19}).
If the \iso{22}Ne($\alpha$,n)\iso{25}Mg reaction rate was instead much lower than currently used in the models, then \iso{182}Hf may be under-produced, shortening the timescale for the Solar System formation, in disagreement with the same constraint derived using \iso{107}Pd,  an isotope unaffected by branching points and dependent almost exclusively on the well-constrained neutron-capture cross section.

Given these consequences for model-observation comparisons, an accurate determination of the low energy $S$-factor of the $^{22}$Ne$(\alpha,n)^{25}$Mg reaction is greatly needed. Likewise, the competing $^{22}$Ne$(\alpha,\gamma)^{26}$Mg $S$-factor also needs to be well characterized~\cite{2016PhRvC..93e5803T}. The challenging target and the high beam intensities ($\gtrsim$50~$\mu$A) are not available at many facilities, thus there are strikingly few measurements. By far the most high resolution and precise measurement available is that of Ref.~\cite{2001PhRvL..87t2501J}, which covers the energy range from $E_\alpha\approx$~800~keV up to 1500~keV. For the reaction rate at $s$-process temperatures, the most important resonance is the strong one observed at $E_\alpha$~=~830~keV. This resonance is also the lowest energy resonance observed in the $^{22}$Ne$(\alpha,\gamma)^{26}$Mg reaction~\cite{2019PhRvC..99d5804H}.

While new measurements of both the $^{22}$Ne$(\alpha,n)^{25}$Mg and $^{22}$Ne$(\alpha,\gamma)^{26}$Mg reactions are in preparation at underground facilities around the world, the only recent measurement has been that of Ref.~\cite{2019PhRvC..99d5804H}, who re-investigated the $(\alpha,\gamma)$ strength of the $E_\alpha$~=~830~keV resonance. The measurement was performed at the LENA facility and used an active shielding setup, which results in room background suppression of several orders of magnitude. Both the resonance energy and strength were measured, with the resonance energy being somewhat higher than that reported by Ref.~\cite{2001PhRvL..87t2501J} and with a reduced uncertainty. The measured strength is also somewhat larger than that of Ref.~\cite{2001PhRvL..87t2501J}, but the two measurements are in good statistical agreement with one another.

Because of the experimental challenges with direct measurements, most recent investigations have taken indirect approaches. 
Talwar {\em et al.} \cite{2016PhRvC..93e5803T} used the $^{22}$Ne$(^6$Li$,d)^{26}$Mg $\alpha$-transfer (82.3~MeV) and the $^{26}$Mg$(\alpha,\alpha')^{26}$Mg (206~MeV) reactions, performed using the Grand Raiden spectrometer at the Research Center for Nuclear Physics in Osaka Japan, to preferentially populate $\alpha$-cluster states in $^{26}$Mg.
Spectroscopic factors and spin-parity assignments were determined, which suggest a substantial increase in the $^{22}$Ne$(\alpha,\gamma)^{26}$Mg rate at low temperatures, and thus a lower neutron flux available for $s$-process nucleosynthesis. 
More recently, Ref.~\cite{2020PhLB..80235267J} have also preformed $^{22}$Ne$(^6$Li$,d)^{26}$Mg and $^{22}$Ne$(^7$Li$,t)^{26}$Mg $\alpha$-transfer reactions, but now at sub-Coulomb energies ($\approx$6~MeV), where the determination of the partial widths is less sensitive to the assumed potential model. 
Similarly, a larger low temperature reaction rate was determined for the $^{22}$Ne$(\alpha,\gamma)^{26}$Mg reaction. 
At around the same time, Ref.~\cite{2020PhLB..80235256O} has reported improved n/$\gamma$ decay branchings for for $^{26}$Mg levels near the $\alpha$-particle threshold also using the $^{22}$Ne$(^6$Li$,d)^{26}$Mg reaction.

\begin{figure} 
  \centering
\includegraphics[width=1.0\columnwidth]{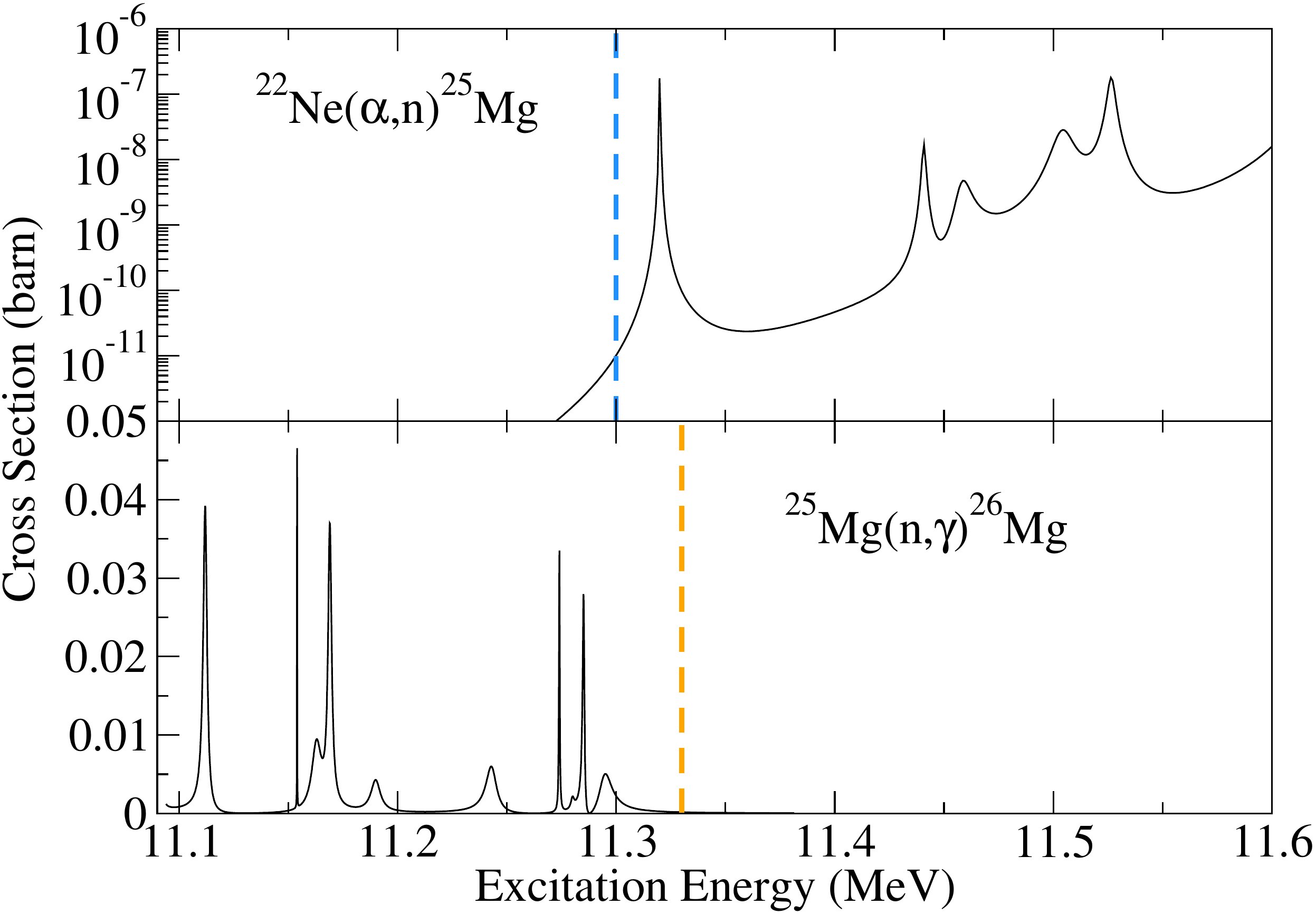}
\caption{$R$-matrix calculations showing the current range of $^{22}$Ne$(\alpha,n)^{25}$Mg and $^{25}$Mg$(n,\gamma)^{26}$Mg reactions (down to approximately the blue dashed line in the upper panel and up to approximately the orange dashed line in the lower panel). Calculations were made with the $R$-matrix code \texttt{AZURE2}~\cite{AZUM2010, EBER2015} based on fits to the data of \cite{2001PhRvL..87t2501J, 2017PhLB..768....1M}. \label{fig:multichannel_picture}} 
\end{figure}

Another indirect method that has been used frequently is the study of the compound nucleus with $^{25}$Mg+$n$ reactions. The most recent study was made using the time-of-flight facility at CERN (CERN-n\_TOF)~\cite{2017PhLB..768....1M}. The experimental yields were analyzed using the $R$-matrix framework and neutron and $\gamma$-ray partial widths were extracted from the data. Spin-parity assignments were also confirmed or revised for several resonances.
Unfortunately, the resonances populated in the $^{25}$Mg$(n,\mathrm{total})$ and $^{25}$Mg$(n,\gamma)^{26}$Mg reactions do not seem to correspond to those populated in $\alpha$-transfer reactions. 
This is likely due to both the difference in underlying nuclear structure (single-particle states versus $\alpha$-cluster states) and the masking of natural parity states by strongly populated unnatural parity states in these reactions. Because of experimental challenges in both cases, the $^{25}$Mg+$n$ and $^{22}$Ne+$\alpha$ reactions only share a small overlap in excitation energy, further hindering the comparison between the different types of measurements (see Fig.~\ref{fig:multichannel_picture}).

With the availability of $\alpha$-particle beams at the LUNA and, more recently, at the CASPAR underground facilities (see Sec.~\ref{sec:DUA}), new measurements of both the $^{22}$Ne$(\alpha,n)^{25}$Mg and $^{22}$Ne$(\alpha,\gamma)^{26}$Mg reactions are likely on the horizon. 
New measurements of both reactions have also been made at TRIUMF's DRAGON facility as described in Sec.~\ref{sec:DRAGON_recent}.

%% file: MassiveStars.tex
\subsection{Carbon burning and supernovae progenitors}
\label{sec:massive-stars}

Carbon burning is a key stage of stellar evolution and extremely important to  understand supernovae (SNe) outcomes. 
SNe play a critical role in astrophysics as they provide a major contribution to the chemical and physical evolution of galaxies, act as distance ladders  to probe the past history of the Universe, and are associated to the formation of  the most compact objects in nature, such as neutron stars and black holes.

Carbon fusion in stars proceeds primarily through the $^{12}$C$(^{12}$C$,\alpha)^{20}$Ne and the $^{12}$C$(^{12}$C$,p)^{23}$Na reactions. 
SN progenitor models require the rates of these reactions to be known down to $E_{\rm cm}\sim 1.2$~MeV. 
Owing to the very small cross sections, direct measurements are challenging already at energies above 2.2 MeV. 
On the other hand, He burning provides the fuel for C burning. 
In this evolutionary phase, carbon is produced by the triple-$\alpha$ reaction and destroyed by $^{12}$C$(\alpha,\gamma)^{16}$O. 
The cross section of the second process, in particular, needs to be known down to $\sim 300$ keV.
More generally, any uncertainty affecting the amount of fuel available for C burning and the rate of the primary carbon fusion process, $^{12}$C$+^{12}$C, hampers our knowledge of the final fate of almost all SN progenitors. 
In the following, we will illustrate some examples of the influence of the C burning on supernova events.

%-------- SNIa ---------------
\subsubsection{Carbon simmering and type Ia SNe.}
 
The current paradigm for SNe Ia is that of a thermonuclear explosion of a mass-accreting carbon-oxygen white dwarf (CO WD) in close binary systems. 
It was early recognized that the composition of the progenitors at the time of explosion, specifically the C/O ratio and the degree of neutronization\footnote{Usually, the neutronization degree of a stellar plasma composed by $\mathcal{N}$ isotopes is defined as: $\eta=\sum_{i}\frac{X_i}{A_i}\left ( A_i-2Z_i \right )$, where $X_i$, $A_i$ and $Z_i$ are, respectively, the mass fraction, the atomic number and the charge number, and $i=$1,..., $\mathcal{N}$.} influences the nucleosynthesis and, in turn, the resulting light curve \cite{hamuy2000,inma2001,cooper2009,howell2009,anderson2015,moreno2016,piersanti2017}. 
In other words, the explosive outcomes may keep memory of the progenitor stars. This occurrence may be exploited to investigate the nature of the exploding WD. 
In particular, abundance measurements of intermediate-mass elements, like Si, and iron-peak elements in the material ejected by a SN  may provide some hints on the neutronization of the exploding WD. 
In addition, the light-curve rise time is sensitive to the pre-explosive C/O ratio.    
 
\begin{figure}
    \centering
     \includegraphics[width=\columnwidth]{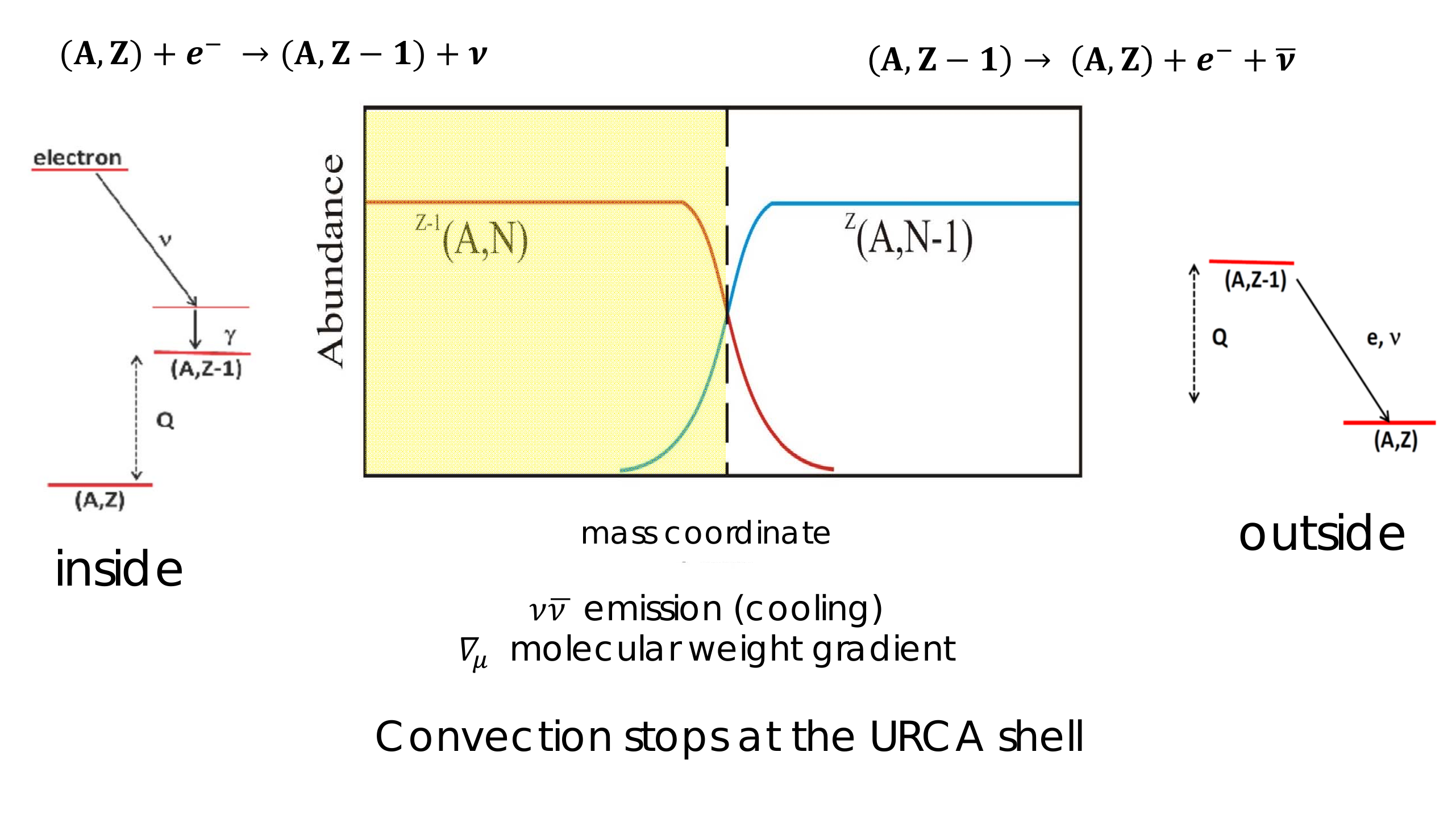}
    \caption{Schematic illustration of a convective urca shell, including a diagram of weak reaction processes that are dominant at radii smaller than (inside) and larger than (outside) the urca shell radius. The yellow area marks the convective core.}
    \label{fig:urca}
\end{figure}

In principle, the neutronization of a WD depends on the progenitor metallicity.
Indeed, the composition of the C-O core of an intermediate-mass star is the result of both the H burning and the subsequent He burning. 
In the first evolutionary phase, the original CNO material is mainly converted into $^{14}$N.  
Later on, during He burning, $^{22}$Ne is produced through the chain  $^{14}$N$(\alpha,\gamma)^{18}$F$(\beta^+,\nu_e) ^{18}$O$(\alpha,\gamma)^{22}$Ne. 
As a consequence, the higher the original CNO content, the larger the $^{22}$Ne abundance in a CO WD and, in turn, the higher the neutronization degree. 
 Is that all? Certainly not, because other  processes can modify the neutronization during the so-called simmering phase, i.e. the  non-explosive C burning taking place during the last $\sim 10000$ yr prior to the explosion. 
Weak interactions, that transform protons into neutrons and viceversa, may accomplish this.  During the accretion phase, the WD is progressively compressed. Then, when the density at the center approaches a few $10^9$ g/cm$^3$, neutron-rich isotopes are efficiently produced there through electron captures. 
In contrast, $\beta$ decays are forbidden, because of the Pauli suppression mechanism. 
At larger radii, owing to the lower electron density, $\beta$ decays are favored, while electron captures are suppressed. 
This occurrence naturally leads to the development of urca shells (see figure \ref{fig:urca}). In a urca shell, first discussed by \cite{gamo1941}, repeated electron-capture and $\beta$-decay reactions give rise to neutrino emissions, thus leading to an effective energy loss.
During the compression phase, urca shells form at the center and, then, they move outside. Owing to the cooling induced by the $\nu_e  \bar{\nu_e}$ emissions, the C ignition will occur at lower temperature and at larger density. Then, C burning causes the development of a convective core that progressively increases its extension. Once the external boundary of the convective core reaches an active
 urca shell, the steep molecular-weight gradient limits a further increase of the convective instability. This occurrence determines the location of the transition layer between the internal core mixed by convection and the external zone whose chemical composition is not modified during the simmering phase. 
 In practice, the presence of active urca shells during simmering implies a larger C consumption and a higher mean neutronization of the core. 
 
 The most efficient urca shells are those associated to the pairs $^{23}{\rm Na}$$\leftrightarrow$$^{23}{\rm Ne}$ and $^{21}{\rm Ne}$$\leftrightarrow$$^{21}{\rm F}$. 
 There is remarkable feedback with the $^{12}$C+$^{12}$C reaction \cite{bravo2011}, as $^{23}$Na is directly produced via the $p$ channel, while $^{21}$Ne is synthesized via $n$ captures on the $^{20}$Ne produced by the $\alpha$ channel.  
 An enhanced value of the $^{12}$C+$^{12}$C astrophysical factor at $E_{\rm cm}<2$ MeV would hasten the C ignition and the C burning will occur at lower density, thus reducing the amount of energy released by the electron captures near the center. 
 As a consequence, a  lower final carbon abundance is expected. 
 Another interesting possibility is that of not-equal rates for the $p$ and the $\alpha$ channels of the $^{12}$C+$^{12}$C reaction. 
 Current models usually assume that both these rates are equal to the 50\% of the total $^{12}$C+$^{12}$C  rate. 
 A different $p/\alpha$ rate ratio would affect the relative influence of the two main urca shells on the extension of the convective core during the simmering phase. 
 
 The C/O ratio also affects the final outcome \cite{inma2001}. In addition to the convective urca shells, this important quantity is also affected by the $^{12}$C$(\alpha,\gamma)^{16}$O rate operating during the He-burning phase. In general, a lower C/O in the inner portion of the exploding WD favors a larger production of $^{56}$Ni, whose decay powers the early-time light curve (first $40-50$ days since the explosion). The light-curve rise time is particularly sensitive to the C/O ratio. On the other hand, a larger C/O in the external layers favors the production of intermediate-mass elements. Since the C/O ratio is mainly determined by the competition between the triple$-\alpha$ and the $^{12}$C$(\alpha,\gamma)^{16}$O reactions during the He burning, the uncertainties affecting these two nuclear processes  inevitably affect our understanding of the SN Ia phenomenon.

% ---- e-capture SNe --------------------
\subsubsection{Electron-capture SNe: thermonuclear explosion or core-collapse and bounce?}

\begin{figure}
    \centering
     \includegraphics[width=\columnwidth]{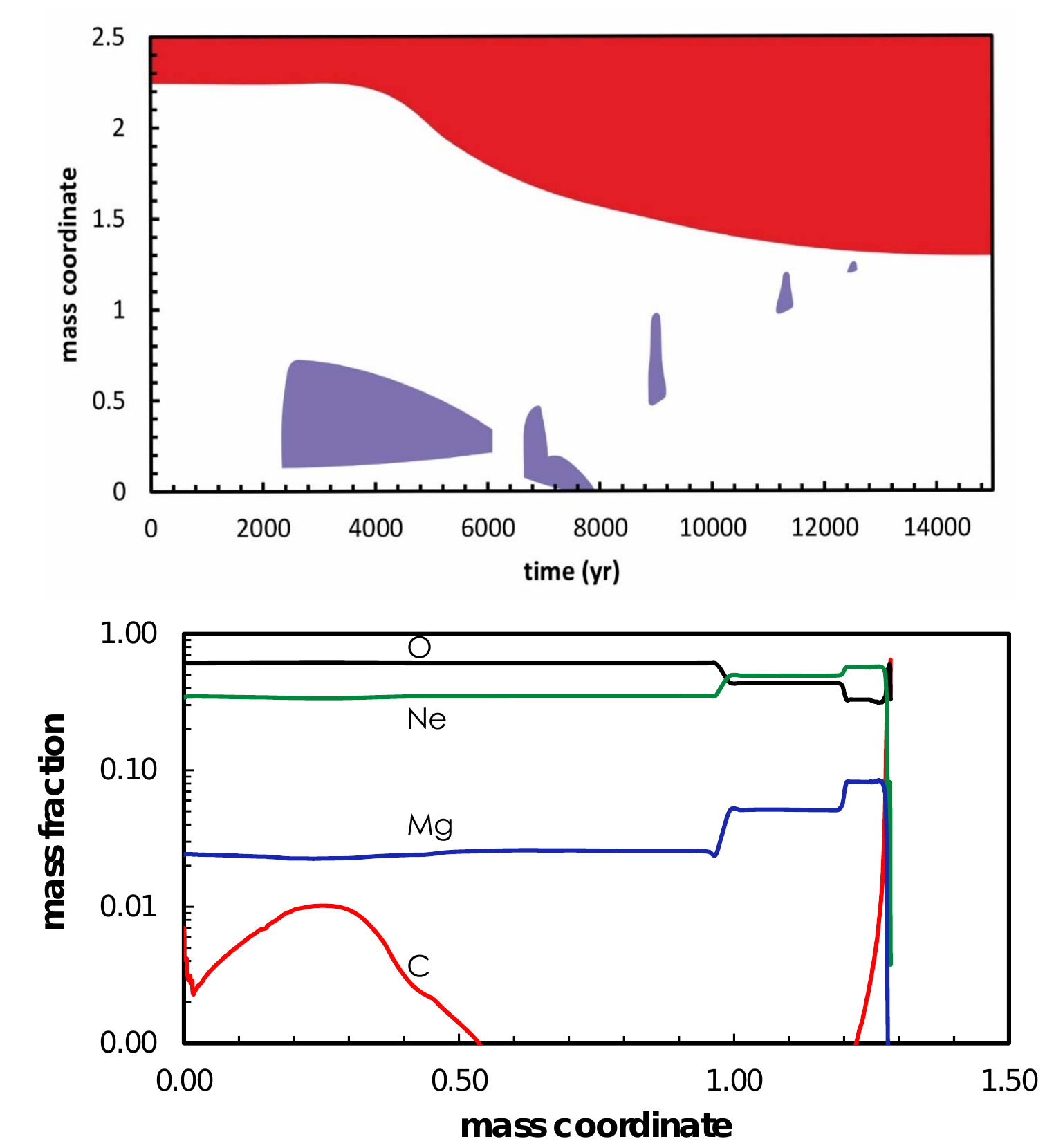}
    \caption{Upper panel: Kippenhahn diagram of the degenerate C burning in the 8.5 M$_\odot$ stellar model. The red region corresponds to the convective envelope, while the violet regions are the convective C-burning episodes. The t=0 point is arbitrary. 
    Lower panel: final core composition: C (red), O (black), Ne (green), Mg (blue).}
    \label{fig:m9}
\end{figure}

Stars with mass $8<$M/M$_\odot<10$ ignite carbon in a degenerate core. As an example,
the C-burning phase of a star with initial mass M$=8.5$ M$_\odot$ and solar composition is illustrated in figure \ref{fig:m9}.
The C burning proceeds through a series of thermonuclear runaways, each one generating a convective zone (figure \ref{fig:m9}, upper panel). 
The resulting core composition at the end of this phase is shown in the lower panel. The main constituents are O, Ne and Mg. Note that the C burning is incomplete and that a non-negligible amount of unburned carbon is left within the innermost 0.5 M$_\odot$. At that time, the mass of the degenerate core is $\sim1.3$ M$_\odot$, which is slightly smaller than the Chandrasekhar limit. Later on, the star enters the super-AGB phase, during which the mass of the degenerate core increases. Meanwhile, an intense mass-loss erodes the external layers. 
If the core will attain the Chandrasekhar mass before the complete erosion of the envelope, a rapid contraction starts. 
Apart from the different core composition, the situation is similar to that already described for the SN Ia progenitors. Also in this case, electron captures are fundamental players. In particular, the contraction starts when the $^{20}$Ne$(e,\nu)^{20}$F is activated (see, e.g. \cite{ecapture2019}).  
Likely, all these stars explode,  but the SN engine, i.e., thermonuclear or core-collapse and bounce, is still a matter of debate \cite{isern1991,jones2016}. 
Indeed, the O ignition requires a much higher temperature and density than the C ignition. Therefore, in case of a pure O-Ne-Mg core, a core-collapse rather then a thermonuclear explosion may take place. 
Also the nature of the resulting compact remnant is unknown. It may be either a neutron star, a peculiar WD or nothing.      
In this context, the presence of some unburned C in the core may favor a thermonuclear runaway.
The existence of this C trigger strongly depends on the $^{12}$C$+^{12}$C low-energy cross section.

% ----- core-collapse SNe -------------
\subsubsection{The final fate of massive stars: ingredients for a successful bounce.}

Stars with $M >11~M_\odot$ ignite carbon in non-degenerate conditions and proceed their evolution through more advanced burning phases, up to the formation of a degenerate iron core\footnote{In very massive stars, those that develop a He core with mass larger than about 40 $M_\odot$,  $e^+e^-$ pair production causes the dynamically
unstable contraction of the O-rich core, which induces an explosive O burning.}. Their final fate is a collapse of the iron core (see, e.g., \cite{woosley2002}). It can be demonstrated that about $10^{53}$ ergs of gravitational energy are released and that most of this energy is spent to produce  neutrinos by weak interactions. Initially, due to the interactions with the in-falling material, these neutrinos remain trapped within a spherical surface called the neutrinosphere. Once a hot proto-neutron star forms at the center, the in-falling  material bounces on its surface and a forward shock starts. However, the kinetic energy acquired by the bounced material is insufficient to bring it to the escape velocity and the shock stalls. Nevertheless, extant models show that on a longer timescale, neutrinos may transfer enough energy to the shock giving rise to a supernova. 
This is called the neutrinos-driven supernova mechanism. 
Various supernova types are likely powered by this engine, among which those classified as type II(P,L,N), Ib and Ic. 
Not in all cases, however, is the result a core collapse a supernova. 
The energy deposited by neutrinos may not be enough to fully sustain the forward shock. 
In such a case, a  black hole forms, either directly or by fallback. 
Several energy-loss processes may contribute to prevent the supernova, among which the photo-disintegration of the matter passing throughout the shock. 
Recent parametric studies of core-collapse  models, revealed the fundamental role played by the progenitor structure \cite{oconnor2011,ugliano2012,ertl2016,Sukhbold2016,ebinger2019}. 
In particular, it was found that the compactness of the pre-supernova core determines if the explosion occurs or fails. 
As first noted by Ref.\cite{imbriani2001} (see also Ref. \cite{sukhbold2020}), the compactness of the pre-supernova core is strictly connected to the efficiency of the C burning.  In particular it was shown that when He burning leaves a larger amount of C, the innermost pre-supernova structures are less compact. As a consequence, a less efficient carbon consumption during He burning, as due to a slow $^{12}$C$(\alpha,\gamma)^{16}$O reaction at $E_{\rm cm}=300$ keV, may favor explosion after core collapse.  

Note that the efficiency of C burning also depends on the  $^{12}$C$+^{12}$C reaction rate. For instance, in the case of low-energy fusion hindrance, a phenomenon that has been observed in reactions involving heavy ions, a significant depression of the $^{12}$C$+^{12}$C reaction rate at  $E_{\rm cm}<2$ MeV would be expected \cite{jiang2007}, with important consequences on the pre-supernova structure and on the resulting nucleosynthesis \cite{Gasques2007}.        

\subsubsection{\texorpdfstring{$^{12}{\rm C}+^{12}{\rm C}$}{12C+12C} fusion}
\label{ssec:c12c12}

\begin{figure} 
\centering
\includegraphics[width=0.9\columnwidth]{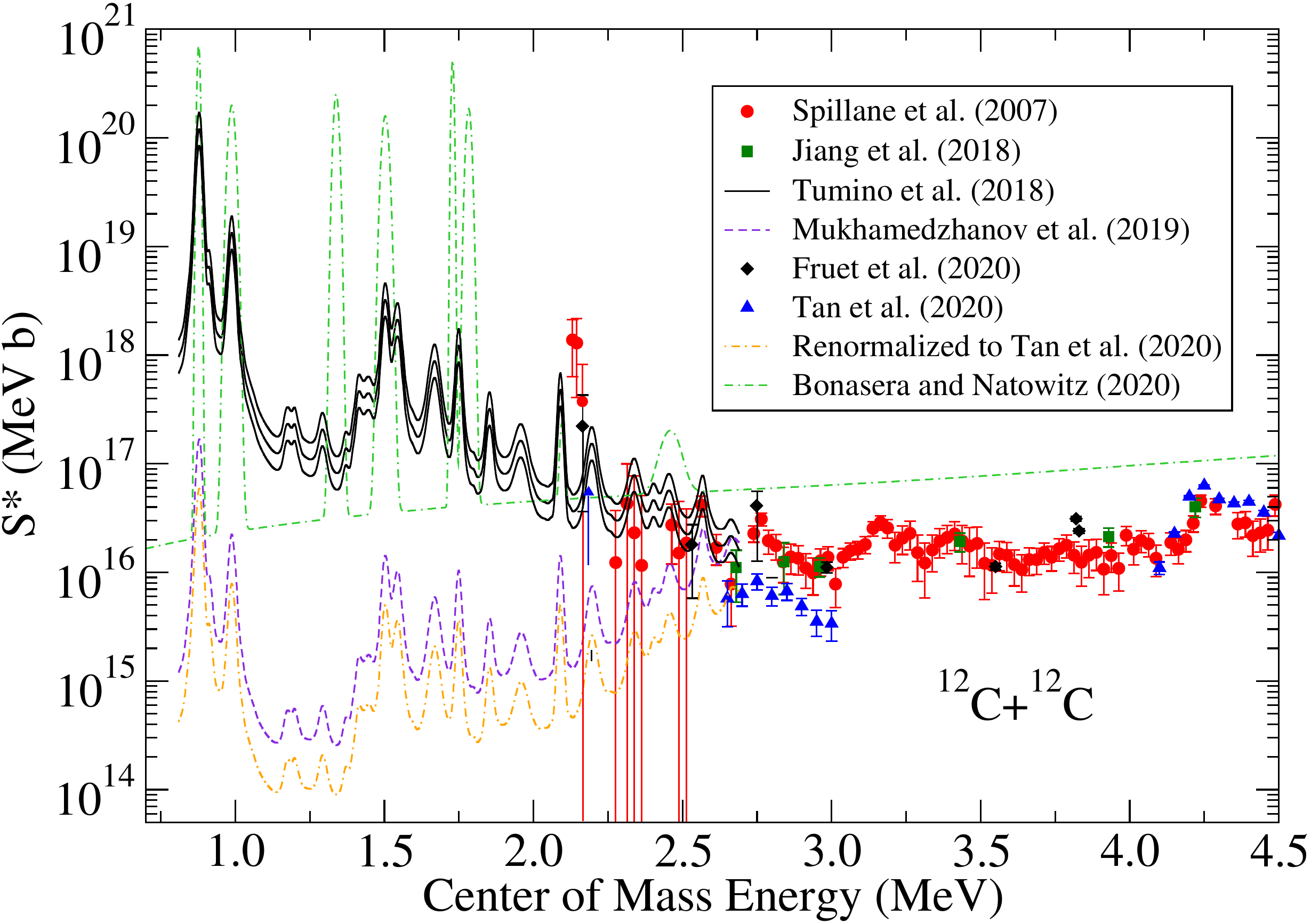}
\caption{Modified $S$-factor ($S^*$) for the $^{12}$C+$^{12}$C fusion reaction from recent experiments: red filled circles from \cite{Spillane07}, green filled squares from \cite{Jiang2018}, black filled diamonds from \cite{2020PhRvL.124s2701T} and blue filled triangles from \cite{2020PhRvL.124s2702T}. The black solid line is from the THM measurements of \cite{2018Natur.557..687T}. The purple dashed line represents suggested corrections to the THM measurements by \cite{PhysRevC.99.064618, Beck2020} and the dashed-dotted orange line represents a renormalization to the data of \cite{2020PhRvL.124s2702T}. Finally, the green dashed-dashed-dotted line represents the recent theory calculations of \cite{Bonasera20}.}\label{S-12C12C} 
\end{figure}

 The relevant channels for $^{12}$C + $^{12}$C fusion at astrophysical energies are those emitting protons and $\alpha$ particles. These channels have been measured by detecting the charged particles and/or the $\gamma$ decay. In particular, the largest branching is for the de-excitation of the first excited states of $^{23}$Na and $^{20}$Ne and for their ground states. 
A reliable measurement of the $^{12}$C+$^{12}$C cross section at low energies is extremely challenging, due to the exponential decrease of the cross section, thus causing a very low counting rate; in this context any natural or beam-induced background must be carefully taken into account for a successful measurement. This was detailed in Ref.~\cite{Spillane07}, reporting the first measurement down to $E_{\rm c.m.}$=2.14 MeV, the lowest energy ever reached for this reaction. The deduced astrophysical $S$-factor exhibits new resonances below 3~MeV, in particular, a strong increase at the lowest energies. This result has triggered several new experimental studies. Here we briefly summarize the recent ones providing the total $S$-factor as a final result. 

The measurement reported in Ref.~\cite{Jiang2018} pushed down to  $E_{\rm c.m.}=2.84$~MeV and 2.96~MeV for the $p$ and $\alpha$ channels, respectively, using a sphere array of 100 Compton-suppressed Ge detectors in coincidence with silicon detectors. 
To overcome the experimental limitations due to the low counting rate, an indirect measurement was performed using the Trojan Horse Method (THM) \cite{2018Natur.557..687T}, covering 
the entire astrophysical region of interest from $E_{\rm c.m.}$=2.7 MeV down to 0.8 MeV and revealing well-resolved resonance structures. 
THM results were normalized to available direct data at $E_{\rm c.m.}$=2.5$-$2.63 MeV.
Following Ref.~\cite{2018Natur.557..687T}, further theory calculations \cite{PhysRevC.99.064618} resulted in large corrections to the initially reported $S$-factors. 
However, these corrections are not the final word and the convergence and numerical stability of calculations involving transfer to the continuum require critical examination.
For example, recent theory calculations using the Feynman path-integral method \cite{Bonasera20} lead to S-factor values that show some agreement with the THM results, but are at odds with the Coulomb-correction to the THM results performed by Ref.~\cite{PhysRevC.99.064618}.

A step forward in the context of direct experiments was achieved in \cite{2020PhRvL.124s2701T}, who reported a measurement down to $E_{\rm c.m.}$=2.2 MeV using the particle-$\gamma$ coincidence technique. Charged particles were detected using annular silicon strip detectors, while $\gamma$-ray detection was accomplished with an array of LaBr3(Ce) scintillators. 
Further recent results were published in \cite{2020PhRvL.124s2702T}, using similar techniques. 
In particular, $p$ and $\alpha$ detection using a silicon detector array, and $\gamma$-ray detection with a high-efficiency HPGe detector. 

Figure \ref{S-12C12C} shows an overall comparison of the modified S-factor, S$^*$, from recent experiments. 
A general agreement within experimental errors is observed in the region of astrophysical relevance, except for the two lowest data points of \cite{Spillane07} and those from \cite{2020PhRvL.124s2702T} in the region $E_{\rm c.m.}=2.7– 3.0$~MeV. 
The current picture calls for additional experimental work in the future in order to corroborate existing results and to push direct measurements down to the astrophysical energies. While the effort towards a direct study of the $^{12}$C + $^{12}$C continues at Notre Dame using an improved version of the SAND detector, a new initiative is under development at the LNGS MV accelerator at the Gran Sasso underground laboratory taking advantage of the cosmic-ray free environment to reduce the background. Complementary to this effort, a new THM approach is planned at the Texas A\&M cyclotron facility using the $^{12}$C($^{13}$C,n)$^{24}$Mg* reaction to minimize the Coulomb interaction in the exit channel and verify the resonance structures observed in the previous $^{12}$C($^{14}$N,d)$^{24}$Mg* study.

%% file: LUNA.tex
Nuclear reaction cross sections are generally extremely small at energies of astrophysical relevance. Therefore, it is often the case that extrapolations guided by nuclear theory must be made from data measured at higher energies~\cite{2017RvMP...89c5007D}. These extrapolations frequently come with large uncertainties, as it is challenging to account for the complex nuclear structure of the nuclides involved. As such, it is important to extend direct nuclear reaction measurements to a wide range of energies, especially down towards those relevant for the astrophysical environment.

Measurements with high-intensity facilities on and near Earth's surface have substantially advanced our understanding of stellar burning over time. For instance, such data largely formed the foundation for our understanding of the solar neutrino flux~\cite{RevModPhys.83.195}. Developments in this area continue, where challenges from background signals are met with increased beam intensities and sophisticated shielding schemes, e.g. using the St. Ana~\cite{2018PhRvC..97f5802L} and LENA~\cite{2020PhRvC.102a4609D} accelerators. Moving to an environment a few tens of meters under rock, as in the Felsenkeller~\cite{2020PhRvD.101l3027G}, provides further background reduction that enables measurements closer to energies relevant for stellar burning. However, to achieve true stellar energies, more drastic measures are necessary. This is where deep underground laboratories enter the picture.

Low-energy studies of thermonuclear reactions in a laboratory at the Earth’s surface are complicated by several sources of background, namely, cosmic rays, environmental radioactivity, and beam-induced nuclear reactions on target impurities.
For a given stellar temperature $T$, nuclear reactions take place mainly inside the Gamow peak, this means that in realistic experimental conditions, the expected counting rate is prohibitively low and the competition with cosmic background strongly hinders obtaining statistically signiﬁcant results. 
The various sources of background result in signals of a different nature and energy, so that each reaction studied needs special attention in suppressing the relevant background component.
In a laboratory on the Earth’s surface, measurements are hampered predominantly by the interaction of cosmic rays in the detectors, leading typically to more than 10 events per hour in common detectors. 
Conventional passive or active shielding around the detectors can only partially reduce the problem. 
Neutron backgrounds require special attention due to the interaction of primary cosmic-ray particles with the Earth's atmosphere. 
The neutron flux is dependent on the geomagnetic latitude and on the phase in the 11-years solar cycle~\cite{Heu1995}.
The flux fluctuations are quite large and the continuous interplay of absorption and new formation in the measuring device is not easy to control~\cite{Bes2016A}. 
The best solution to attenuate the muon and neutron flux is to install an accelerator facility in a laboratory deep underground, as also done for solar neutrino detectors.
The ﬁrst example of this approach has been realised in the experimental halls of Laboratori Nazionali del Gran Sasso (LNGS) in Italy.
Similar approaches are presently exploited in USA (CASPAR) and China (JUNA).

Thermonuclear reactions induced by charged particles are mainly studied by detecting the $\gamma$-ray and/or particle emission accompanying the reaction. This poses an experimental challenge, as natural backgrounds are plentiful. The natural $\gamma$-ray background derives from radioactive decay of long-lived nuclides, e.g. $^{40}{\rm K}$ or the uranium and thorium decay series, at $\gamma$-ray energies below $\sim$3.5~MeV. For energies above $\sim$2.6~MeV, signals from the muons and neutrons produced by cosmic-ray interactions are the primary $\gamma$-detector background source. As such, moving a laboratory underground significantly reduces the higher-energy background, leading to considerable benefits for high $Q$-value reactions. Reducing the $\gamma$-ray background at lower energies requires choosing a location with low natural radioactivity, e.g. due to the local rock composition. A further complication at $\gamma$-ray energies below 2.6~MeV arises due to $(n,\gamma)$ reactions enabled by the $(\alpha,n)$ neutrons that result from $\alpha$-decaying nuclides. One mitigation tactic is to house the detector in a positive pressure environment, so as to flush any $^{222}{\rm Rn}$ that may have escaped the surrounding rock and building materials after being produced in the uranium decay series.
In the following we will present existing and upcoming underground accelerator facilities as a powerful tool to determine nuclear cross sections inside the Gamow peak.

\subsection{LUNA: Laboratory for Underground Nuclear Astrophysics}

The world first underground accelerator was set up by the LUNA Collaboration inside the Laboratori Nazionali del Gran Sasso (LNGS) which is part of the Italian Istituto Nazionale di Fisica Nucleare. 
The underground site of LNGS is covered by a 1400~m thick overburden of rock (3.800~meter of water equivalent (m.w.e.)) which reduces the cosmic muon flux by six orders of magnitude. Easy horizontal access and numerous user facilities attract an international scientific community of more than 950 scientists~\cite{LNGS2011}.  

In this context, the LUNA collaboration has established underground nuclear physics as a powerful tool for determining nuclear reaction rates at Gamow peak energies, paving the way for this experimental approach during thirty years of continuous work. Activities started in 1992 with the installation of a home-made 50~kV accelerator. This pioneering work excluded a resonance in the $^3$He($^3$He,2p)$^4$He reaction at solar energies, which had been suggested as a possible nuclear physics based explanation of the results of solar neutrino measurements, without having to invoke physics beyond the standard model~\cite{Junker1998}. 

The 400~kV Singletron$^{\tiny{\textregistered}}$ accelerator {\em LUNA-400} has been in operation since 2000~\cite{Formicola2003}.
One of the first results obtained using this machine was the measurement of the $^{14}$N(p,$\gamma$)$^{15}$O reaction rate, which was found to be a factor of two slower than expected~\cite{Formicola2004}. This result had enormous consequences such as increasing the age of globular clusters by about 1~Gy~\cite{Imbriani2004} and reducing the CNO solar neutrinos by a factor of two~\cite{Formicola2004}.

During the last 15 years several processes belonging to CNO, MgAl and NeNa cycles have been measured, contributing for example to the understanding of the origin of meteoric stardust~\cite{Lugaro2017}. Also the D(p,$\gamma$)$^3$H reaction has been studied covering the whole energy of interest for the Big Bang nucleosythesis (BBN). LUNA results reached an unprecedented precision, settling the most uncertain nuclear physics input to BBN calculations and obtaining an accurate determination of the density of baryonic matter at the end of BBN~\cite{Mossa2020}.
More recently, thanks to the intense He beam available, the prolific neutron source from the $^{13}$C($\alpha$,n)$^{16}$O reaction has been measured directly inside the Gamow peak (lower energy measured $E_{cm}$=245keV), largely reducing the uncertainty of the cross section determination \cite{CSEDREKI2021165081}.

The LNGS-INFN is currently expanding the accelerator laboratory with special funding of the Italian Ministry of Research,  installing a new 3.5~MV Singletron$^{\tiny{\textregistered}}$ machine designed and set up by High Voltage Engineering Europe (HVEE)~\cite{Sen2019}. The 3.5~MV machine will be equipped with two independent beamlines which can be operated with solid and a gas target systems.  Acceptance tests at HVEE proved that the machine can deliver intense proton, helium and carbon beams (1, 0.5 and 0.15~mA respectively) with well defined energy resolution (0.01~\% of TV) and stability (0.001~\%h$^{-1}$ of TV)~\cite{DiLeva2020}.
 The new 3.5~MV accelerator will be situated only a few meters away from experiments searching for Dark Matter and Neutrinos Double Beta Decay. These projects require that the beam induced neutron flux at their locations is lower than the natural neutron flux inside the underground laboratory. As has been shown in specific  audits, this is achieved by installing the machine and all experimental setups inside 80~cm thick concrete shielding, by careful accelerator design, and by specific procedures for accelerator operation. As the existing  400~kV Singletron$^{\tiny{\textregistered}}$ accelerator still is the perfect blend for the study of most of the proton-capture reactions involved in the stellar H burning it will be moved close to the new 3.5~MV accelerator. 
 
 The two accelerators will be the heart of the {\em LNGS Underground Accelerator Facility} which will be operated by LNGS as a user facility to provide intense  $p$, $\alpha$, and carbon beams in an energy range reaching from a few tens of keV up to MeV. This will enable further study of the key reactions of helium and carbon burning (namely $^{12}$C + $^{12}$C fusion and $^{12}$C($\alpha$,$\gamma$)$^{16}$O).
A first experimental proposal presented by the LUNA-Collaboration focuses on measurements of the reactions  $^{14}$N(p,$\gamma$)$^{15}$O, $^{12}$C + $^{12}$C, $^{13}$C($\alpha$,n)$^{16}$O and $^{22}$Ne($\alpha$,n)$^{25}$Mg, the latter being in the context of the ERC Starting Grant {\em SHADES}.

%% file: CASPAR.tex
\subsection{CASPAR: Compact Accelerator System for Performing Astrophysical Research}

The CASPAR (Compact Accelerator System for Performing Astrophysical Research) laboratory is the only US-based deep underground accelerator facility and is operated by a collaboration of the University of Notre Dame and the South Dakota School of Mines and Technology \cite{ROBE16}. The accelerator system has been fully operational since 2018 and is located 4850 feet underground at the Sanford Underground Research Facility (SURF) \footnote{\url{http://www.sanfordlab.org}} in Lead, South Dakota, formerly the Homestake gold mine. The rock overburden results in a 4300 m.w.e shielding effect, significantly decreasing cosmic ray induced background with a muon flux level of 0.4 $\times$ 10$^{-8}$/cm$^{2}$/s. The residual neutron flux consists of primarily low-energy ($<$10 MeV) neutrons generated by ($\alpha,n$) reactions induced through the decay of naturally occurring uranium and thorium decay-chain isotopes in the surrounding rock, and is generally on the order of 10$^{-6}$ neutrons/cm$^{2}$/s \cite{MEI06,MEI09}.   

The low-background environment and unique location has made SURF an underground science hub ever since 1965, when Ray Davis installed his Noble Prize winning neutrino detector and observed what would become labelled as the solar neutrino problem. A continued expansion for science has resulted in extensive infrastructure available for low-background experiments, including the MaJorana Demonstrator experiment (MJD) \cite{alvi19}, the LUX-ZEPLIN (LZ) dark matter detector \cite{MOUNT17} and the soon to be established Deep Underground Neutrino Experiment (DUNE) system for the Long-Baseline Neutrino Facility (LBNF) \cite{ABI20}. CASPAR is located along side these at the 4850 main science level of SURF.

The CASPAR laboratory is currently aligned towards the measurement of ({$\alpha$,$\gamma$}) and ({$\alpha$,n}) reactions, including those of relevance for the production of neutrons in core helium burning of massive red giant stars (weak s-process) \cite{PIGN10} and shell or inter-shell helium burning of low-mass AGB stars \cite{KAPP11} (see Section~\ref{sec:sprocess}). 

Building on previous above ground work, the CASPAR system consists of a 1 MV Van de Graff style JN accelerator with a 150 kV to 1100 kV operational range well suited for overlap with higher energy measurements. The system is focused on the production of proton and $\alpha$ beams up to $\sim$ 250 $\mu$A on target. To extend the measurements of low-energy reactions, the combined underground environment for background suppression and high intensity ion beam delivery, is further enhanced through the use of high efficiency detection systems. Amongst the standard use of high-purity Germanium (150\%) detectors, CASPAR takes advantage of high efficiency 4$\pi$ detectors such as an array of 20 $^3$He gas filled tubes for neutron detection \cite{FALA13} and the High EffiCiency TOtal absoRption NaI array (HECTOR) \cite{REIN19} for $\gamma$-ray detection. Both detector systems demonstrate up to 50\% efficiency, with an additional benefit for $\gamma$ detection of utilizing the summing technique for excellent peak identification and separation for higher $Q$-value reactions.         

Measurements of interest so far have included the primordial stellar burning reactions $^{11}$B($\alpha$,n)$^{14}$N and $^{7}$Li($\alpha$,$\gamma$)$^{11}$B~\cite{Liu20}, as well as reactions resulting in or competing with $s$-process neutron production, $^{18}$O($\alpha$,$\gamma$)$^{22}$Ne, $^{22}$Ne($\alpha$,$\gamma$)$^{26}$Mg and $^{22}$Ne($\alpha$,n)$^{25}$Mg~\cite{PhysRevC.87.045806}. The program continues to explore stellar neutron sources and expands the present studies into the magnesium range probing the alpha capture reactions on the $^{24}$Mg, $^{25}$Mg, and $^{26}$Mg isotopes. A new program has been initiated to explore the endpoint of nova nucleosynthesis, studying proton capture reactions in the Ar to Fe range. CASPAR is well suited for these measurements, but will be complemented by a new low energy machine, presently under development at Notre Dame. 

%% file: JUNA.tex
\subsection{JUNA: Jinping Underground Nuclear Astrophysics} \label{sec:JUNA}

China Jinping Underground Laboratory (CJPL) was established on the site of hydro-power plants in the Jinping mountain, Sichuan, China \cite{chen,cheng}. The facility, located near the middle of a traffic tunnel, is shielded by 2400 m of mainly marble overburden (6720 m.w.e.), with radioactively quiet rock. 
CJPL phase I (CJPL-I) now houses the CDEX \cite{2013PhRvD..88e2004Z} and PandaX dark matter experiments. 

\begin{figure}[ht] 
  \centering
\includegraphics[width=1.0\columnwidth]{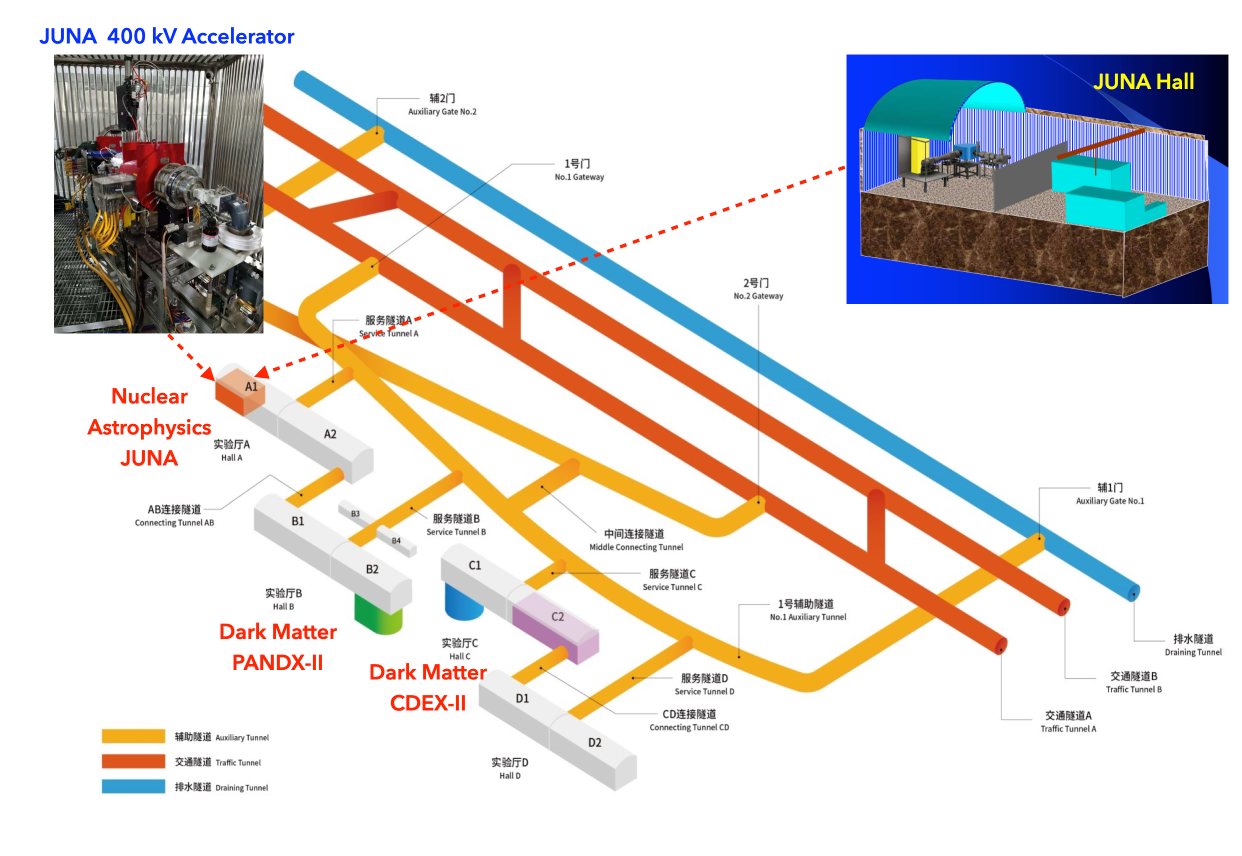}
\caption{The layout of JUNA in CJPL-II. \label{JUNAlayoutene}} 
\end{figure}

CJPL phase II \cite{2014Sci...346.1041N} (CJPL-II) is the expansion followed by the success of CJPL-I.
It has much larger scale underground experiments space (300,000 m$^3$ volume), planned to house CDEX-II, PandaX-II, and JUNA \cite{2016SCPMA..59d5785L}. The layout of JUNA in CJPL-II is shown in Fig.~\ref{JUNAlayoutene}. The complete commissioning of CJPL-II is scheduled for March 2023.
In December of 2020, the JUNA collaboration installed the accelerator in CJPL-II before the long-period of construction. Four reactions have been studied in the first quarter of 2021. Some preliminary results are presented in the following. 

$^{12}$C($\alpha$,$\gamma$)$^{16}$O measurements were performed with a 1~emA $^{4}$He$^{2+}$ beam impinging on a pure $^{12}{\rm C}$ ($<10^{-5}$ $^{13}{\rm C}$ contamination) target surrounded by BGO and LaBr detectors. This resulted in an upper limit of 10$^{-13}$ barn at E$_{c.m.}$ = 538 keV, the most sensitive to date. Further work is ongoing to improve signal-to-noise issues related to the beam intensity, target purity, vacuum, and uncertain sources of background.

$^{13}$C($\alpha$,n)$^{16}$O was measured using two different neutron detector configurations. The first configuration of the detector consists of 24 $^{3}$He proportional counters, distributed in concentric rings, surrounded by a cylindrical plastic scintillator used to suppress background. The second configuration replaces the scintillator with a plastic moderator shielded by borated polyethylene. $^{4}$He$^{1+}$ and $^{4}$He$^{2+}$ beams with intensities from 0.1-2.0~pmA were impinged on 2~mm thick $^{13}{\rm C}$ targets over the energy range E$_{c.m.}$ = 230-600 keV. The experiment results will be published in the near future.

$^{25}$Mg(p,$\gamma$)$^{26}$Al measurements have focused on precision width determinations for the of 92 and 189 keV resonances.  These are two of the key resonances for this reaction rate at temperatures experienced in the hydrogen-burning cores of massive stars, ultimately influencing the nucleosynthesis of cosmic magnesium and aluminum, including the short-lived radio nuclide $^{26}{\rm Al}$~\cite{STRIEDER201260}.
Thick target yields were obtained using a 2 pmA proton beam and 4$\pi$ BGO $\gamma$-ray detector. The 92~keV measurements, which will provide a precise resonance width and ground-state feeding fraction, resulted in $\sim$200~events/day, to be compared to the background rate $<5$/day, over two weeks. These results and the data obtained for 189~keV were combined with an indirect constraint on the 58~keV resonance~\cite{2015SCPMA..58.2002L} and above-ground measurements at 304~keV to result in the most precise determination of this reaction rate to date. Final results will be published soon.

$^{19}$F($p$,$\alpha\gamma$)$^{16}$O and $^{19}$F($p$,$\gamma$)$^{20}$Ne, where the latter connects CNO cycles and the former competes with this connection~\cite{2021PhRvC.103e5815D}, were measured using ground-based test runs and underground runs at JUNA. The measurements employed a CaF$_2$ target surrounded by an HPGe array at $E_{c.m.}>140$~keV and a BGO array at lower energies for high efficiency. The ground based studies, mainly done on the 320 kV platform at IMP Lanzhou, resulted in optimized fluorine targets: first, implanting fluorine ions into the pure Fe backings with an implantation energy of 40 keV, and then sputtering a 50 nm thick Cr layer to further prevent the fluorine material loss.
Underground measurements extended $^{19}$F($p$,$\alpha\gamma$)$^{16}$O results down to $E_{\rm c.m.}$ = 72~keV and $^{19}$F($p$,$\gamma$)$^{20}$Ne down to 188~keV. The latter resulted in the observation of a new resonance at 225~keV, which enhances the rate significantly. This increases the leakage from the CNO cycle and may help explain the Ca abundance in the first generation Pop III stars~\cite{Zhan21}.

The JUNA accelerator includes a 2.45 GHz ECR source, developed for the China ADS project (CIADS). This ion source delivers up to  10 emA proton, 6 emA $^{4}$He$^{+}$, and 2 emA $^{4}$He$^{2+}$ (by a separate ion source). The maximum beam energy out of the ion source is 50 keV/q with emittance less than 0.2 $\pi\cdot$mm$\cdot$mrad. The Low Energy Beam Transport line (LEBT) minimizes space charge effects and improves the beam transport efficiency, where the beam is accelerated before being focused with two solenoids. To keep the LEBT as short as possible, all the steering magnets are built inside the solenoids. Since $^{4}$He$^{2+}$ beam is expected to be mixed with a large fraction of the $^{4}$He$^{1+}$ beam. A 30 $\deg$ magnet will be added between the two solenoids to filter out the intense $^{4}$He$^{1+}$ to reduce the burden of the acceleration tube and to purify the beam.

%% file: intro-rings.tex
Recoil separators play an important role in a number of current research areas, including nuclear astrophysics and the study of nuclei far from stability. 
In these areas, they have made it possible to perform otherwise unfeasible measurements between light projectiles and radioactive nuclei with too short a lifetime to be used as a target. 
The approach requires  measurements to be done in inverse kinematics, {\em i.e.} using the radioactive species as a beam onto hydrogen or $\alpha$-particles as a target (either solid or gas). 
With radioactive ion beam facilities, such as ISOL~\cite{2006LNP...700.....A} and fragmentation~\cite{2004LNP...651.....A} producing accelerated beams of relatively high intensity, the inverse kinematics technique brings 
several advantages to traditional measurements, such as unique capabilities for recoil identification, as long as the primary beam can be significantly suppressed~\cite{2014EPJA...50...99R}. 
In essence, recoil separators separate reaction products from unreacted beam and disperse them according to their mass-to-charge-state ratio. 
This is achieved using a combination of electric and magnetic fields in devices such as Wien filters, dipoles, and quadrupoles. 
In conjunction with a suite of focal plane detectors they can provide full identification of the recoiling reaction products.

In a radiative capture reaction, the fusion of projectile and target nuclei produces a nucleus that recoils in the laboratory system with an average momentum $p_{\rm r0}$ very close to that of the projectile. The momentum of a given recoil nucleus depends on the emission angle for the corresponding $\gamma$-ray. 
As a result, the trajectories of the recoils lay within a cone centered on the beam axis with an opening angle $\theta = \arctan(E_\gamma /(c~p_{\rm r0}))$, where $E_\gamma $ represents the energy of the emitted gamma ray and $c$ is the speed of light. 
Correspondingly, the momentum of the recoils varies in a range equal to $2 E_\gamma /c$ around $p_{\rm r0}$. 

The principle of Recoil Mass Separator (RMS) measurements is to determine the reaction yield, and thus the cross section, by directly detecting the recoils produced in a reaction. This requires that the reaction measurement take place in inverse kinematics, where the heavier reaction species impinges on the lighter one, in order to maximize $p_{\rm r0}$. The reaction target must also be thin enough for recoils, which are forward-focused by reaction kinematics, to escape. The escaping recoils and unreacted beam nuclides are emitted with similar angles and energies, where one of the former is emitted for every $\sim10^{10}-10^{17}$ of the latter. This staggering difference in statistics is overcome with an ion optical system that is tuned to select the recoil species while suppressing the beam. Typical ion optical elements include magnetic dipoles for momentum selection, electrostatic analyzers or cross field Wien filters for velocity selection, along with focusing elements to keep a selected charge-state of recoils within the system. The ion optical system terminates in a focal plane consisting of one or more end detectors used to measure properties such as the species energy and time-of-flight in order to identify the detected nuclides and provide further suppression of unreacted beam. The rate observed in the focal plane 
$R$ is related to the total reaction cross section $\sigma$
by the following expression: 
\begin{equation}
R=N_{\rm b} N_{\rm t} \sigma \epsilon T \Phi_r,   
\end{equation}
where $N_{\rm b}$ is the number of projectiles impinging on the target with an areal density $N_{\rm t}$; $\epsilon$ is the detection efficiency of the end detector; $T$ and $\Phi(q_r) $ are the transmission and the charge state probability for the selected charge state $q_r$ of the recoils, respectively. 

The target thickness required to operate a RMS is usually not sufficient to reach the equilibrium charge state distribution. As a consequence, the charge state distribution of the recoils at the exit of the target is determined by their charge state distribution at their formation and the reaction coordinate along the target. Since its prediction is quite complex and uncertain, a post-stripper consisting of a foil or a windowless gas-cell is often used to reach charge state equilibrium.  
Occasionally, the use of a post-stripper can be avoided measuring the reaction yield in all possible charge states for the recoils.  

The transmission factor $T$ requires a trade-off between an analyzing power sufficiently high to suppress unreacted beam and an acceptance large enough to collect all recoils and achieve $T=1$. Achieving $T=1$ is critical, as the exact momentum and angle distribution of recoil nuclides depends on the angular distribution of $\gamma$-rays relative to the beam direction, along with further modifications from straggling within the target and post-stripper. Therefore, correcting for $T<1$ is extremely complicated and generally involves large systematic uncertainties~\cite{GIAL11}.

When compared to more traditional experiments in direct kinematics, the use of recoil separators can further help to control sources of systematic errors, and, importantly, to suppress additional experimental background. 
Recoil separators such as DRAGON, ERNA, St George, have been successfully exploited for measurements with both stable and radioactive ion beams and key studies are presented in the following sections.

In what follows, we review the techniques and performance of some of the separators used to address experimental challenges of nuclear astrophysics, and highlight some of the important advances in the field.

%% file: ERNA.tex
\subsection{ERNA: European Recoil Separator for Nuclear Astrophysics}

In the late 1990s,  the RMS ERNA was built and installed at the Dynamitron Tandem Laboratorium of the Ruhr University Bochum, Germany, based on the experience of the NABONA (NAples-BOchum Nuclear Astrophysics) RMS~\cite{GIAL96}. NABONA managed the first measurements of a radiative capture reaction, $\rm ^7Be(p,\gamma)^8B$, by means of the direct detection of the recoiling nuclei, without the condition of a coincident detection of the prompt gamma rays to suppress the  background originating from the beam ions leaking to the final detector~\cite{GIAL00}. 
The design of ERNA aimed at the necessary acceptance to measure $^{12}$C($\alpha,\gamma$)$^{16}$O down to $E_{\rm cm}=700$~keV exploiting the peculiar recoil emittance caused by the angle-energy correlation induced by momentum conservation in the gamma ray emission. 
The result was achieved using crossed electric and magnetic field velocity filters (Wien Filters, WFs), that allow varying the analyzing power, at the cost of a complicated optics because of the difficult matching of the electric and magnetic fields. 
The extension of the gas target limited the actual acceptance to $E_{\rm cm}=1.3$~MeV \cite{SCHU04}. 
In fact, measurements were further limited to $E_{\rm cm}=1.9$~MeV~\cite{SCHU05}, because of an unexpected $^{16}$O background and a drastic reduction of the beam suppression at lower energy. 
This issue turned out to be determined by the over-focussing of beam ions in a charge state higher than the one selected for the recoils in the lens directly following the target. 
A solution to this problem was the modification of the RMS layout, introducing a dipole magnet that selects a single charge state for both recoil and beam ions entering the first lens. 
The design of this Charge State Selection Magnet (CSSM) was rather complex, since its effective length had to be kept as short as possible not to increase the distance of the lens to the target too much. 
Due to the short length a significant fraction of the magnetic field strength is in the fringe field, that needed to be accurately tailored. 
This solution was implemented upon the transfer of ERNA to the Italian laboratory CIRCE (Center for Isotopic research on Cultural and Environmental heritage) of the Department of Mathematics and Physics, University of Campania, Caserta, Italy.

\subsubsection{Setup}
\begin{figure}[ht]
    \centering
     \includegraphics[width=1.0\columnwidth]{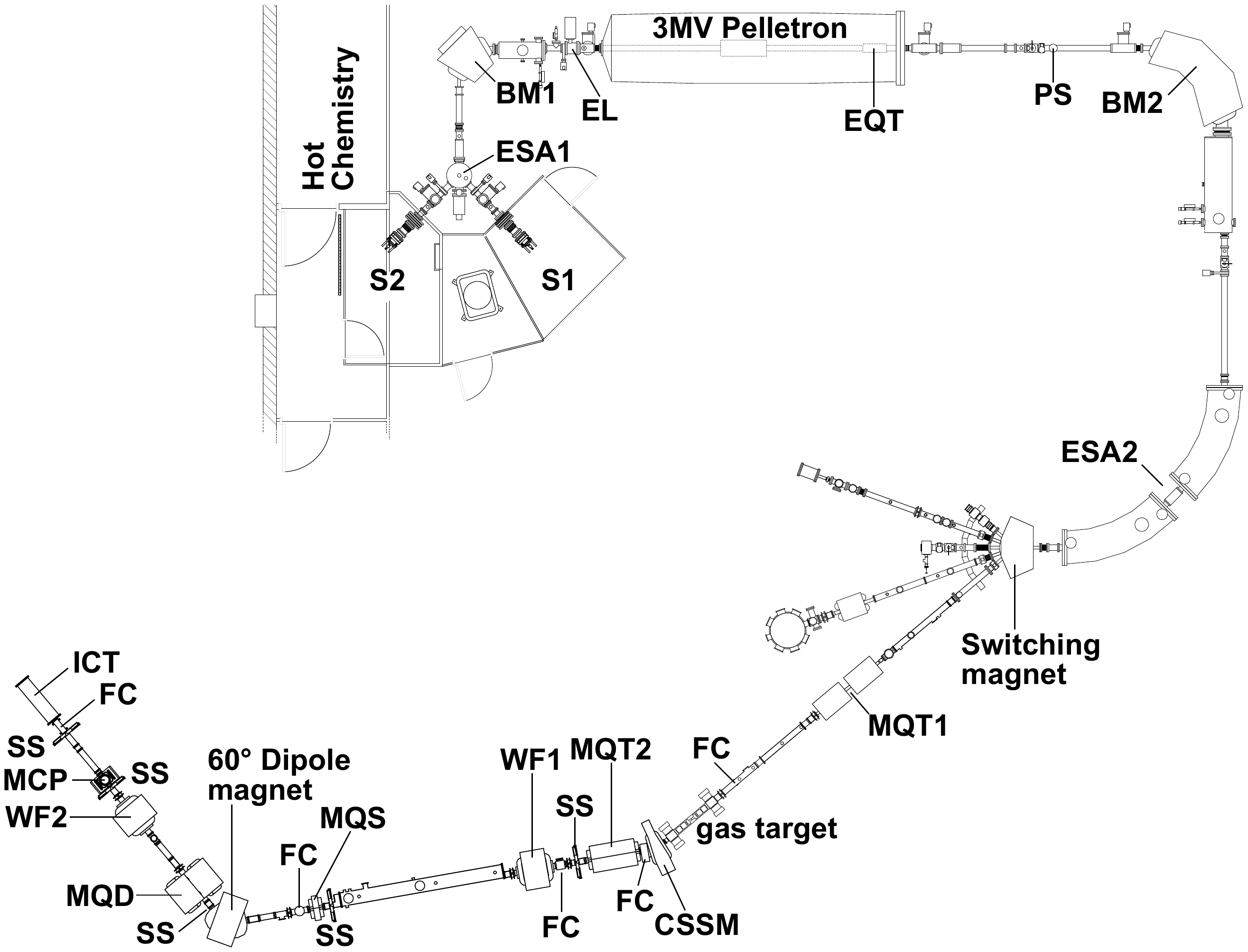}
    \centering 
    \caption{Scheme of the ERNA RMS at the Tandem Accelerator Laboratory of CIRCE, Caserta, Italy. s=ion Source; ESA=ElectroStatic Analyzer; BM= Bending Magnet; EL = Einzel Lens; EQT = Electric Quandrupole Triplet; PS = Post stripper; MQT = Magnetic Quadrupole Triplet; FC = Faraday Cup; SS = Slit System; WF = Wien Filter; MQT = Magnetic Quadrupole Singlet; MQT = Magnetic Quadrupole Doublet; MCP = Multi Channel Plate; ICT = Ionizazion Chamber Telescope.  The drawing is adapted from Ref.~\cite{BUOM18}}
    \label{ERNA:fig1}
\end{figure}

Fig.\ref{ERNA:fig1} shows the layout of the recoil separator ERNA. 
Briefly, a negative ion beam is produced by the two available ion sources, S1 for stable ions, S2 for medium to long lived radioactive ion beams. 
The beam is extracted with an energy up to 90 keV, analyzed by a combination of an electrostatic analyzer and a dipole magnet.
The beam component with the selected mass is injected into the 3MV Tandem accelerator in the selected mass. 
Positive ions are formed in the HV terminal, where both a solid and a gas stripper systems are available. 
The beam emerging from the accelerator is analyzed by a combination of a 90 degree bending magnet and an electrostatic analyzer, that is designed for AMS applications. 
It is worth noting that the high mass resolution of the injection system combined with the high analyzing power of the AMS component reduces the contamination of recoil-like ions in the beam to a negligible amount, thus not requiring an additional purification stage as reported in Ref.~\cite{1999EPJA....6..471R}. 
Finally the beam is transported and focused onto the ERNA target system. 

There are two options for the target system: an extended windowless gas target~\cite{2013EPJA...49...80S} and a gas jet target~\cite{2017NIMPB.407..217R}. Both systems are equipped with a post-stripper cell allowing the recoils to reach an equilibrium charge state distribution regardless of the position in the target where they are formed. 
Recoils emerge from the target accompanied by the intense beam ions in all possible charge states and an overlapping linear momentum spectrum. 
The CSSM immediately after the target selects a single charge state for both recoils and beam ions, as mentioned above. 
Subsequently, a series of focusing and analyzing elements provides the necessary beam suppression. 
The new layout of ERNA strongly reduces the probability for the beam ions to reach the end detector by multiple scattering, since different charge states are captured at different locations with optimal separation.
 Different end detectors are available: a two stage ionization chamber for $\Delta E-E$~\cite{1999NIMPA.437..266R} and a TOF-E detector~\cite{2008JPhG...35a4021D} capable of charge and mass identification, respectively,  of the detected particles. Recently, a new position sensitive TOF-E setup has been realized. 
 In the first phase of its operation, ERNA was equipped with rather simple gamma-ray detection setups, consisting in an array of 3 or 6 NaI detectors \cite{DILE08,SCHU11} along the extended gas target.  
 Recently, a new array has been realized and commissioned. By the end of 2021 ERNA will be equipped with the new array, consisting of 18 NaI detectors around the jet target in a geometry optimized to measure angular distributions.
\subsection{Recent measurements}

$\rm ^{15} N(\alpha,\gamma)^{19}F$ was the first reaction studied with the new layout of ERNA at CIRCE \cite{DILE17}. Subsequently, the reaction $\rm ^7Be(p,\gamma)$ has been approached \cite{BUOM18}, exploiting the intense $^7{\rm Be}$ beam available at CIRCE \cite{LIMA17}. Thus ERNA at CIRCE could be used for the first time to measure proton capture reactions, that in the layout at Bochum could never be investigated because of insufficient beam suppression. 
 
 \subsubsection{Future plans}
 The current and near-future science program is focused on pushing direct measurements of $\rm ^{12} C(\alpha,\gamma)^{16}O$ to center of mass energies below $\rm E_{cm}$= 1 MeV, including total cross section measurements and angular distributions. The envisioned goal of the experiments is a more robust estimate of the stellar rate, verifying some inconsistencies of previous measurements of both E1 and E2 ground state transitions, the main amplitudes contributing to the total cross section. In parallel, a broad program exploiting the $\rm ^7Be$ beam available at CIRCE to study both proton and electron captures on $\rm ^7Be$ is planned.

%% file: DRAGON.tex
\subsection{DRAGON: Detector of Recoils And Gammas of Nuclear Reactions}

\subsubsection{Setup}
The DRAGON ({\bf D}etector of {\bf R}ecoils {\bf A}nd {\bf G}ammas {\bf O}f {\bf N}uclear reactions) recoil separator located at the ISAC ({\bf I}sotope {\bf S}eparator and {\bf A}Ccelerator) beam facility at TRIUMF, Vancouver, Canada's Accelerator Center, was designed for direct measurements of radiative capture reactions on protons and $\alpha$ particles~\cite{HUTCHEON2003190}. Post-accelerated radioactive ion beams produced by the ISAC facility as well as stable ion beams from OLIS ({\bf O}ff-{\bf L}ine {\bf I}on {\bf S}ource) are delivered to DRAGON at energies between $\sim$0.15~A~MeV and 1.8~A~MeV.

DRAGON  consists of three main sections: (1) a windowless, differentially pumped, recirculated gas target, with an effective length of 12.3~cm, surrounded by a high-efficiency $\gamma$-detector array consisting of 30 BGO detectors; (2) a high-suppression electromagnetic mass separator consisting of two stages of charge and mass selection by means of magnetic and electrical dipoles and (3) a variable heavy ion detection system with unique capabilities for recoil identification in combination with two micro-channel plate (MCP) based timing detectors allowing for time-of-flight measurements. The recoil detection system either consists of an isobutane-filled ionization chamber with a segmented anode, a double-sided silicon strip detector (DSSSD) or a recently implemented hybrid detector system, combining the advantages of both detector types.

Prior to a measurement, the energy loss across the gas-filled target is determined by measuring the incoming and outgoing beam energy. This allows for determining stopping powers ($\epsilon$) based on the measured energy loss, gas density derived from continuously monitored and read-out pressure and temperature, and the effective target length~\cite{HUTCHEON2003190}. This eliminates uncertainties induced by the commonly used software packages SRIM~\cite{ZIEGLER20101818} and LISE~\cite{Kuchera_2015}.
To  further reduce systematic uncertainties, charge-state fractions for the relevant recoil energies are measured using a stable beam of the isotope of interest impinging on the target.

DRAGON's capabilities have further become more versatile with the commissioning of the SONIK ({\bf S}cattering {\bf O}f {\bf N}uclei in {\bf I}nverse {\bf K}inematics) scattering chamber, which can be installed in place of the regular DRAGON gas target. SONIK is a windowless, extended gas target, surrounded by 30 ion implanted charged-particle detectors mounted on doubly collimated telescopes at precisely defined angles. The design allows for measuring scattering cross sections at three different energies at a given incident energy. SONIK was successfully commissioned in 2018 performing a measurement of the  $^{3}$He + $\alpha$ elastic scattering cross section down to 0.4~MeV in the center-of-mass frame~\cite{Pan20}.

\subsubsection{Recent measurements} \label{sec:DRAGON_recent}
Recent measurements at DRAGON utilized the high-intensity stable beams delivered by TRIUMF's Offline Ion Source (OLIS). One of these measurements concerns the $^{22}$Ne($\alpha$,$\gamma$)$^{26}$Mg reaction, whose importance is covered in Section~\ref{sec:sprocess}. The data taken for resonances in the $^{22}$Ne($\alpha$,$\gamma$)$^{26}$Mg reaction in 2019 are expected to be published in 2021.
Further, the results from stable beam measurements to investigate reactions such as $^{22}$Ne({\it p},$\gamma$)$^{23}$Na~\cite{LENNARZ2020135539, PhysRevC.102.035801}, $^{76}$Se($\alpha$,$\gamma$)~\cite{Fall20}, $^{34}$S({\it p},$\gamma$)~\cite{Love21} and $^{19}$F({\it p},$\gamma$)~\cite{Will21} have recently been published.

\subsubsection{Future plans} \label{sec:DRAGON_future}
To extend DRAGON's capabilities, it is planned to replace the present BGO array with a LaBr$_{3}$ array for superior timing (sub-ns) and energy resolution. This will allow for a faster and more precise method to determine the resonance energy by correlating the beam bunch arrival time (accelerator RF) with a prompt $\gamma$-detection in the array. 
As the beam traverses the gas target, it loses energy and eventually reaches resonance energy. Beam energy and gas target pressure are chosen in such a way that the beam becomes ``on-resonance'' in the target center. However, with the present BGO hit pattern method, this requires knowledge of resonance energies and stopping powers prior to the measurement as collecting sufficient statistics for an accurate determination requires long measurement times for weak resonances. A first test of this new method was carried out in 2020, where 5 LaBr$_{3}$ detectors were used to successfully demonstrate the feasibility of the new approach. Another new development involves the coupling of HPGe clover detectors from the GRIFFIN array with the DRAGON gas target, which will provide a significant addition to DRAGON's capabilities.

Upcoming experiments involve the SONIK scattering chamber to perform measurements of the $^{7}$Be + p and $^{7}$Be + $\alpha$ scattering cross section towards a more accurate multi-channel R-matrix description. Additionally, with improved capabilities of ISAC's beam delivery, the planned measurement campaign on $^{11}$C + p with DRAGON and TUDA is now within reach.

%% file: StGeorge.tex
\subsection{St. GEORGE: Strong Gradient Electromagnetic Online Recoil Separator for Capture Gamma-Ray Experiments}

The St.~George recoil separator \cite{COUD08} is dedicated to the study of ($\alpha, \gamma$) reactions of astrophysical interest. High-intensity ion beams for elements up to mass $A\approx50$ are delivered to St.~George by the Santa Ana single-ended Pelletron accelerator, which has an electron cyclotron resonance (ECR) ion source in the terminal. The target for ($\alpha, \gamma$) reactions with St.~George is the HIPPO gas-jet target, which is well characterized \cite{KONT13,2016NIMPA.828....8M}.
St.~George is designed to have an angular and energy acceptance of $\theta=\pm$40~mrad and $\Delta E/E=\pm7.5\%$, respectively.
Separation of recoils from unreacted beam is accomplished with St.~George by using six dipole magnets, a Wien filter, and a focal plane detection system.
The focal plane detection system achieves particle identification using time-of-flight and energy-loss measurements, which are appropriate for the type of beam/recoil energies found in typical experiments. The time-of-flight measurements rely on micro-channel plate detectors in an $\vec{E}\times \vec{B}$ configuration to maximize transmission. 

An experimental demonstration of the energy acceptance at 0$^\circ$ is presented in \cite{MEIS17}. Measurements at larger angles yielded a more limited acceptance of $\theta=\pm$30~mrad and $\Delta E/E=\pm4\%$. Efforts are underway to understand and optimize these results.

Two commissioning experiments, well within the measured acceptance, have been performed to validate the whole system. The first experiment was the study of a resonance doublet at 5603 and 5604~keV excitation energy in $^{18}$F with the $^{14}$N($\alpha,\gamma$)$^{18}$F reaction, and the second experiment was the study of three well-known resonances in the $^{20}$Ne($\alpha,\gamma$)$^{24}$Mg (10.916, 11.015 and 11.216~MeV in $^{24}$Mg). Beams of $\approx$100~pnA were used for both reactions. With St.~George tuned for maximum transmission, a count rate of $\approx$1000~pps was measured in the focal plane detector, yielding a beam suppression with the separator alone of less than 1.6$\times$10$^{-9}$. The added separation provided by the detection system, ~10$^{-4}$, was sufficient to achieve identification of the recoils. The final analysis of the resonance strengths for each of the reactions is underway, but preliminary results agree with literature values.

While the preliminary results are encouraging, measurements off-resonance demonstrated the presence of beam contamination by ions of the same mass and p/q as the products of the reaction of interest at a level of ~$10^{-13}$ recoil-like contaminant per incoming beam particle. There is no contaminant rejection solution within St.~George or the focal plane detector. A Wien filter was recently installed on the beamline upstream of HIPPO to remove the contamination and allow for the measurement of small cross sections.

The upcoming research program of St.~George will be dedicated to the study of the reaction chain $^{14}\mathrm{N}(\alpha,\gamma)^{18}\mathrm{F}(\beta^{+}\nu)^{18}\mathrm{O}(\alpha,\gamma)^{22}\mathrm{Ne}(\alpha,\gamma)^{26}\mathrm{Mg}$ or $^{22}\mathrm{Ne}(\alpha,n)^{25}\mathrm{Mg}$, whose importance is described in Section~\ref{sec:sprocess}. The reaction $^{15}\mathrm{N}(\alpha,\gamma)^{19}\mathrm{F}$, responsible for the production of $^{19}$F, will also be investigated.

%% file: SECAR.tex
\subsection{SECAR: Separator for Capture Reactions}

SECAR is designed \cite{BERG18} to study ($p$,$\gamma$) and ($\alpha$,$\gamma$) reactions induced by beams from the ReA3 reaccelerator \cite{KEST10} at FRIB. It will address long-standing questions associated with explosive stellar environments thanks to the significant increase in radioactive beam production capabilities of FRIB. It will also allow for stable-beam induced reaction studies with ReA3 in standalone mode. SECAR consists of three parts: (i) JENSA the windowless gas target \cite{2014NIMPA.763..553C,BARD16,2018NIMPA.911....1S}, that can be reconfigured in extended mode, is surrounded by a BGO detector array for $\gamma$-ray detection; (ii) an electromagnetic mass separator composed of a charge-state selection section, two Wien filters for high beam suppression and a final magnetic rigidity analysis, and; (iii) at the focal plane, final beam discrimination is achieved with a modular detection system composed of two MCP systems 1.4~m apart for time-of-flight measurements, an ion chamber for $\Delta E-E$ measurements, and a large silicon detector for total energy measurements, with each of the three detection system parts being position sensitive.

SECAR construction is completed (as of March 2021) and commissioning has already started. Early commissioning has demonstrated the reproducibility of the ion optics settings at various magnetic rigidity. Beam was delivered to the focal plane (the second Wien filter was replaced with a beam pipe) and machine learning techniques were developed, and used, to tune the beam on target within the spot size and exit angle requirements to maximize transmission and guarantee that the beam is properly centered in the various magnetic elements.

A large open collaboration\footnote{\url{http://secar.space/\#collaboration}} has now submitted a broad set of proposals to the first FRIB program advisory committee \cite{FRIBPAC2021}. In those proposals, in addition to the core goals of SECAR, new experimental directions are being explored such as: the study of ($\alpha,n$) reactions in the context of the weak r-process; ($p,n$) reaction measurements for supernovae nucleosynthesis studies; ($d,n$) reactions to indirectly evaluate ($p,\gamma$) reactions with lower beam intensity requirements; and ($d,p$) surrogate reactions for indirect ($n,\gamma$) measurements. With a large open collaboration supporting a broad scientific program, SECAR has an exciting future for direct and indirect reaction studies.

%% file: intro-rings2.tex
A general limitation of low-energy nuclear astrophysics studies comes from the extremely low cross sections, which translate into small yields and low signal-to-noise ratios.
This means that whenever an ion beam impinges on a target, only rarely will a nuclear reaction take place: most of the beam remains un-reacted and goes to waste either in the target itself or in a beam dump downstream of the target. 
This issue becomes especially critical for radioactive ion beams whose intensities are typically many orders of magnitude lower than for stable beams.

A solution to overcome beam intensity limitations comes from the use of storage rings, where the beam is recirculated many times and therefore has repeated chances to interact with the target.
Storage rings therefore also hold the potential to access more exotic reactions far from stability. 
Pioneering measurements have already been performed at the \textsl{Experimental Storage Ring} ESR at GSI in Darmstadt, Germany~\cite{1987NIMPB..24...18F}: the $^{96}$Ru(p,$\gamma$)$^{97}$Rh radiative capture reaction~\cite{2015PhRvC..92c5803M}, albeit at energies much higher than those of astrophysical interest and, later, the $^{124}$Xe(p,$\gamma$)~\cite{2019PhRvL.122i2701G} at energies approaching the Gamow window. In both cases, the beam consisted of stable ions. While, at the moment, (p,$\gamma$) reactions on stored \textsl{radioactive beams} in a certain lifetime range are studied at the ESR showing promising first results, exciting new developments are also opening up a new era for nuclear astrophysics with storage rings in the near future. The recent installation of an additional storage ring at GSI, CRYRING, enables the first proton-capture measurement at energies relevant for hydrostatic as well as explosive astrophysical scenarios. Moreover, new and more dedicated storage rings for nuclear astrophysics are planned at other facilities~\cite{2020PrPNP.11503811S}.

%% file: ESR.tex
\subsection{ESR \& CRYRING: Reaction Studies on Stored Exotic Beams}

The Experimental Storage Ring (ESR) is coupled to the rare ion beam facility FRS at GSI and therefore plays a pioneering role in nuclear physics, which is obvious from the unique nature of exotic decay studies and the rich spectrum of mass measurements of exotic nuclei accomplished in the past \cite{2013NIMPB.317..603L}. 
Only recently, with the initiative for the so-called proton-capture campaign, the focus has been shifted towards direct studies of nuclear reactions at low energies motivated by nuclear astrophysics and with the final goal to address radioactive nuclei. 

\begin{figure}[ht]
    \centering
     \includegraphics[width=1.0\columnwidth]{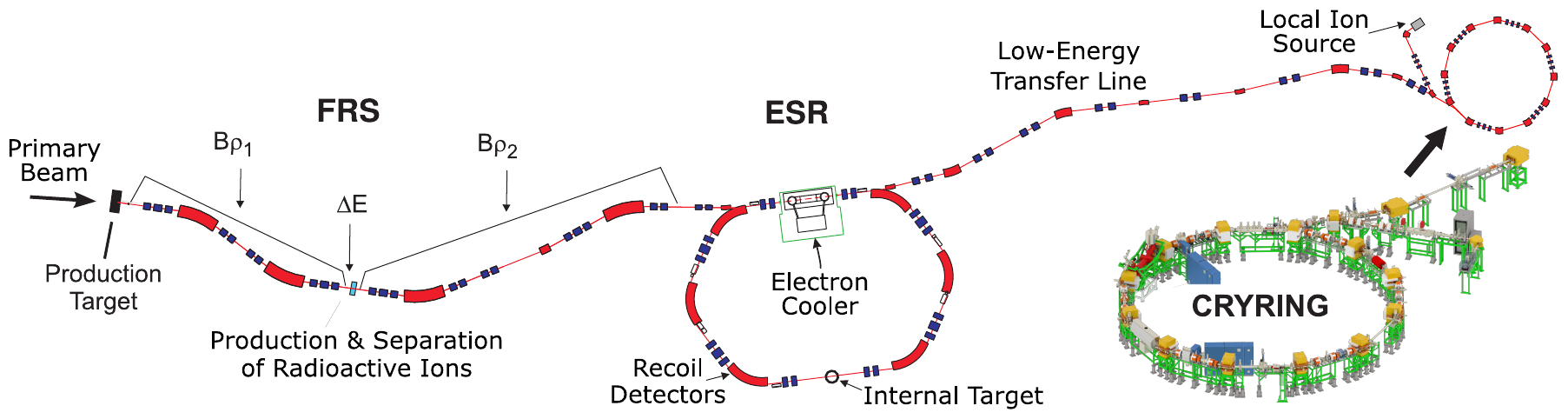}
    \caption{Shown is the layout of the FRS-ESR-CRYRING complex at GSI. Fragment beams produced at relativistic energies in the FRS can be stored, manipulated and investigated in the connected rings. The ESR can store ions at energies as low as 3 MeV/$u$ and it can serve as an injector for CRYRING covering energies down to about 100 AkeV. The picture is taken from \cite{2020PrPNP.11503811S}}
    \label{ESR:fig1}
\end{figure}

The ESR can store any ion beam within an energy range of 500 MeV/$u$ to 3 MeV/$u$ and provide clean and brilliant (exotic) beams for reaction studies using, e.g., the internal jet gas target \cite{1987NIMPB..24...18F}. 
With the newly installed CRYRING facility GSI inherited a dedicated low-energy storage ring from Stockholm University \cite{2016EPJST.225..797L}. After its initial recommissioning the smaller ring now serves as a low-energy extension for ESR beams and also as a standalone machine with a local ion source. After in-flight production in the FRS at relativistic energies, rare ions can now be cooled, post-decelerated and stored in the full range down to about 100 keV/$u$. 

\subsubsection{Technique} \hfill\\

%Beam Intensity and Reaction Luminosity
When studying stored exotic ions the beam intensity is usually the main challenge due to the limited production in the FRS. 
The issue of disturbing contaminants in the fragment beam is, however, slightly simplified by injecting to a storage ring, because many initial contaminants are out of acceptance of the ring and will not be stored.
Additionally, a post-stripper can be used, which dominantly converts all products to bare ions and thins out the number of potentially disturbing m/q-values.

The stored ions circulate in the ring and hit the interaction zone at a frequency of several 100 kHz, which boosts the reaction luminosity by about 5 orders of magnitude. 
If the directly injected beam intensity is still too low to accomplish a certain experiment, as for instance expected for radioactive beams with low production efficiency in the FRS, several subsequent beam bunches can be accumulated in the ring. 
This procedure is called stacking and can fill up the available phase space in the ring step-by-step. 
In extreme cases several tens of bunches can be accumulated within minutes in order to reach the desired intensity.

As an example from the recent past, about $10^{7}$ stable $^{124}$Xe ions could be stored at energies as low as 5.5 MeV/$u$, which is close to the space charge limit. These ions hit the target at a frequency of about 300kHz, where the H$_2$ jet provided densities of about $10^{14}$ atoms per cm$^{2}$. 
In total a luminosity of about $10^{26}$ cm$^{-1}$s$^{-1}$ could be reached for the study of $^{124}$Xe(p,$\gamma$) \cite{2019PhRvL.122i2701G}. 
For lighter ions the space charge limit allows more intensity; however, usually the energies of interest and the related cross sections are significantly lower. As a result the luminosity challenge remains unchanged or becomes even more critical.

%in-ring beam manipulation
One of the major advantages of a heavy ion storage ring as the ESR is the provided flexibility regarding beam manipulation \cite{1987NIMPB..24...18F,2020PrPNP.11503811S}. A central feature is beam cooling, which ensures a narrow momentum distribution on the order of $\Delta$p/p = $10^{-4}$ or below. 
In the ESR two complementary cooling techniques are employed: Stochastic cooling and electron cooling. The first is extremely useful to rapidly cool down a hot radiobeam injected from FRS, while the latter is commonly used continuously during the measurement to counteract the steady energy loss and beam expansion caused by interactions at, e.g., the gas target.

In combination with beam cooling, a set of slits and scrapers allows a certain fragment beam to be singled out among still disturbing contaminants, given that the spatial separation of the fragments, i.e. the difference in ion mass, is large enough.

For low-energy studies another key ingredient is the post-deceleration of the stored beam. 
In the ESR the ions can be slowed down to about 3 MeV/$u$ using RF cavities, while simultaneously ramping the entire magnetic system of the ring in order to keep the beam on a central orbit. 
This technique is especially powerful for radioactive ion beams, because it enables the combination of high energy in-flight production and low-energy measurements available nowhere else in the world.

The key to all beam manipulations in the ring is a detailed beam diagnosis. 
With a recycling beam the powerful, non-destructive technology of Schottky noise detection is available and used as the central diagnosis in ESR as well as CRYRING. 
The tiny pickup signal any ion leaves in the Schottky cavity is used to generate a frequency spectrum covering a large bandwidth \cite{2020RScI...91h3303S}. 
This spectrum reveals the revolving frequency of any stored ion and, as a result, beam operations such as orbit changes, deceleration, beam cooling, and many more can be monitored in a unique way.
For exotic decay studies  and mass measurements the Schottky technique is often used as the main detection system, because it is extremely flexible and applicable to any beam and intensities down to single ions.

%ultra-high vacuum, ion detection and storage times
For many experiments in the ring the goal is to sustain storage for minutes, hours or even days on a high intensity level.
Unfortunately, the storage time is limited by the extent of interactions leading to losses. 
At low energies this is typically dominated by the atomic processes of recombination and ionization. 
The three main sources for such beam losses in the ring are the electron cooler, the gas target and the residual gas in the beam pipes. 
For reaction studies, continuous cooling as well as the use of the gas target can only be optimized for efficiency but never fully avoided. 
For this reason it is vital to ensure that vacuum quality and composition do not dominate these losses and cause severely reduced storage times. At low beam energies where atomic cross sections can reach the megabarn regime, this is one of the main challenges at the rings. 

The vacuum system in the ESR is designed for $10^{-11}$ mbar, which dictates a highly restrictive list of ultra-high vacuum (UHV) compatible materials and leads to the design of standard ion detection systems located behind a vacuum barrier, usually a stainless steel foil \cite{2003NIMPB.204..553K}. 
For low-energy ion detection, however, this solution is not feasible anymore, since heavy ions below ~10 MeV/$u$ will be stopped inside the foils.
In the last decades huge efforts have been made to design a versatile detection system compatible to the UHV environment. For heavy recoil detection in ESR there are now double-sided silicon strip detectors (DSSDs) available, which comply with the strict UHV conditions and provide full performance, i.e., position, energy and time resolution for ions impinging at moderate rates \cite{2019PhRvL.122i2701G}. 
This recent achievement finally facilitated direct reaction studies at energies relevant for nuclear astrophysics and the wide field of low-energy applications.

\subsubsection{Proton-Capture Measurements} 
\begin{figure}[ht]
    \centering
     \includegraphics[width=1.0\columnwidth]{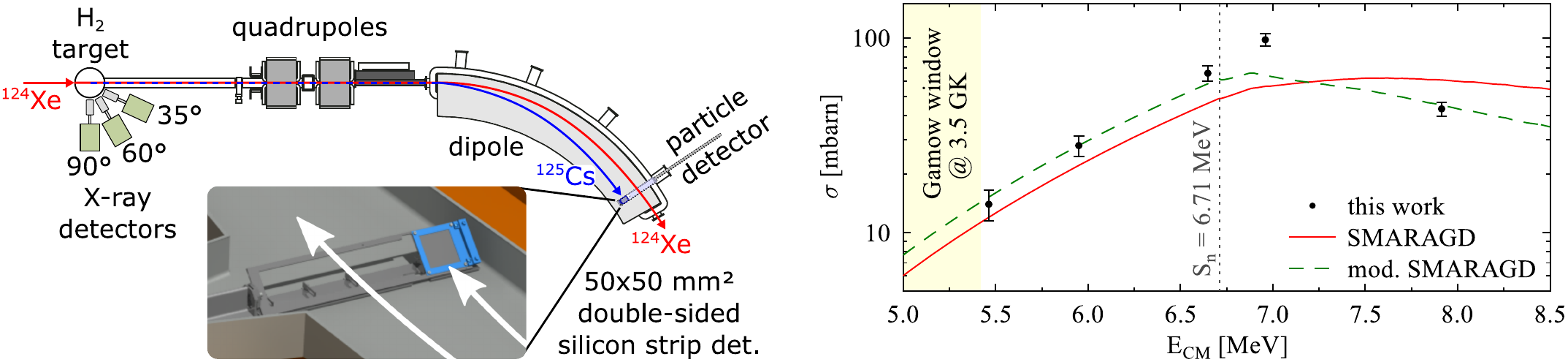}
    \centering 
    \caption{Left: The experimental setup for proton-capture measurement on $^{124}$Xe in ESR is shown, which comprises X-ray detectors surrounding the H$_2$ target and the UHV recoil detection system inside the dipole. Right: Cross section results for $^{124}$Xe(p,$\gamma$) measured just above the Gamow window for explosive nucleosynthesis. Both pictures are taken from \cite{2019PhRvL.122i2701G}.}
    \label{ESR:fig2}
\end{figure}

\hfill\\
Proton-capture reactions are an important ingredient for element synthesis and often involve unstable nuclei, especially for scenarios of explosive nucleosynthesis as in the $\gamma$ or \textit{rp} process \cite{2013RPPh...76f6201R}. 
Inverse kinematic investigations are the only feasible technique for direct measurement in such cases and the storage rings at GSI provide a novel approach in this respect. The method is rather simple and based on the fact that a stored heavy ion that undergoes radiative capture of a proton at the H$_2$ target, does in good approximation keep its initial momentum. 
This leads to a beam-like focus of reaction products, which can be straightforwardly separated from the main beam in a dipole field and intercepted by appropriate detectors. 
As indicated in figure \ref{ESR:fig2} the UHV recoil detection system in ESR is located inside the first dipole after the gas target and can be moved close to the beam.

It is important to note that proton-capture and electronic ionization, due to the comparable momentum and equal charge states of the recoils, would leave nearly indistinguishable signatures in the detector in terms of energy and position. 
Because of the large atomic cross sections, a crucial condition for this technique is to utilize bare ions for which ionization is excluded, otherwise the proton-capture signal would be hidden below an overwhelming background. 
Further background contributions are to be expected from elastic scattering and other open nuclear channels, e.g. (p,n) or (p,$\alpha$). 

The nuclear cross section can be measured relative to well-known radiative recombination cross sections by employing X-ray spectroscopy around the gas target. 
In many cases the obvious choice is to concentrate on the electron capture into the empty K-shell of the orbiting ions, which is the inverse process of the photo-effect and can be predicted theoretically with very low uncertainties \cite{2007PhR...439....1E,2010PhRvA..82b2716A}. Such atomic physics techniques are well established at the GSI storage rings and help to avoid the large uncertainties inherent to a classical luminosity determination via target density, beam intensity, and their mutual overlap. 
Additionally, the feasibility of normalizing to the Rutherford scattering distribution, which usually dominates the background below the (p,$\gamma$) signature, has been demonstrated recently at a comparable level of uncertainty \cite{2020NIMPA.98264367X}.

Even for stable beams the injection energy of the beam has to be on the order of 100 MeV/$u$ to ensure efficient stripping of the ions. 
Once the beam is stored and potentially stacked, the next steps are cooling and deceleration. 
After obtaining a brilliant low-energy beam, the H$_2$ target is switched on and data taking can start until the stored intensity drops below a certain value, at which point the ESR is reset and the entire cycle starts over again.    

\subsubsection{Major Achievements and Recent Developments} \hfill\\
The first experiment of the proton-capture campaign at the ESR, about a decade ago, was the proof-of-principle study for the technique aiming to measure $^{96}$Ru(p,$\gamma$). Recoil detection was accomplished at that time by using standard DSSDs behind a stainless steel foil of 25 $\mu$m, which prevented a measurement below $E_{\rm cm}=9$~MeV. Three data points for $^{96}$Ru(p,$\gamma$) between 9 MeV and 11 MeV have been published and the method was assessed to be worthy of refinement \cite{2015PhRvC..92c5803M}. 
The main difficulty in the data analysis in this experiment was to disentangle the various signatures of different reaction channels measured by the recoil detector, which was accomplished by employing detailed Monte-Carlo simulations. 

After a long shutdown period at GSI and extensive development of the new UHV detection system, another pilot experiment was launched. 
This time, by addressing the reaction $^{124}$Xe(p,$\gamma$), the main goal was to measure the cross section at energies close to the Gamow window for explosive nucleosynthesis. 
Enabled by the new in-vacuum recoil detector used in the configuration shown in Figure \ref{ESR:fig2}, five data points could be measured between $E_{\rm cm}=5.5-8.0$~MeV, just above the Gamow window, demonstrating the applicability for astrophysically motivated measurements~\cite{2019PhRvL.122i2701G}. 
At these lower energies competing nuclear channels can be mostly neglected and the focus in data analysis moved to deal with the broad background distribution below the (p,$\gamma$) signature originating from Rutherford scattering of $^{124}$Xe off the hydrogen target.

In fact the signal-to-background ratio goes down rapidly when approaching the Coulomb barrier more closely because of the divergent behaviour of the involved cross sections. 
Since the ambitious final goal of the experimental campaign is to measure with radioactive ion beams of limited intensity, a new approach to increase the overall sensitivity of the method has been pursued. 
The main idea is based on blocking the scattering distribution directly in front of the separating dipole magnet, where the Rutherford cone has already a sizable extension, while the proton-capture products are still on the central orbit of the ring \cite{2020JPhCS1668a2046V}. 
In accordance with the simulations the preliminary results of a very recent measurement show that this blocking scheme indeed seems to work well and will maximize the sensitivity of the technique as expected. 

\subsubsection{Prospects for Direct Measurements at ESR and CRYRING} \hfill\\
For the future, the proton-capture campaign at ESR will be continued with radioactive beam measurements relevant for the $\gamma$ and \textit{rp} process.
Moreover, it is envisioned to advance the experimental technique in order to address other nuclear channels highly relevant for nuclear astrophysics, such as (p,n) and ($\alpha$, $\gamma$). 
There are ideas to establish prompt $\gamma$-ray detection at the gas target or to extract the beam-like reaction products from the ring entirely. Finally, the first proton-capture measurement in CRYRING is intended to be realized as soon as possible, which will eventually allow full coverage of the Gamow window even for stellar scenarios of lower temperature.

The combination of ESR and CRYRING provides some additional benefits for direct measurements, in particular for beams that need complicated treatment in the ESR and for which only short storage times can be realized. 
In such a case the measurement in CRYRING can run, while the next load of ions is prepared in the ESR simultaneously, which strongly improves the duty cycle of an experiment.

For CRYRING there are several projects on-going to realize indirect and direct measurements with astrophysical motivation. Two prominent examples are the direct determination $^{44}$Ti($\alpha$,p) as well as the investigation of deuterium destruction during the Big Bang by addressing the reaction D(p,$\gamma$)$^3$He.

%% file: NeutronActivation.tex
Together with accurate stellar models, neutron-induced reaction
rates (particularly neutron capture) are needed to understand the
abundances of the heavy elements ($60<A<210$).  
Because there is
no Coulomb barrier to overcome, the energies at which the reactions
proceed are determined by the thermalized energy distribution, so
the neutron energies of interest match the stellar temperature when
the neutrons are being produced.   
At these low energies, neutron
capture is the dominant reaction channel that is open.  The
Maxwellian-Averaged Cross Section (MACS), the differential cross section
folded with a temperature-dependent Maxwellian weight, is needed at
temperatures determined by the active stellar burning process.
The two
major \emph{s}-process contributors, the \emph{main} and \emph{weak}
s~processes, as discussed in section \ref{sec:sprocess}, 
are responsible for \emph{s}-process contributions to $90<A<210${}
and $60<A<90$, respectively.

For the main \emph{s}~process, both $^{13}$C$(\alpha,n)${} and
$^{22}$Ne$(\alpha,n)${} serve as neutron sources in low-mass AGB
stars, however, the two neutron sources operate at different
temperatures.  $^{13}$C$(\alpha,n)${} activates following the formation
of a $^{13}$C pocket following the proton ingestion in the He-shell,
operating at temperatures of kT$\sim$8~keV.  In contrast, the
$^{22}$Ne source activates briefly but intensely during the He-shell
flash at a much higher temperature of kT$\sim$30~keV.  As a result,
for the main \emph{s}~process, cross sections covering neutron
energies from $E_{n}\sim$500~eV to $E_{n}\sim$200~keV are needed.

The weak \emph{s}~process, producing \emph{s}-process isotopes in the
$60<A<90${} mass range, operates in massive stars.
$^{22}$Ne$(\alpha,n)${} is the primary neutron source, activating both in core
He-burning, at temperatures of kT$\sim$30~keV and in C-shell burning, at
kT$\sim$90~keV.  With this expanded temperature range, for the weak
\emph{s}~process neutron capture cross sections are needed up to
~$E_{n}=$500~keV.

A Hauser-Feshbach (HF) approach is typically employed to
calculate neutron capture cross sections \cite{RaT01}. HF parameters for $(n,\gamma)$ reactions on stable nuclides have been able to be determined for many nuclei
such that the calculational accuracy is typically 25-30\%.
Unfortunately, for many \emph{s}-process studies, higher fidelity is
needed.  Further, the predictive capability for HF approached for
neutron capture moving away from stability, where the nuclear structure
is less well studied, can be suspect~\cite{2020arXiv201001698N}.  Finally, in nuclei with low level
densities, the statistical assumptions needed for the HF approach to be
reliable are not achieved until higher temperatures are reached.
Together, these issues all argue for the continued need for detailed
measurements.

There are two basic classes of approach to measuring
these cross section.  The first is to do an integral measurement in a
neutron spectrum that is similar to the Maxwellian neutron distribution 
and will be discussed in section~\ref{neutron_activation}.
The second, discussed in section~\ref{neutron_tof},
is to perform a neutron-energy differential cross section
measurement and then calculate the MACS from a Maxwellian weighting.

\subsection{Recent Accomplishments}

Before discussing the details of the experimental techniques and 
facilities, we first highlight both a broad effort and several
individual measurements, representative new measurements, and advances
of the last decade.

\begin{figure}
  \includegraphics[width=\textwidth]{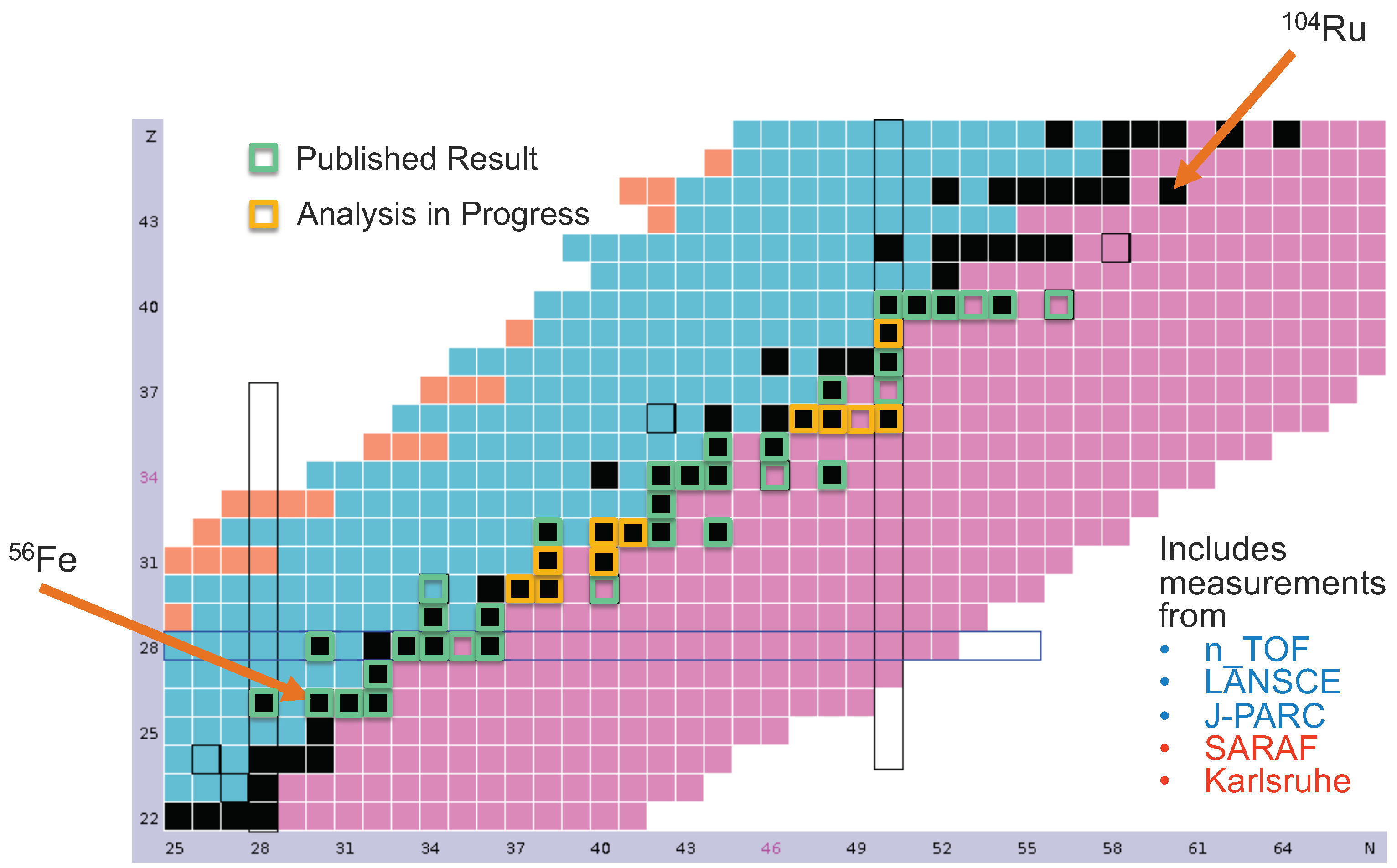}
  \caption{Highlighted in green are those isotope where new neutron
           capture measurements have been performed.  Indicated in
           orange are those cases where new measurements have been
           performed and analysis is in progress.
           \label{weak_capture_expt}
           }
\end{figure}

The weak \emph{s}~process, operating in the mass $60<A<90${} region,
involves nuclei that tend toward smaller neutron capture cross section
with relatively large background reactions from neutron scattering.
They exhibit the low level densities that make HF calculations
challenging.  In what started over a decade of concerted effort, Nassar
\emph{et al.} \cite{NPA05} showed that the $^{62}$Ni$(n,\gamma)${} cross section was
a factor of two smaller than previously thought \emph{and} that this
change impacted the nucleosynthesis of approximately 20 other isotopes
produced following capture on $^{62}$Ni \cite{NPA05}.  
A subsequent study
demonstrated that this propagation effect generally held in the weak
\emph{s}~process for nuclei with
a cross section of $\lesssim 100$~mb \cite{PGH10}.  
As a result, this
entire mass region was actively measured over the last decade,
discovering that while the prior understanding of some of the cross
sections was correct, many were off by 30-50\%, as was the case with
$^{62}$Ni.  In Fig.~\ref{weak_capture_expt}, the range of new
measurements is shown.  As can be seen, almost all of the isotopes in
the weak \emph{s}~process have been remeasured in last two decades.
These measurements have benefited from improvements in neutron
sources as well as, most importantly, detector systems and
data-acquisition systems.  

\begin{figure}
    \centering
    \includegraphics[width=\textwidth]{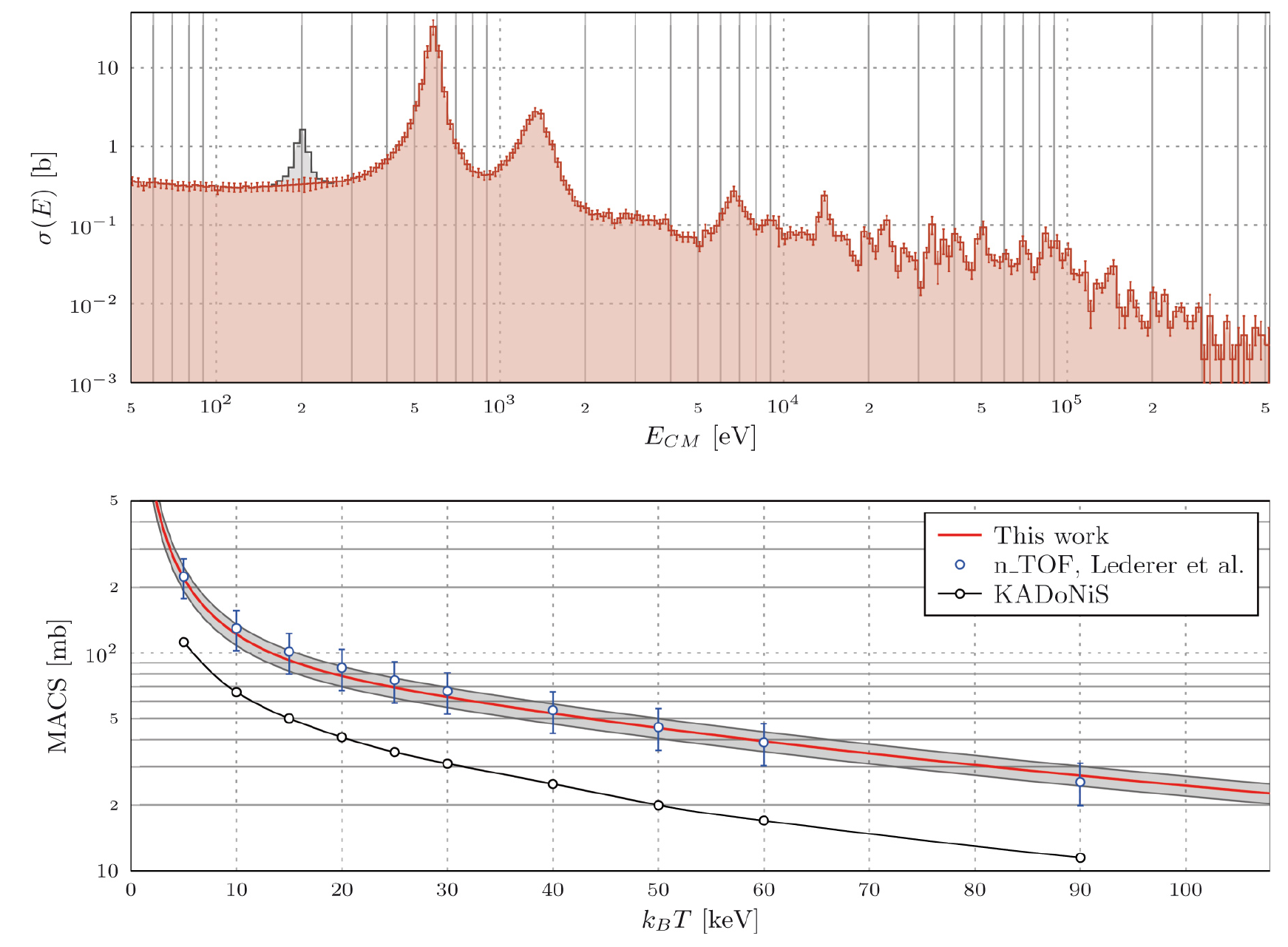}
    \caption{In the upper panel, the differential neutron capture cross section 
             for $^{63}$Ni is shown as measured with the DANCE instrument at Los 
             Alamos.  Shown below is the calculated MACS, together with the results
             from the n\_TOF measurement and prior estimates (figure used with
             permission from \cite{WBC15}).
             }
    \label{ni63_ng}
\end{figure}

As one highlight of this work, time-of-flight measurements (see Section~\ref{neutron_tof}) were performed on $^{63}$Ni, a branch-point nucleus at the onset of the weak 
\emph{s}~process \cite{LMA13,WBC15}.  While cross sections for 
$^{63}$Ni$(n,\gamma)${} have long been desired, acquiring adequate sample material has been an enduring challenge.  
The increased fluxes
that modern TOF facilities provide made measurement possible on much 
smaller, and less pure, samples than historical facilities could handle.  The 
measured cross section from \cite{WBC15} is shown in Fig.~\ref{ni63_ng}.

\begin{figure}
    \centering
    \includegraphics[width=0.8\textwidth]{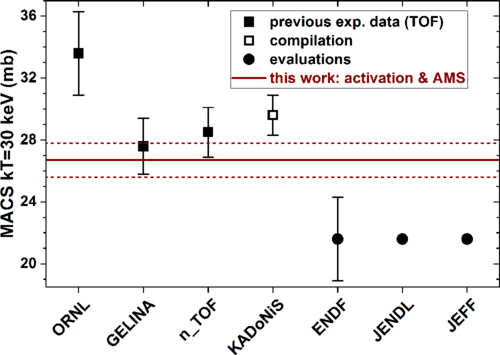}
    \caption{Shown above are the cross section results from the Wallner \emph{et al.} 
             activation measurement of $^{54}$Fe$(n,\gamma)${} compared to prior work
             and evaluations. (figure used with permission from \cite{WBB17}). }
    \label{fe54_ng}
\end{figure}

Similarly, activation techniques have advanced, making strides to perform
measurements on unstable isotopes as well as advance detection techniques to make
a wider range of isotopes amenable to activation techniques, even when the half-lives
make simple counting difficult.  The recent measurement of Wallner \emph{et al.}
of neutron capture on $^{54}$Fe$(n,\gamma)${} illustrates how these techniques have
advanced \cite{WBB17}.  Capture on $^{54}$Fe produces $^{55}$Fe, with a 2.7~y 
half-life, which decays 100\% to the groundstate of $^{55}$Mn by electron capture.
A range of neutron sources sources was used to perform activations across a wide
range of energies.  The produced $^{55}$Fe was counted by Accelerator Mass 
Spectroscopy (AMS), using the Vienna Environmental Research Accelerator (VERA).
The deduced MACS cross section at $kT =30$~keV is shown in Fig.~\ref{fe54_ng}.  Because
of the potential role non-resonant capture can play in $^{54}$Fe, the alternate
systematic approach offered by activation was particularly important for resolving
discrepancies between past evaluations.

\subsection{Neutron Activation}
\label{neutron_activation}

%Before discussing the details of current measurement facilities, it is worth highlighting some recent successes in improved neutron capture measurements, with a focus on work performed for the weak \emph{s}~process.  

Neutron activation measurement facilities offer powerful, highly
selective, integral measurements when activation is possible.  This
typically requires that, to perform a measurement on $^{A}Z$,  the product
$^{A+1}Z${} have a half-life that is amenable to counting---typically
hours to 100s of days---and a decay radiation that can be reliably and
definitively measured.  When $\gamma$-rays are produced and the decay
branchings are well known, counting with high-purity germanium detectors gives clear, unique signatures that allow cross section
determinations even with low-enrichment or even unenriched samples.  In
addition to decay counting, AMS techniques
have been used to count the number of product atoms produced in cases
where the half-lives are too long for direct counting.

\subsubsection{The Karlsruhe Facility}
Forschungszentrum Karlsruhe, now part of the Karlsruhe Institute of
Technology (KIT), established techniques to use the $^{7}$Li$(p,n)${}
reaction to produce a pseudo-Maxwellian spectrum at $kT\approx 25$~keV
\cite{BeK80}.  By employing a proton beam with energy slightly above the
reaction threshold and a solid lithium deposit target, the neutron spectrum generated is peaked at 25~keV
and extends up to approximately 100~keV.  The high-energy deviation can
be corrected by complementing the measurement with theory at high
energies.  In addition to activation measurements, the Karlsruhe
accelerator could be operated in pulsed mode, providing low-resolution
differential cross section measurements within the $^{7}$Li$(p,n)${}
spectrum.  The Karlsruhe group also designed and built a BaF$_2${}
calorimeter \cite{WGK90} that proved to be a workhorse for neutron capture
measurements of astrophysical interest and inspired other calorimeters
at n\_TOF and LANSCE.  The neutron fluence that could be
achieved was ultimately limited by the stability of the solid lithium
target under proton irradiation.  After decades of contributions to
understanding \emph{s}-process nucleosynthesis, the facility was
shut down in the last decade.

\subsubsection{The Soreq Applied Research Accelerator Facility (SARAF)}
The Soreq Applied Research Accelerator Facility has presently completed
the SARAF-I project, which delivered an extremely high-intensity,
low-energy RFQ-based accelerator delivering up to 2~mA of proton current
with a maximum proton energy of 4~MeV \cite{MAA18}.  This intensity is
far beyond what the Karlsruhe lithium target could withstand.  Instead,
a windowless, flowing liquid lithium target was designed and implemented
(LiLiT) \cite{PTF19}.  Since the LiLiT target is liquid and flowing, it
repairs itself as the the proton beam interacts, allowing more than on
order of magnitude higher intensity than could be achieved at Karlsruhe,
with neutron fluences of up to $6\times10^{10}$~n/s, assuming a 2~mA
beam current, for a pseudo-Maxwellian spectrum.  Though higher fluxes are
available with higher energy protons or even deutron beams, these
spectra are generally less relevant for stellar burning scenarios.
While the measurements at SARAF are presently limited to activation as
no prompt detection capabilities exist, SARAF measurements have pushed
to employ a wide range of post-activation measurement techniques
including, $\gamma$-, $\alpha$-, and $\beta$-counting as well as AMS counting
\cite{WTA17,STP17,TPH18}.

\subsubsection{Frankfurt Neutron Source}
The Frankfurt Neutron
Source (FRANZ), installed at the Goethe of Frankfurt, employs
neutron production via $^{7}$Li$(p,n)$. The current version of the facility uses a 3~MV Van de Graaff
accelerator to produce proton beams, with maximum DC beam currents of 
20~$\mu$A. The facility has capabilities for off-line activation
counting of $\gamma${} and $\alpha${} activities \cite{RBB19}. The current focus are activations with short-lived products and neutron spectra different than the $kT$=25~keV \cite{Rei18,Kha21}. In the future, a RFQ-based driver with DC and pulsed-beam capabilities and currents in the regime of several mA is anticipated \cite{Rei09}.

\subsubsection{Future Developments}
The future for activation measurements is bright, with the rapid
advances in neutron source brightness enabling new ranges of
measurements.  The SARAF facility is presently undergoing an upgrade to
SARAF-II, which is expected to double the proton beam on target,
drastically expand the energy range for the proton and deuteron beams,
and may add prompt measurement capabilities.  At FRANZ, there is a planned
accelerator upgrade to a RFQ that is expected to deliver mA proton beams 
for neutron production via $^{7}$Li$(p,n)$.
While a liquid lithium target is not planned at FRANZ, lithium target
advances should allow measurements at much higher intensities than
achieved at Karlsruhe.  Further, the BaF$_{2}${} array from Karlsruhe
has been moved to Frankfurt, which will enable prompt measurement of
neutron capture with this much brighter source.  Finally, both of these
facilities offer intensities high enough to consider in situ
radio-isotope production and then neutron capture on the produced
species, a novel approach to addressing the challenge of sample material
of unstable isotopes.

%% file: NeutronTOF.tex
\subsection{Neutron Time-of-Flight}
\label{neutron_tof}

\begin{figure}
  \includegraphics[width=\textwidth]{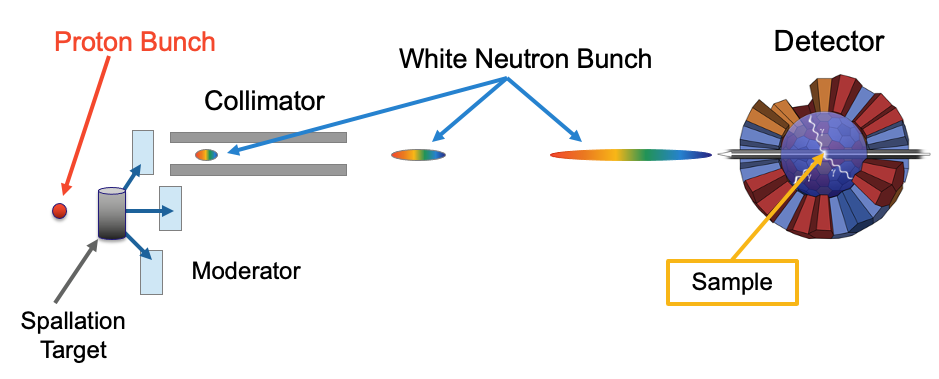}
  \caption{Shown above the basic concept for a pulsed, proton-driven, white
           spallation neutron source.  Protons hitting the spallation
           target liberate neutrons which may interact with a moderator
           before continuing down a flightpath.  Collimators define the
           beam shape and size before it arrives at the sample position.
           The time between the arrival of the proton pulse and an event
           in the detector defines the neutron time-of-flight, and thus,
           the neutron energy.
           \label{tof_schematic}
          }
\end{figure}

In contrast to activation measurements, which are limited to a single or
a small number of energies, time-of-flight (TOF) measurements typically employ
a so-called ``white'' neutron source as a wide range of neutron energies
are covered in a single measurement.  While there are a
range of techniques that can be used, the three facilities
providing the majority of neutron capture measurements for nuclear
astrophysics are all using proton spallation on a heavy, dense target as
the neutron source.  This offers the advantage of good timing and very
high intensity.  In addition to cross section measurements, TOF
facilities can measure neutron resonance properties, which are often
important input for HF calculations.  In some cases, this allows
reliable determination of reaction cross sections even when direct
measurements are not possible at astrophysically relevant energies. The basic concept of a TOF facility is illustrated in
Figure~\ref{tof_schematic}.

\subsubsection{The Los Alamos Neutron Science Center (LANSCE)}
The LANSCE facility TOF neutron sources employ an 800~MeV proton beam to
produce neutrons from tungsten spallation on two different spallation
targets, the Weapons Neutron Research (WNR) bare tungsten target and the
Luj\'{a}n Center moderated tungsten spallation target
\cite{LBR90,LiS06}.  The time structure for the WNR facility makes it
ideal for studies with neutron energies above 1~MeV, so the majority of
astrophysical research at LANSCE is performed at the Luj\'{a}n Center.
Both the Detector for Advanced Neutron Capture Experiments (DANCE) and
the Device for Indirect Capture Experiments on Radionuclides (DICER) are
located on Luj\'{a}n Center flightpaths \cite{HRF01,KUC21}.

The Luj\'{a}n Center spallation target receives a nominal 100~$\mu$A,
800~MeV proton beam at a 20~Hz repetition rate and has the capability to serve
up to 16 independent flight paths.  The beam is accumulated in a storage
ring prior to delivery to the spallation target.  The proton beam is
delivered to the spallation target in a single, $\sim$125~ns FWHM spill.
Together with the moderation time, this sets the ultimate resolution
achievable \cite{MoM13}.

The Detector for Advanced Neutron Capture Experiments is situated on a
$\approx$20~m flight-path at the LANSCE Luj\'{a}n Center.  A
calorimeter for neutron capture, DANCE consists of 160~BaF$_{2}${} scintillators arranged in a
spherical geometry, with each detector subtending the same solid angle.
The goal of this design was to enable measurements on radioactive
samples where the individual detector instantaneous rate is a major
concern.  By using Q-value gating, DANCE can perform measurements on a
wide range of radioactive targets once samples are available
\cite{CoR07,ERM19}.  Recent measurements have focused on isotopes for
the weak \emph{s}~process, \emph{e.g.}, $^{63}$Ni \cite{WBC15}, $^{65}$Cu \cite{PCJ19}, and $^{63}$Cu \cite{WBC17}.

\subsubsection{The CERN n\_TOF Facility}

The n\_TOF facility at CERN began operation in 2001.  Neutrons are
produced by delivering 20~GeV protons from the CERN Proton Synchrotron
onto a lead spallation target \cite{GTB13}.  Because of the very low repetition rate
($\sim$0.4~Hz) with high instantaneous intensity, n\_TOF can use
a nominal 185~m flight path for neutron capture measurements, which
offers exceptional neutron energy resolution.  There are two independent
detector arrays designed for neutron capture experiments.  The first is
a custom-designed, low-background C$_{6}$D$_{6}${} array consisting of
deuterated liquid benzene detectors with a carbon-fiber
superstructure to minimize interactions with scattered neutrons
\cite{PHK03}.  C$_{6}$D$_{6}${} offers the advantage of exceptionally
low neutron interaction cross sections.  The pulse-height weighting
technique is used to correct for the variation in the gamma-ray
efficiency as a function of neutron energy \cite{AAA04}.  A second
capture detector system, the Total Absorption Calorimeter (TAC), is
40-element BaF$_{2}${} array based on the original Karlsruhe calorimeter
design and optimized for use with a spallation neutron source
\cite{GAA09}.  The TAC system offers the advantage of Q-value separation
to separate capture on different isotopes or materials in the sample.
Together, resonance analysis and cross section measurements have come
from the n\_TOF neutron capture program on a wide range of isotopes for
nuclear astrophysics, $^{70}$Ge \cite{GLA19} and $^{171}$Tm \cite{GLP20}.

A second flightpath---EAR2---has been constructed at n\_TOF.  The EAR2
flightpath is nominally 20~m instead of the almost 200~m of EAR1.  While
this decreases the TOF resolution, it drastically enhances the neutron
flux, making measurements possible on much smaller samples \cite{CGC15}.  While the
present focus has been on fission cross section measurements, this will
offer extended reach for neutron capture and neutron-induced charged
particle reactions.

\subsubsection{The J-PARC Facility}
The final major spallation neutron source to discuss is the Materials
and Life Science Experimental Facility (MLF) and the Japan Proton
Accelerator Reseach Complex (J-PARC).  Similarly to the Luj\'{a}n Center
at LANSCE, the MLF is a neutron facility where the moderated neutron
source produces neutron beams for both nuclear physics and
material science.  Neutron production is again driven by a 3~GeV,
333~$\mu$A proton
beam (1 MW) impinging on a mercury target at 25~Hz \cite{KFH10,MHO10}.  

The Accurate Neutron-Nucleus Reaction measurement Instrument (ANNRI) is
a two-station flightpath with both a HPGe spectrometer and a NaI(Tl)
scintillator array at $\sim$22 and $\sim$28~m, respectively.  This is
the only spallation facility utilizing an HPGe array for capture
measurements.  This offers resolved gamma spectroscopy, but does not
offer the advantages of Q-value gating.  In addition to focused work on
transuranium isotopes, measurements on ANNRI has focused on measurements
for the main \emph{s}-process, including branch point $^{99}$Tc
\cite{KMI17} and $^{243}$Am \cite{KMT19}.

\subsubsection{Other Neutron TOF Facilities}

While these three white spallation sources have been briefly discussed,
it is worth noting several other facilities where work has been done or
continues.  Most notable is the Oak Ridge Electron Linear Accelerator,
which has now been shut down, but had provided neutron capture and
transmission measurements for over three decades \cite{BCW90}.  The GELINA
time-of-flight facility at the IRMM in Belgium supports a wide range of
neutron time-of flight measurements, also driven by an electron linac
and photo-induced neutron production \cite{MoS06}.  The Gaerttner
Laboratory at Rensseller Polytechnic Institute produces time of flight
neutron beams, primarily for nuclear engineering applications
\cite{rpi-web}.  As
mentioned above, the Karlsruhe facility, while still operational,
produced low-resolution time-of-flight beams over the 1-100~keV neutron
energy range.   The nELBE time-of-flight facility at HZDR (Dresden)
offers time-of-flight neutrons in the range of hundreds of keV to a few
MeV \cite{ABF07}.  While these facilities often do not have the
combination of the range of energy, neutron intensity, or detector systems of the
spallation sources discussed in more detail above, they still offer key
capabilities to address outstanding questions for neutron reaction
studies.

\subsubsection{Future Advances}

For TOF facilities, major facility advances generally come in the form
of higher intensity, improved resolution, or new detector systems.  In
addition, a capability driver for all facilities remains availability
and suitability of high-quality sample material, particularly for
radioactive samples.  Advances and collaborations in target preparation
have made unique measurements possible from existing facilities, whether
by mining old spallation targets for remaining radio-isotopes
\cite{URS09}, combining measurements across facilities to use the best
of each \cite{GLP20}, or pursuing dedicated sample development and
fabrication funding \cite{KFF12}.  

One example of a detector system upgrade is the Device for Indirect
Capture Experiment on Radionuclides (DICER), a newly developed instrument
at LANSCE, focused on neutron transmission measurements on extremely
small samples \cite{KUC21}.  While the concept of neutron transmission measurements
is not new, such measurements have typically required gram-sized samples
for measurements at keV energies.  DICER is designed to take advantage
of the fact that transmission only depends on the sample
\emph{thickness}, not the total number of atoms.  The brightness of the
Luj\'{a}n source at LANSCE makes it possible to perform measurements
with extremely small collimation---nominally 0.1~mm---to complete
measurements on $\mu$g-sized samples.  First measurements with DICER
have begun, and measurements on radio-isotopes are expected in the
coming years.

In concert with detector upgrades, source upgrades are also in the
works.  The Luj\'{a}n spallation target at LANSCE is scheduled to be
replaced with a redesigned target in 2022, offering a line-of-sight to
the W spallation target, which is expected to increase the keV flux and
resolution for both DANCE and DICER instruments \cite{ZMK18}.  The
planned FRANZ upgrades, in addition to activation measurements, will
also allow low resolution time-of-flight measurements, similar to those
performed at Karlsruhe, but with much great source intensities
\cite{RCH09}.  Finally, the planned SARAF-II upgrades include the
possibility of time-of-flight measurement capability, expanding the
reach of possible measurements and this extremely bright source beyond
activation measurements \cite{MAA18}.

%% file: Photon-Induced.tex
The nuclei in stellar burning environments exist within a bath of photons, given the plasma conditions associated with such extreme temperatures. Photodisintegration is therefore an important component of astrophysical nuclear reaction networks. However, $\gamma$-induced reactions in the laboratory often provide a limited picture of the $\gamma$-induced reactions occurring in stellar environments~\cite{2013RPPh...76f6201R}. While the laboratory reaction proceeds through the ground state of the target nucleus, a substantial fraction of stellar photodisintegration rates proceed through the excited states. As such, the role of photon beams in direct measurements for nuclear astrophysics is somewhat limited, often consisting of constraining statistical model parameters, e.g. Ref.~\cite{2019PhRvC..99b5802B}. Nonetheless, photon beams are an essential component of the nuclear astrophysics experimental toolbox, filling gaps left by direct measurements.

As shown in the earlier sections of this manuscript, the cross-sections for charged-particle capture reactions at energies relevant to the stellar nucleosynthesis are very small, in many cases too small to be measured through direct reactions. An alternative approach is to measure the inverse, photo-induced reaction and deduce the capture cross section via the principle of detailed balance. In such case, the reaction phase space provides an enhancement factor to the measured cross section of even two orders of magnitude. Additionally, photo-induced reactions result in much lower background than their charged particle equivalents, which further improves the resolving power of the experimental setup. 

At facilities such as HI$\gamma$S and ELI-NP, described in detail in the following subsections, the $\gamma$-ray flux is produced through Compton scattering of intense laser light with an electron beam. As a result a high-intensity flux $\gamma$-rays is produced at energies between 1-100 MeV. At energies relevant to stellar nucleosynthesis, additional collimation systems allow for precise selection of the beam energy allowing for selective excitation of nuclear levels.

\subsection{HI\texorpdfstring{$\gamma$}{g}S: High-Intensity \texorpdfstring{$\gamma$}{g}-ray Source}\label{sec:HIgS}

The High Intensity $\gamma$-ray Source (HI$\gamma$S) operated by the Triangle Universities National Laboratory (TUNL) is currently the world-leading $\gamma$-ray beam facility, producing an intense ($10^{3}$ photons/s/eV), nearly mono-energetic (bandwidth of 3-5$\%$), maximum energy of 100 MeV, highly polarized  $\gamma$-ray source dedicated to low and medium energy nuclear physics research \cite{Weller2009257}. The $\gamma$-ray beams are produced via the Compton backscattering process in which photons generated with a free-electron laser collide with high-energy electrons. Circularly or linearly polarized $\gamma$-ray beam bunches can be produced at HI$\gamma$S with a repetition rate of 5.58 MHz.

The HI$\gamma$S facility supports a broad research program in nuclear physics, including nuclear structure, nuclear astrophysics, and industrial applications. The main focus of the nuclear astrophysics program is measuring the astrophysical S-factor for the $^{12}$C($\alpha,\gamma$)$^{16}$O reaction closer to helium-burning energies, the importance of which is discussed in Section~\ref{sec:motivation}. 

The HI$\gamma$S optical time-projection chamber (O-TPC), aimed at studying the photo-dissociation of $^{16}$O and $^{12}$C, was commissioned more than 10 years ago \cite{2010JInst...5P2004G}. The device filled with CO$_2$ was used successfully to measure angular distributions for the $^{12}$C($\gamma,\alpha$)$^{8}$B reaction at several energies and identify a 2$^+$ resonance at 10.03 MeV in $^{12}$C \cite{2013PhRvL.110o2502Z}. Measurements of the $^{16}$O($\gamma,\alpha$)$^{12}$C reaction were carried out at $\gamma$-ray beam energies between 9.1 and 10.7 MeV during 2008-2009 \cite{2012JPhCS.337a2054G}. The gas target consisted of a mixture of CO$_2$(80$\%$) + N$_2$(20$\%$) at 100 Torr. The O-TPC has demonstrated the capability to measure angular distributions on an event-by-event basis, which is essential for separating the E1 and E2 contributions of the cross section. However, the analysis of the $^{16}$O($\gamma,\alpha$)$^{12}$C data is still in progress. The future effort to measure the $^{16}$O($\gamma,\alpha$)$^{12}$C reaction will be based on the recently proposed HI$\gamma$S-TPC, similar to the ELI-TPC described in Sec. \ref{sec:ELI-NP}.

Several other measurements, relevant to nuclear astrophysics, were carried out at HI$\gamma$S over the last few years. The Inventory-Sample Neutron Detector (INVS) consists of eighteen $^3$He proportional counters arranged in two concentric rings with radii of 7.24 and 10.60 cm \cite{2018PhRvC..98e4609S}. The INVS was recently upgraded to allow the readout of the proportional counters to be individually recorded with a sixteen-channel 500 MHz digitizer. In a recent experiment, Banu {\it et al.} measured the ($\gamma$,n) excitation function on the p-nuclei $^{94}$Mo and $^{90}$Zr from the neutron emission thresholds to about 13.5 MeV \cite{2019PhRvC..99b5802B}. \\

\subsubsection{Recent Highlights}

A measurement of the $^{7}$Li($\gamma,t$)$^{4}$He ground-state cross section between $E_\gamma$ = 4.4 and 10 MeV was recently performed at HI$\gamma$S \cite{2020PhRvC.101e5801M}. This was the first time a large-area segmented silicon detector array was used for a direct measurement with mono-energetic $\gamma$-ray beams. Considerable theoretical interest was shown over the last decade to calculate the capture cross section in the mirror $\alpha$-capture reactions $^{3}$H($\alpha,\gamma$)$^{7}$Li and $^{3}$He($\alpha,\gamma$)$^{7}$Be \cite{2011PhRvL.106d2502N,2016PhLB..757..430D}. However, while for $^{3}$He($\alpha,\gamma$)$^{7}$Be the calculations are in good agreement with recent measurements below 2.5 MeV center-of-mass energy \cite{2006PhRvL..97l2502B,2009PhRvL.102w2502D}, the calculated capture cross section for the mirror $^{3}$H($\alpha,\gamma$)$^{7}$Li reaction does not agree with the experimental data of Brune {\it et al.} \cite{1994PhRvC..50.2205B}.

The tritons and $\alpha$ particles resulting from the photodisintegration of $^7$Li were detected in coincidence by the SIDAR array of segmented silicon detectors. SIDAR was arranged in a lampshade configuration \cite{2001PhRvC..63f5802B} with twelve YY1 silicon detectors of 300, 500, and 1000 $\mu$m thickness. The coincidences were clearly separated from the beam-induced background for $\gamma$-ray beam energies above 6 MeV. However, at energies between 4.4 and 6 MeV, coincidences were identified only in a subset of the thinner detectors. 

\begin{figure}
\centering
\includegraphics[width=0.65\columnwidth]{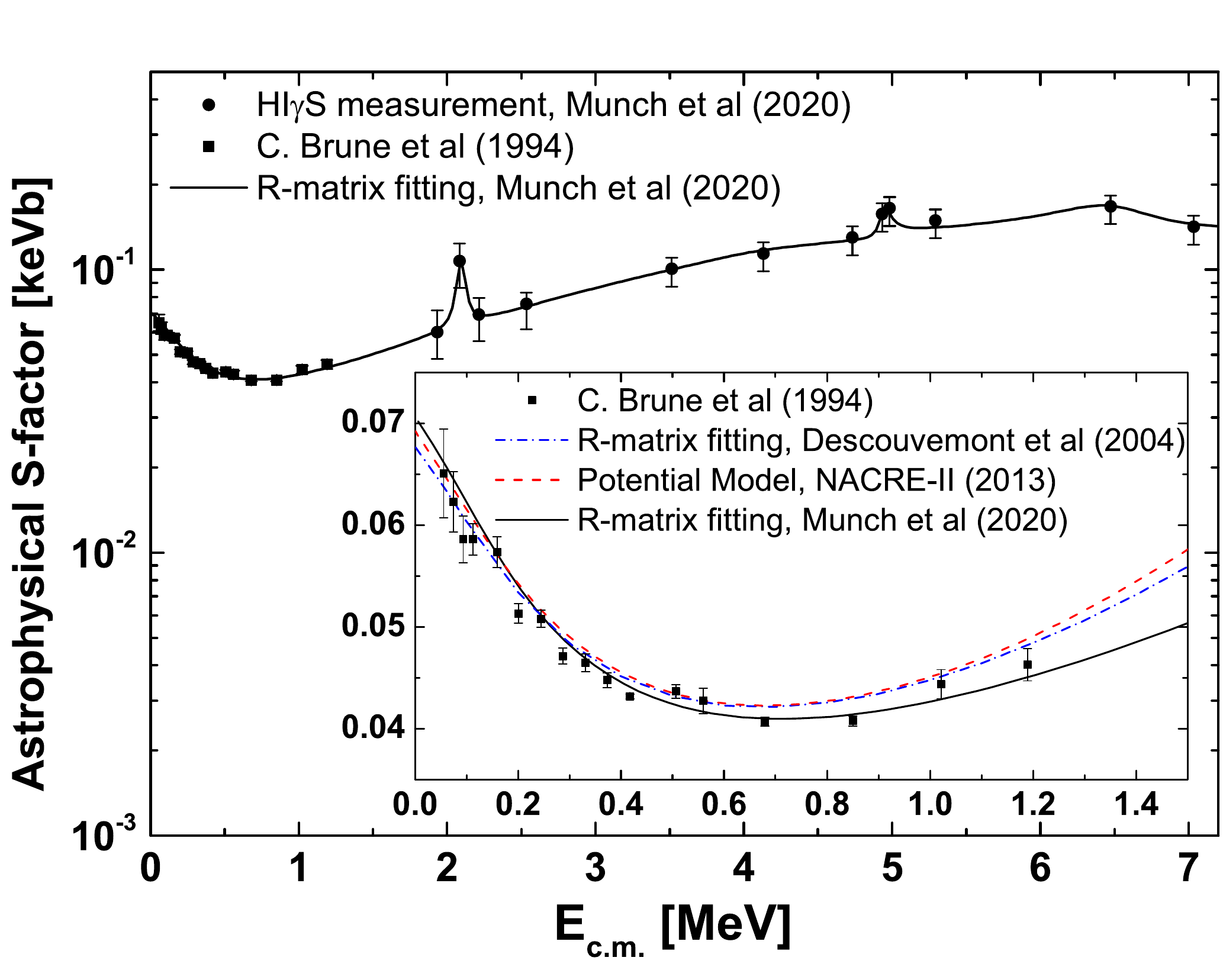}
 \caption{$R$-matrix fit of the ground state $S$ factor data from the recent measurement by Munch {\it et al.}. Inset: Details of the $R$-matrix calculation below $E_{c.m.}$ = 1.4 and comparison with experimental data of Brune {\it et al.} \cite{1994PhRvC..50.2205B} and the potential models of Descouvemont {\it et al.} \cite{2004ADNDT..88..203D} and NACRE-II \cite{2013NuPhA.918...61X}.} \label{fig:higs2}
\end{figure}

The calculated $^{7}$Li($\gamma,t$)$^{4}$He ground-state cross section from this measurement doesn't agree with previous bremsstrahlung experiments which were carried out in the 6 to 10 MeV energy range. The experimental astrophysical $S$ factor of $^{3}$H($\alpha,\gamma$) calculated from the present data was analyzed within the $R$-matrix formalism. The $R$-matrix extrapolation shown in Fig. \ref{fig:higs2} agrees with several potential model calculations and lower energy experimental data but its reliability below $E_{cm}$ = 1.2 MeV is limited due to large uncertainties in the experimental data. A thinner $^{7}$Li target and using silicon detectors of 100 $\mu$m thickness would allow the detection of $\alpha$ particles and triton coincidences down to previously measured data around $E_\gamma$ = 3.65 MeV.

A new measurement campaign using silicon-strip detectors of several reactions relevant to nuclear astrophysics was approved by the HI$\gamma$S Program Advisory Committee in 2019. The approved measurements include the $^{7}$Li($\gamma,t$)$^{4}$He ground-state cross section below $E_\gamma$ = 4.4 MeV and photodisintegration studies of the p-process nuclei $^{112}$Sn and $^{102}$Pd.  

\subsection{ELI-NP VEGA System}\label{sec:ELI-NP}

The Extreme Light Infrastructure-Nuclear Physics (ELI-NP) aims to use extreme electromagnetic fields for nuclear physics research \cite{2018RPPh...81i4301G}. The facility will operate two major installations: the 10 PW High Power Laser System (HPLS) and the Variable Energy Gamma (VEGA) system. The VEGA System, which will be operational in 2023, is based on a room-temperature linear accelerator coupled to a storage ring and a high-finesse Fabry-Perot cavity. Mono-energetic photon beams are produced via Compton backscattering of a laser beam off a relativistic electron beam. The high-brilliance narrow-bandwidth $\gamma$-ray beam will be delivered with energies up to 19.5 MeV, a spectral density higher than 5 $\times 10^{3}$ photons/s/eV, bandwidth of 0.5$\%$, and linear polarization higher than 95$\%$. Precise and accurate measurements of the $\gamma$-ray beam properties are required for the delivery of the beam within the design parameters but also to facilitate the ELI-NP scientific program. Several instruments to measure the spatial, spectral, and power properties of the $\gamma$-ray beam are in different stages of implementation at ELI-NP \cite{2016JInst..11P5025M,2019NIMPA.921...27T}.

ELI-NP will provide unique opportunities to experimentally study the photon-induced  ($\gamma$,n), ($\gamma$,p) and ($\gamma$,$\alpha$) reactions \cite{cpd2015,gant2015} with implications in a wide range of astrophysical scenarios, from Big Bang nucleosynthesis to explosive burning in the last stage of a massive star existence, to the elusive p-process nucleosynthesis. 

A flat-efficiency 4$\pi$ triple-ring detector, ELIGANT-TN, based on $^3$He proportional counters for measuring ($\gamma,n$) reactions for p-process nuclei was recently finalized at ELI-NP \cite{SODERSTROM2021109441}. The detection elements are 28 cylindrical counters (2.54 cm diameter and 49.5 cm length) of $^3$He at 12 bar of pressure. The counters are placed equally spaced in three concentric rings of 120, 260 and 310 mm diameter, containing 4, 8, and 16 detectors, respectively. The neutron moderator is a cube of 66 $\times$ 66 $\times$ 75 cm$^3$ made of high density polyethylene. GEANT4 and MCNP simulations have been used to calculate an efficiency around 38 $\%$ for neutrons below 2 MeV. Among the p-process nuclei, $^{180}$Ta and $^{138}$La will be measured with the highest priority \cite{gant2015}.

ELISSA (ELI Silicon Strip Array) is a silicon detector array in the final stages of implementation at ELI-NP. The array consists of 36 X3 position-sensitive silicon-strip detectors arranged into a three-ring barrel configuration \cite{cpd2015}. The X3 are 4-strip detectors 4 cm wide, position sensitive along the longitudinal axis (7.5 cm long), leading to an energy-dependent position resolution better than 1 mm \cite{2018JInst..13.5006C}. The angular coverage is extended by using an assembly of four QQQ3 or MMM segmented end-cap detectors. The experimental program with ELISSA includes studies of photodisintegration of light nuclei ($^{2}$H, $^{6}$Li, $^{7}$Li), and heavier nuclei for stellar burning ($^{24}$Mg) and p-process ($^{74}$Se, $^{78}$Kr, $^{84}$Sr, $^{92}$Mo, $^{96}$Ru) \cite{2018PhRvC..98e4601L}. 

An electronic-readout time projection chamber, ELITPC, is planned for studies of the multi $\alpha$-particle decay of light nuclei such as $^{12}$C and $^{16}$O and measurements of the cross section of astrophysically-relevant ($\gamma,p$) or ($\gamma,\alpha$) reactions. The chamber has an active length of 33 cm and a square cross-section of 20 cm $\times$ 20 cm, centered around the beam axis with thin windows for the $\gamma$-ray beam entry and exit. The electron amplification stage is achieved by three Gas Electron Multiplier (GEM) foils. The electronic readout is formed by three groups of non-orthogonal (u-v-w) grids with a total of 1024 read-out channels \cite{2020NIMPA.95461779G}. An FPGA-based customized DAQ module will read digitized signals from four GET electronics ASAD boards.\\

\subsubsection{The case for $^{16}$O($\gamma,\alpha$)$^{12}$C at ELI-NP}

Several measurements of the $^{12}$C($\alpha,\gamma$)$^{16}$O reaction were carried out in the vicinity of $E_{c.m.}$ = 1.0 MeV. However, the E1/E2 S-factors were determined with large uncertainties. The goal of the proposed measurements at ELI-NP with the ELITPC will be to measure detailed cross sections and angular distributions not only below $E_{c.m.}$ = 1.0 MeV but also at higher energies. One advantage of measuring the $^{16}$O($\gamma,\alpha$)$^{12}$C reaction is the enhancement by a factor of 100 of the cross section with respect to the inverse $^{12}$C($\alpha,\gamma$)$^{16}$O process \cite{XU2007866,2019PhRvC.100f5802H} at the same $E_{c.m.}$.

A simulation of the $^{16}$O($\gamma,\alpha$)$^{12}$C experiment was performed based on the the $\gamma$-ray beam parameters of the VEGA System at ELI-NP and the configuration of the ELITPC. Using the detailed balance principle, the cross section of the  $^{16}$O($\gamma,\alpha$)$^{12}$C reaction implemented in the simulation was calculated from the $^{12}$C($\alpha,\gamma$)$^{16}$O capture cross section extracted from a comprehensive R-matrix analysis \cite{2017RvMP...89c5007D}. Note that only the ground state of $^{16}$O is accessible in a measurement of the $^{16}$O($\gamma,\alpha$)$^{12}$C reaction. Both the E1 and E2 contributions were considered. The simulation was performed with an energy bin width of 0.1 MeV, for an experiment running for one week, and a 100$\%$ efficiency for particle detection. 

\begin{figure*}
\centering
\includegraphics[width=0.45\columnwidth]{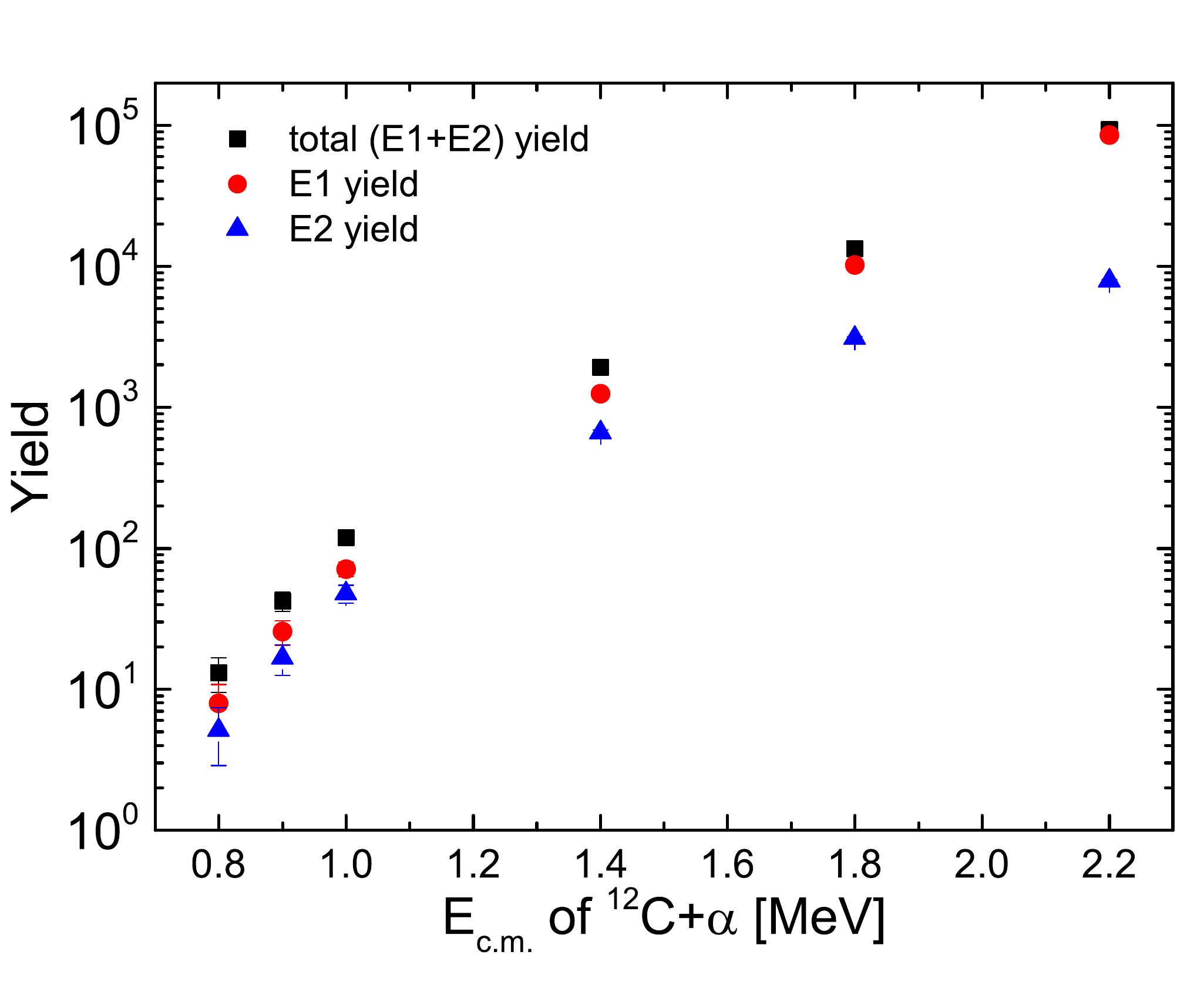}
\includegraphics[width=0.45\columnwidth]{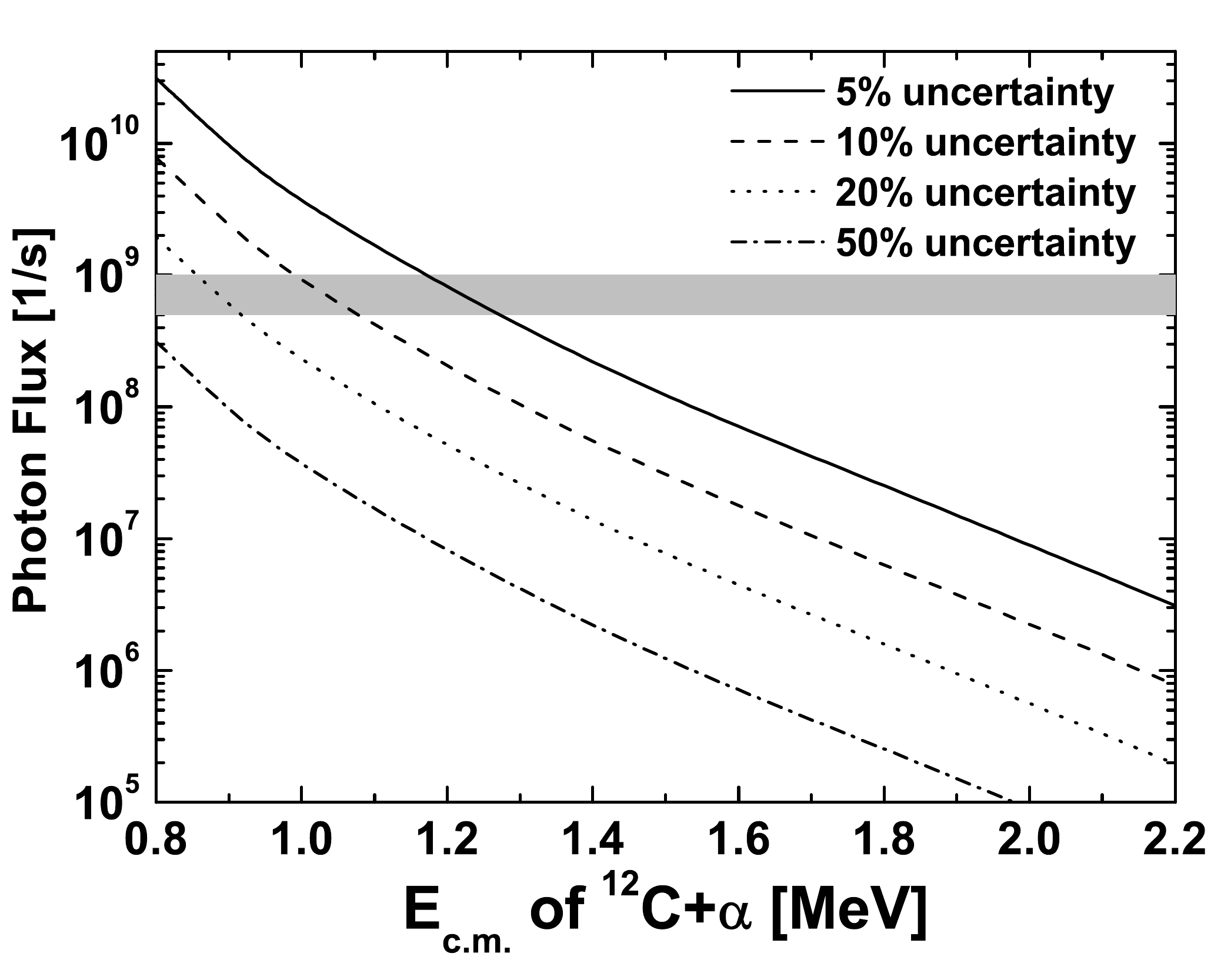}
 \caption{a) The $\alpha$-particle simulated experimental yields from the $^{16}$O($\gamma,\alpha$)$^{12}$C experiment at ELI-NP in terms of $E_{c.m.}$ of the $^{12}$C($\alpha,\gamma$)$^{16}$O reaction. The results are given in terms of E1, E2 and total (E1+E2) contributions. b) The total flux of the $\gamma$-ray beam at ELI-NP and different statistical uncertainties (5$\%$, 10$\%$, 20$\%$ and 50$\%$) for the calculated $^{12}$C($\alpha,\gamma$)$^{16}$O reaction cross section} \label{fig:eli-np1}
\end{figure*}

The $\alpha$-particle simulated experimental yields from $^{16}$O($\gamma,\alpha$)$^{12}$C at ELI-NP were obtained in terms of the incident $\gamma$-ray beam energies $E_{\gamma}$ that are converted to the corresponding $E_{c.m.}$ of $^{12}$C($\alpha,\gamma$)$^{16}$O using $E_{c.m.}$ = $E_{\gamma}$ - $Q$ ($Q$ the Q-value of $^{12}$C($\alpha,\gamma$)$^{16}$O). The simulation results are plotted in Fig. \ref{fig:eli-np1} for the E1, E2 and total (E1+E2) contributions. For one week of beam time, the total yield of the $\alpha$ particles is anticipated to reach about 10 at $E_{\gamma}$ = 7.96 MeV and 100 at $E_{\gamma}$ = 8.16 MeV. These two incident $\gamma$-ray beam energies correspond to $E_{c.m.}$ = 0.8 MeV and $E_{c.m.}$ = 1.0 MeV for $^{12}$C($\alpha,\gamma$)$^{16}$O, respectively.

The simulation of the $\alpha$-particle experimental yields from the $^{16}$O($\gamma,\alpha$)$^{12}$C experiment at ELI-NP also reveals the achievable statistical uncertainty with the $\gamma$-ray beam flux produced by the VEGA System. The gray band in Fig. \ref{fig:eli-np1}, indicating the range of the total photon flux available at the VEGA System at ELI-NP with a 0.5$\%$ bandwidth, intersects different statistical uncertainties (5$\%$, 10$\%$, 20$\%$ and 50$\%$) calculated for the $^{16}$O($\gamma,\alpha$)$^{12}$C reaction. Fig.~\ref{fig:eli-np1} shows that the experimental cross section of $^{12}$C($\alpha,\gamma$)$^{16}$O reaction can be measured with 20$\%$ statistical uncertainty at $E_{c.m.}$ = 0.9 MeV during one week of beam time. Furthermore, the beam time can also be increased to lower the statistical uncertainty or reach a statistical uncertainty less than 50$\%$ at $E_{c.m.}$ = 0.8 MeV. This will be a great significance because no experimental data of $^{12}$C($\alpha,\gamma$)$^{16}$O is available below $E_{c.m.}$ = 0.9 MeV (See Section~\ref{sec:HeBurning}). However, continuously increasing the running time is not a practical approach for $E_{c.m.}$ below 0.8 MeV. Overall, a measurement of the $^{16}$O($\gamma,\alpha$)$^{12}$C reaction down to $E_{c.m.}$=0.8 with the ELITPC at ELI-NP could mark a significant improvement in our understanding of the $^{12}$C($\alpha,\gamma$)$^{16}$O reaction at the lower energy range.

Although the majority of the $^{12}$C($\alpha,\gamma$)$^{16}$O experimental and theoretical work over the last 50 years has concentrated on the energy region below $E_{c.m.}$ = 5 MeV, availability of accurate $^{12}$C($\alpha,\gamma$)$^{16}$O experimental data at higher energies up to $E_{c.m.}$ = 10 MeV would allow to include more states in the R-matrix analysis and to reduce the uncertainty of the extrapolation at lower energies. Therefore, another aim of the $^{16}$O($\gamma,\alpha$)$^{12}$C measurement based on the ELITPC at ELI-NP would be to determine the angular distributions and the cross sections for $^{12}$C($\alpha,\gamma$)$^{16}$O between known resonances at $E_{c.m.}$ ranging from 3 to 10 MeV. For example, in order to determine the $^{12}$C($\alpha,\gamma$)$^{16}$O cross sections between the 9.84 MeV ($2^{+}$) state and the 11.52 MeV ($2^{+}$) state in $^{16}O$ with the statistical precision of 3$\%$, a beam time of ten hours is required for the measurement of $^{16}$O($\gamma,\alpha$)$^{12}$C.

%% file: Conclusion.tex
Direct measurements continue to play an essential role in nuclear astrophysics, reaching down into the Gamow window and, even when this is not possible, providing important data for extrapolation from higher reaction energies. As the previous sections illustrate, these measurements involve a diverse and complementary set of tools and techniques. Underground laboratories provide shelter from sea-level backgrounds, enabling many of the lowest-energy measurements. Recoil separators circumvent the background problem altogether by focusing on the recoil in lieu of particle and $\gamma$-ray detection. Storage rings circumvent the beam intensity limitations of some recoil separator measurements by recycling unreacted beam. Neutron beams provide the only direct path to measure the neutron-induced reactions that play such a prominent role in heavy-element nucleosynthesis. Meanwhile $\gamma$-beams provide a unique avenue to access inverse reactions of astrophysical interest. Together, along with more traditional sea-level direct measurement facilities that we do not highlight in this article, these approaches provide the gold-standard nuclear physics data desired in astrophysics model calculations.

Most of the facilities and techniques described above have come online only in the last couple of decades and have yet to be optimized to their full potential. We can expect several exciting developments in the near future. 

First science from the CASPAR facility has demonstrated that the higher intensity and higher voltage accelerator concept DIANA would be a major leap forward for underground nuclear astrophysics, extending the range of possible measurement energies to both push further into astrophysically relevant energies and extend to the higher energies probed in sea-level labs. The recently realized JUNA laboratory in Sichuan and LUNA-MV upgrade to the laboratory in Gran Sasso will assume these tasks, while CASPAR will focus on developing new complementary techniques. The three laboratories will continue to foster the collaborative, healthy international competition that ensures high-fidelity physics results that address the many open questions in stellar burning.

The DRAGON recoil separator continues to push the boundaries of what this device can accomplish, expanding to higher masses and different reaction types than were originally anticipated. Meanwhile, the world's other recoil separators are beginning to deliver first results. The pioneering device ERNA has recently been resurrected at the CIRCE lab in a new an improved form. In the Midwestern United States, St. George has completed proof of principle measurements and the final upgrades required to meet design performance, while its descendent SECAR is entering the final phases of commissioning. This suite of separators will feature complementary capabilities, each specializing in reactions involving different characteristics for the ion beams, recoils, and associated light ejectiles.

The expansion of the storage ring ESR's capabilities from one of the world's most prolific precision mass measurement devices to include in-ring reaction measurements demonstrates the advances that can be made by ingenious upgrades to existing scientific equipment. ESR has advanced the science of in-ring reaction measurements to the point of approaching the Gamow window. The downstream storage ring CRYRING will build on these achievements by enabling measurements well within the Gamow window. By recycling the incident ion beam, the intensity gains achieved by these ring measurements complement the wider recoil acceptance achieved by separators.

Neutron beams for nuclear astrophysics have been around for more than 50 years, with a correspondingly large collection of physics results. Present and near-future advances at the FRANZ and SARAF facilities will achieve unprecedented beam intensities, enabling measurements on minute sample sizes for rare and short-lived samples. With the same end in mind, detector development at LANSCE will play a substantial complementary role. Ultimately these approaches combined will help close the door on the last remaining questions involving the origin of heavy elements in $s$-process nucleosynthesis. 

HI$\gamma$S pioneered the use of intense photon beams as a unique probe of astrophysically interesting nuclear reactions, making important contributions to nuclear astrophysics as well as detector technology. This pioneering work continues. Meanwhile, the ELI-NP facility comes closer to completion in Romania. The expanded energy range and increased intensity relative to its predecessor promise to provide key input to stellar burning reactions that are resistant to experimental progress, such as $^{12}{\rm C}(\alpha,\gamma)^{16}{\rm O}$.

New developments in direct measurements will result in great strides towards answering major open questions in stellar burning, some of which were touched on in Section ~\ref{sec:motivation}. Though the main story of stellar burning and nucleosynthesis has long been established, our current view is more of a general outline than a detailed description. In the following we recapitulate some major open questions and make connections to near-future developments in nuclear physics experiment.

What are the reaction rates for the core-fusion reactions $^{12}{\rm C}(\alpha,\gamma)$ and $^{12}{\rm C}+^{12}{\rm C}$? While decades of research has come a long way toward providing the answer, it has also made it clear that further dedicated additional high-precision and low-background studies will be required. New underground laboratories and the availability of higher-intensity $\gamma$-beams will be indispensable in this regard.

What is the full set of reactions that provides neutrons for the $s$, $i$, $n$, and weak-$r$ alphabet soup of nucleosynthesis processes? As important, what are the strengths of competing reactions and neutron sinks, each of which would rob an environment of a robust neutron flux? Virtually all of the techniques mentioned in this article will play an indispensable role here. Underground laboratories will provide the low backgrounds needed to directly measure neutron detection for light ion reactions of interest, while separators and rings will work to circumvent the background problem  and use complementary information to arrive at a reaction cross section. Neutron sources will enable direct measurements of the neutron-induced reactions requiring further refinement, while intense $\gamma$-beams will provide a complementary probe in the inverse direction. It is hard to imagine that such a multi-pronged international approach will not transform our understanding of these astrophysical phenomena in the coming decades.

Where are the elements heavier than iron made? The majority of this nucleosynthesis is split between the $s$-process and $r$-process, but precise yield estimates are not yet possible and these yield estimates are intertwined. Inferring the $r$-process pattern from our Sun requires precise knowledge of $s$-process yields that our models have not quite achieved. Higher precision nuclear data from neutron and $\gamma$-beam facilities for $s$-process branching points will break the degeneracy between astrophysics model calculations and shed light on the relevant astrophysical conditions.

Progress in direct measurements for stellar burning will inform the nuclear science and nuclear measurement techniques associated with explosive burning processes. Indeed, nearly all of the facilities and techniques mentioned above have branched into the measurement of radioactive ions of interest for various astrophysical processes operating near stability, such as the $rp$-process of X-ray bursts and novae, shock-driven nucleosynthesis and the $p$-process of core collapse supernovae, and the $i$-process of rapidly-accreting white dwarf stars. The synergy will continue to be particularly strong for medium-mass nuclides, where a solid foundation of statistical nuclear properties on and near the valley of $\beta$-stability is necessary to make accurate inferences for more exotic nuclides. Thus progress in stellar burning will help inform questions such as: what powers transient phenomena in the night sky and how can we explain the fine features of the cosmic abundances not explained by the main nucleosynthesis processes?

Answering these and related questions will only be possible through concerted international efforts. Coordination helps avoid the repetition of mistakes while also ensuring the repetition required for validation. Collaboration supports the intellectual stimulation needed to bring creative solutions to bear on long-standing problems. Though international connections can naturally occur in a piecemeal fashion, such a chance-based approach lacks the inclusivity, staying power, and broad communication power of a formal international research network.
Networks provide the glue needed to ensure that the latest data, complemented by state-of-the-art theory interpretations, are incorporated into comprehensive evaluations, results from evaluations are incorporated into the latest astrophysics model calculations, and measurements are in turn motivated by the latest questions posed by model calculation results. Supported by these networks, nuclear astrophysics direct measurements will continue to solve the mysteries presented by the stars in our universe.

%% file: Acknowledgements.tex
We thank the participants of the 2020 IReNA Virtual Workshop on Stellar Burning for valuable discussions and the research networks that made this possible: International Research Network for Nuclear Astrophysics (IReNA), Joint Institute for Nuclear Astrophysics -- Center for the Evolution of the Elements (JINA-CEE), Chemical Elements as Tracers of the Evolution of the Cosmos (ChETEC), the ExtreMe Matter Institute (EMMI), and Nuclear Astrophysics at Rings (NucAR).
The research described in this report was made possible in part by funding from the US Department of Energy through Grants No. DE-FG02-88ER40387 and DE-SC0019042; the US National Nuclear Security Agency through Grant No. DE-NA0003909; the US National Science Foundation through grants No. PHY-1430152 (JINA-CEE) and OISE-1927130 (IReNA); 
%the Deutsche Forschungsgemeinschaft through Contracts No. \dots, \dots;
%the European Research Commission through Grants No. \dots, \dots; 
 and the European Cooperation in Science and Technology through COST Action CA16117 (ChETEC).
MA acknowledges financial support by EMMI for a Visiting Professorship at GSI and Goethe Universit\"at Frankfurt.
A.C. acknowledges support by the Laboratory Directed Research and
Development program of Los Alamos National Laboratory under project
number 20190021DR.  Los Alamos National Laboratory is operated by Triad
National Security, LLC, for the National Nuclear Security Administration
of U.S. Department of Energy (Contract No. 89233218CNA000001). R.R. acknowledges support from the Hessian collaborative research cluster ELEMENTS and from the European Union (ChETEC-INFRA, project no. 101008324). A.T. acknowledges support from the European Union (ChETEC-INFRA, project no. 101008324). Y. X. acknowledges supports from the Romanian Ministry of Research
and Innovation (Nucleu, PN 19 06 01 05) and the Institute of Atomic
Physics (ELI-RO, ELI 15/16.10.2020). 
YAL and JG acknowledge support by the European Research Council
(ERC) under the European Union's Horizon 2020 research and innovation programme (grant agreement No 682841 ``ASTRUm'').